# METHODOLOGIES FOR IMAGING A USED NUCLEAR FUEL DRY STORAGE CASK WITH COSMIC RAY MUON COMPUTED TOMOGRAPHY

A Dissertation Presented for the

Doctor of Philosophy

Degree

The University of Tennessee, Knoxville

Zhengzhi Liu

August 2018

# ACKNOWLEDGEMENTS

Life is a journey of ups and downs. This is also true of doing research. This dissertation would not have come into being without the kind help and support received from many people. My first and foremost appreciation goes to my mentor and academic advisor Prof. Jason P Hayward for guiding me through the tough times of my graduate study. As a Chinese proverb said "Give a man a fish and you feed him for a day. Teach a man to fish and you feed him for a lifetime." Prof. Hayward's incisive and heuristic intellectual guidance deeply inspired me to unearth the truth and find the resolution to the problems. His scientifically rigorous research attitude shaped my way to treat studies, experiments and research. His unwavering patience and accessible personality restored my confidence again and again at challenging times, just like a father who would never forsake his son. I am forever grateful for Prof. Haywards' teaching, guidance, help and the innumerous amount of time he has spent with me.

I would also like to sincerely thank my co-mentor Dr. Chatzidakis Stylianos for his selfless guidance and help. He encouraged me to think out of the box, looking at the problems from novel angles. I am deeply grateful for his encouragement and pushing me forward, which significantly fostered my intellectual growth. In life, he is one of my best friends. I am honored to be friends with both him and his wife Sonia.

Thanks goes to my other committee members, Prof. Laurence F. Miller, Prof Lawrence H. Heilbronn, and Prof. Hairong Qi for teaching me in relevant classes and providing academic advice and input in this dissertation.

Thanks also to Prof. Haori Yang and graduate student Can Liao from Oregon State University for providing me a summer internship chance in 2016 on large area muon position sensitive detector design and fabrication, and also for the enjoyable collaboration on this research project. Also, thanks to Dr. Chatzidakis Stylianos at Oak Ridge National Laboratory for the precious summer internship opportunity in 2017.

Additionally, thanks to many graduate students inside and outside of our research group who helped me on different fronts but especially to: Dr. John Sparger, Mitchell Laubach and Alireza Rahimpour. Thanks to University of Tennessee and the Department of Energy DE-NE0008292 for providing me with monetary support during this work.

Finally, and most importantly, I want to especially thank my friend Shixiao Yu from Oregon State University for her encouragement, love and teaching me how to be mature. I will forever remember every second we spent together, including the morning runs in the village, our study together on the bank of the Corvallis river, our camping and sky watching on the San Juan and Orcas islands, and the kayaking we did in the North Pacific Ocean, which was the most enjoyable time so far in my life. These memories will always be fresh. She has helped to transform me to be a better man since the second I met her. Thank you, Alicia (Shixiao's English name), and I wish you all well.

# ABSTRACT

It's important to the International Atomic Energy Agency (IAEA) to develop a nondestructive assay technique that may that be used to verify the presence of the used nuclear fuel stored in a dry storage cask once continuity of knowledge has been lost. X-rays and neutrons are not good candidates for assay because they do not penetrate dry storage casks with high probability, and gammas and neutrons are also emitted by the used nuclear fuel. In contrast, cosmic ray muons are naturally occurring highly penetrating particles. Muons interact with matter via two major interaction mechanisms: ionization and radioactive process, and multiple Coulomb scattering leading to energy loss and trajectory deflection, respectively. For a monoenergetic muon beam crossing an object, the scattering angle follows a Gaussian distribution with a zero mean value and a variance that depends on the atomic number of the material object it traversed. Thus, the measured scattering angle may be used to reconstruct the geometrical and material information of the contents inside the dry storage cask.

In traditional X-ray computed tomography, the projection information used to reconstruct the attenuation map of the imaged objects is the negative natural logarithm of the transmission rate of the X-rays, which is equal to the linear summation of the X-ray attenuation coefficients along the incident path. Similarly, the variance of the muon scattering angle is also the linear integral of the scattering density of the objects crossed by the muons. Thus, a muon CT image can be built by equating scattering density with attenuation coefficient. However, muon CT faces some unique challenges including: 1) long measurement times due to low cosmic muon flux, 2) insufficiently accurate muon path models, and 3) the inability to precisely measuring muon momentum.

In this work, three different muon path models, two different projection methods, and two different reconstruction methods were investigated for use in muon CT of dry storage casks. The investigation was conducted in a validated Geant4 workspace, both in an ideal case and with relevant engineering restrictions considered. The results of these investigations and the expected benefits for fuel cask monitoring are reported herein.

# TABLE OF CONTENTS

# LIST OF TABLES

# LIST OF FIGURES

# 1   INTRODUCTION

## 1.1   Background and motivation

Spent nuclear fuel (SNF) is nuclear fuel that has been irradiated in nuclear reactor. 96% of the content in SNF remains to be the original $^{238}$U and normally $^{235}$U accounts for less than 0.83%. Although spent nuclear fuel is not able to continue sustaining nuclear reactor, the fission products $^{235}$U and $^{239}$Pu may be separated and purified for nuclear weapons use. After being dismantled from a nuclear reactor, spent nuclear fuels are still highly radioactive and continue to decay generating heat and radioactive rays. They are typically moved to a pool of more than twenty feet of water to cool down on-site at the nuclear power plant after being removed from the nuclear reactor. The water is kept circulating to take the decay heat away from the spent nuclear fuel. Because of the lack of permanent geological repository for existing spent nuclear fuel in United States and the policy decision to ban reprocessing of commercial spent nuclear fuel [1], nuclear reactor owners have to store spent nuclear fuel in water pool on the reactor sites. Due to the growth of nuclear power industry, spent nuclear fuel inventory on commercial reactors site has been increasing and approached its limitation of the capacity of wet storage. On top of that, safety concern is another main reason to divert spent nuclear fuel from water pool. Malfunction, natural disaster or terrorist attack may cause the leakage of water in the pool. If it is not replenished with water quickly enough, the spent nuclear fuel will be uncovered eventually. Then decay heat can't be taken away and is very likely to cause meltdown. Thus, after cooling down, spent nuclear fuel would be moved out of water pool, placed in dry storage cask (DSC) [2] on-site or transferred to independent spent fuel storage installations (ISFSIs). The ISFSIs were initially licensed as temporary facilities for ~20 years period. Given the cancellation of the Yucca Mountain project [3]and no clear path forward, extension of dry cask storage period (~100 yr.) at ISFSIs is very likely to happen. From the point of view of the nuclear material protection, accountability and control technologies (MPACT) campaign, it is important to ensure that special nuclear material (SNM) in SNF facilities is not stolen or diverted from civilian facilities for other use during the transportation and extended storage period. Once dry storage cask is sealed, it is prohibitively expensive to open it for visual check of spent nuclear fuel inside and then reseal it. Thus, it's imperative to develop a nondestructive assay technique (NDA) to find out whether there is spent nuclear fuel missing or replaced with dummy material in the dry storage cask, especially when the continuity knowledge gets lost.

## 1.2   Previous work on imaging dry storage casks

There has been a couple of methods put forward to image dry storage cask using different particles to achieve the goal of detecting spent nuclear assemblies in the dry storage casks so far. A lot of the work has been simulation-based. According to the source term, these methods can be roughly categorized into two catalogs: 1) methods using manmade source, like x-ray, neutron, gamma beam, and 2) methods utilizing cosmic ray muons. The Missouri University of Science and Technology simulated the imaging spent nuclear fuel dry storage cask with high energy X-ray using computed tomography technique [4]. The University of Florida did high energy neutron transmission analysis of a dry cask storage using MCNP simulated data [5]. Purdue University did the muon radiography simulation on dry storage cask with partial fuel assemblies missing using simple Point of Closest Approach (PoCA) method [6]. Los Alamos National Laboratory (LANL) has actually done the first physical experiments of measuring dry cask [7][8] at Idaho National Laboratory (INL) using muon radiography. Later LANL also did muon computed tomography[9] simulation of a MC-10 dry storage cask.

### 1.2.1 Imaging dry storage casks with high energy X-ray computed tomography

At Missouri University of Science and Technology, researchers proposed using high energy X-ray to image dry storage cask based on traditional x-ray computed tomography technique. A 6 MV Linac can be used to generate 4 MeV bremsstrahlung X-ray at a flux rate of $1\times10^{15}$ photons/s. In their simulation, an X-ray source and a scintillator-based detector array were placed on the lateral side of a TN-24 dry storage cask and rotated simultaneously for a certainty number of times to cover 360 degrees. The overall experimental configuration is shown in Figure 1.1

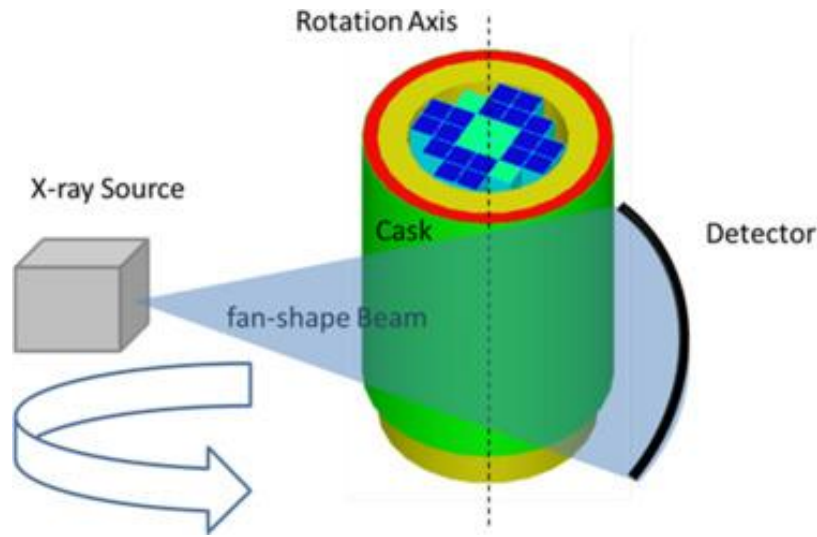

Figure 1.1        Configuration of imaging dry storage cask with high energy x-ray (from [4], used with permission)

To simplify the simulation process, the authors did not use any Monte Carlo based simulation package to simulate the process of high energy X-ray crossing the dry storage cask, instead they built numerical phantom with the dimensions and materials of the dry storage cask to simulate the transmission rate of 4 MeV X-rays. Perfect conditions were assumed, i.e., no scatter or statistical noise. The X-ray source and detector were rotated simultaneously 1000 times at an incremental angle of 0.36 degree to register the number of traversed X-ray photons at each view. Each view took about 1 hour. Both filtered back-projection and iterative method were used to reconstruct the phantom with registered data. An 'anatomical' reconstructed image of a TN-24 metal cask is shown in Figure 1.2. The reconstructed image is expected to enable one to identify all major components in the dry storage cask, even can resolve fuel rods. However, the exposure time could be longer than 41 days. In order to shorten the exposure time, the most straightforward way is to reduce the number of views.

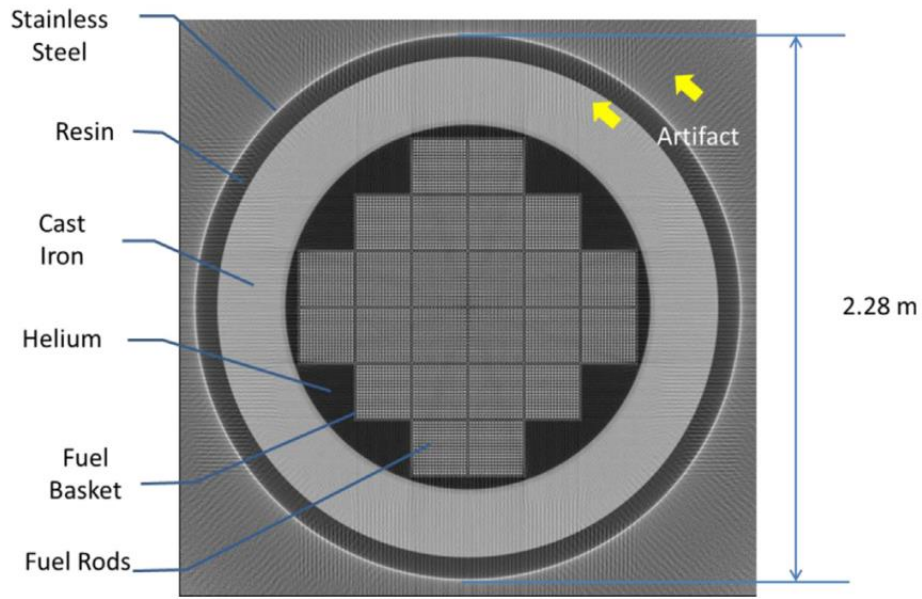

Figure 1.2        Tomographic reconstructed cross-section view of a TN-24 metal cask. The reconstruction is performed with 1000 views of 4 MeV X-ray beam. (from [4], used with permission).

Region of interest (ROI) scans and iterative reconstruction algorithm [10][11] can be alone or together to achieve a maximum reduction in the number of views without losing reconstructed image quality. Due to the existence of prior knowledge of the dry storage cask, Prior Image Constrained Compressed Sensing (PICCS) [12] can be adopted to accurately reconstruct the dry storage cask with significantly fewer views. A comparison of image quality of reconstructed dry storage cask with fuel rod damage using regular iterative reconstruction algorithm and PICCS algorithm is shown in Figure 1.3. Figure 1.3 (a), (b), (c), (d) represent the true image of the cask cross section with a fuel rod, prior image – a cask image under normal condition, reconstructed image with 180 views using traditional iterative reconstruction algorithm, and reconstructed image with 180 views using PICCS algorithm. It can be seen that Figure 1.3 (d) has better contrast resolution than Figure 1.3 (c) due to the increased signal to noise ratio, which indicates that PICCS reconstruction algorithm can effectively reduce the number of views and improve image quality. A quantity of 180 views corresponding to 7.5 days exposure are expected to be enough to resolve fuel rod damage when the PICCS algorithm is used.

After incorporating the Compton scattering and photon nuclear effect, the result may be significantly degraded even with anti-scatter grid. In order to achieve similar reconstructed image quality, either more views or a long exposure for each view will be required. Due to the strong X-ray and neutron radiation emitted from spent nuclear fuel assemblies, the long exposure time, and the use of expensive Linac, a cheaper and faster way would be much preferred to image spent nuclear fuel dry storage casks.

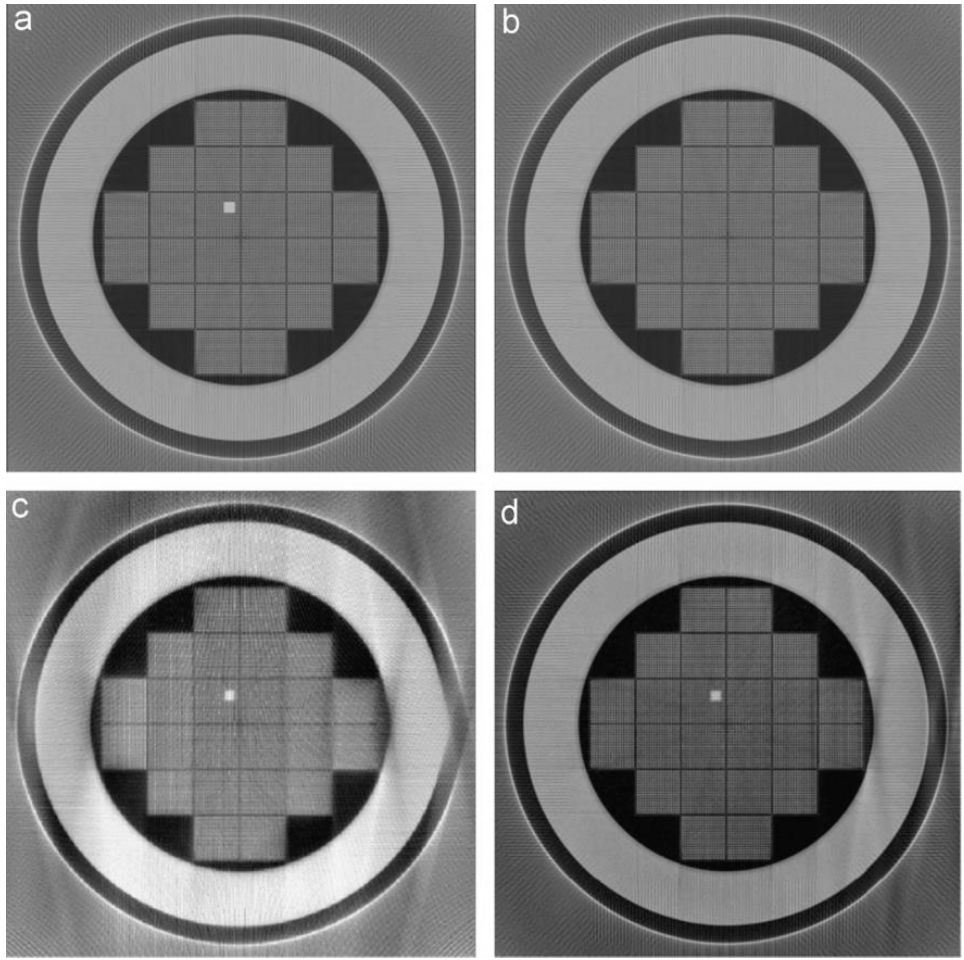

Figure 1.3       Comparison of image quality for regular iterative reconstruction algorithm and PICCS algorithm (from [4], used with permission).

## 1.2.2   Imaging dry storage casks with muon radiography

LANL did a muon radiography experiment on a MC-10 dry storage cask at INL with drift tube detectors [7][8]. Each drift tube is about 121.92 cm wide and long, and 60.96 cm thick. Two identical drift tube tracking detectors were placed on opposite side of the MC-10 dry storage cask with one detector elevated by 1.2 m relative to the other to increase the muon flux rate crossing two muon trackers. The experimental configuration is shown in Figure 1.4. Two muon tracking detectors were aligned with fuel assemblies. The dry storage cask loading profile is shown in Figure 1.5.

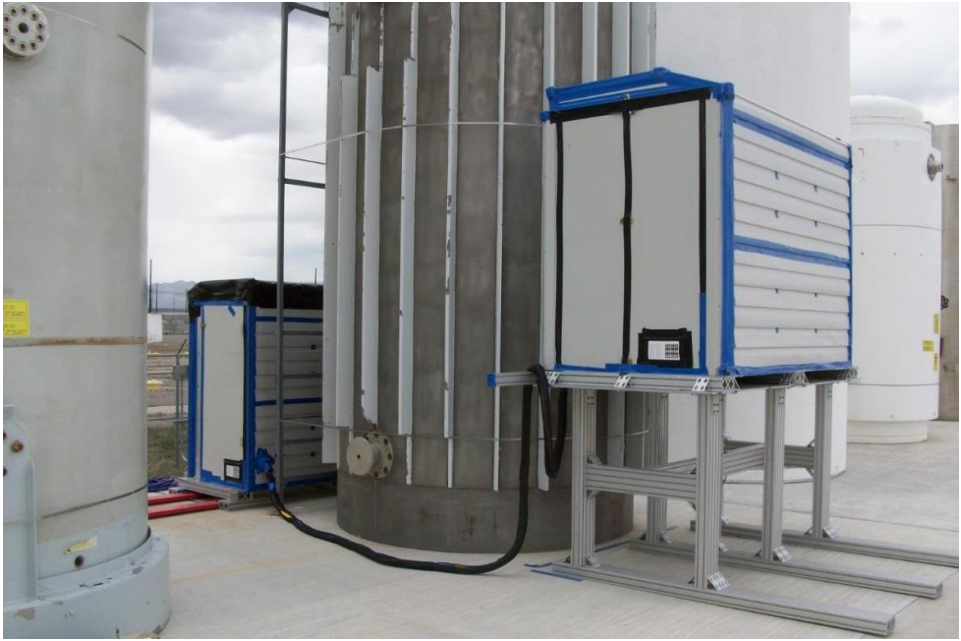

Figure 1.4        Setup of the physical experiment at INL by LANL (from [7], used with permission)

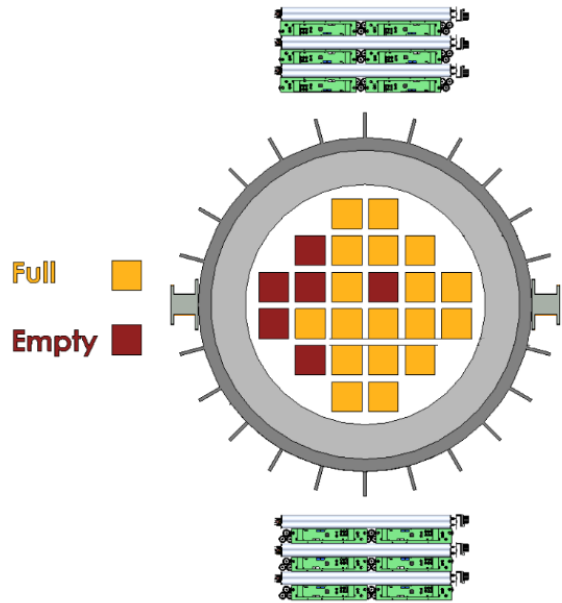

Figure 1.5        Spent nuclear fuel missing pattern on the left and reconstructed image on the right (from [7], used with permission).

From left to right, there are 0, 1, 6, 5, 4 and 2 spent nuclear fuel assemblies in column 1 to 6. Because the width of the muon tracking detectors is smaller than the diameter of the MC-10 cask, the field of view of these two muon tracking detectors is limited to columns 2, 3, 4 and 5. Despite the heavy shielding of the dry storage cask wall, there is still a strong radiation field outside the cask. About 10 mrem/hr of neutrons and 10 mrem/hr of gamma rays activity exist on contact at the cask surface.

To reduce the false muon coincidence rate, a trigger was applied to the data acquisition system that required hits in these two muon trackers within a time window of 600 ns to be counted as a true muon event. The exposure lasted about 200 hours and $1.62 \times 10^5$ muons were registered passing both detectors. The scattering angers of these muons were projected into a plane parallel to both muon detectors near the center of the dry storage cask. The plane was discretized into 2 cm $\times$ 2 cm pixels. Then multigroup method [13] was used to fit the scattering angle in each pixel to extract the areal density of the corresponding volume in the dry storage cask. The areal density of the dry storage cask within the field of the muon detectors is shown at left in Figure 1.6. Because both spent nuclear fuel assemblies and the MC-10 dry storage cask are nearly uniform along the vertical direction, the areal density can be summed up to enlarge the differences between neighboring columns. The vertically integrated areal density is shown at right in Figure 1.6. From the image of cask in terms of areal density, it can be seen that there are fuel assemblies missing in column 2, but one cannot tell how many fuel assemblies were missing in the column. Even from the vertically integrated areal density, it is not very clear whether there was one spent nuclear fuel assembly missing in column 4.

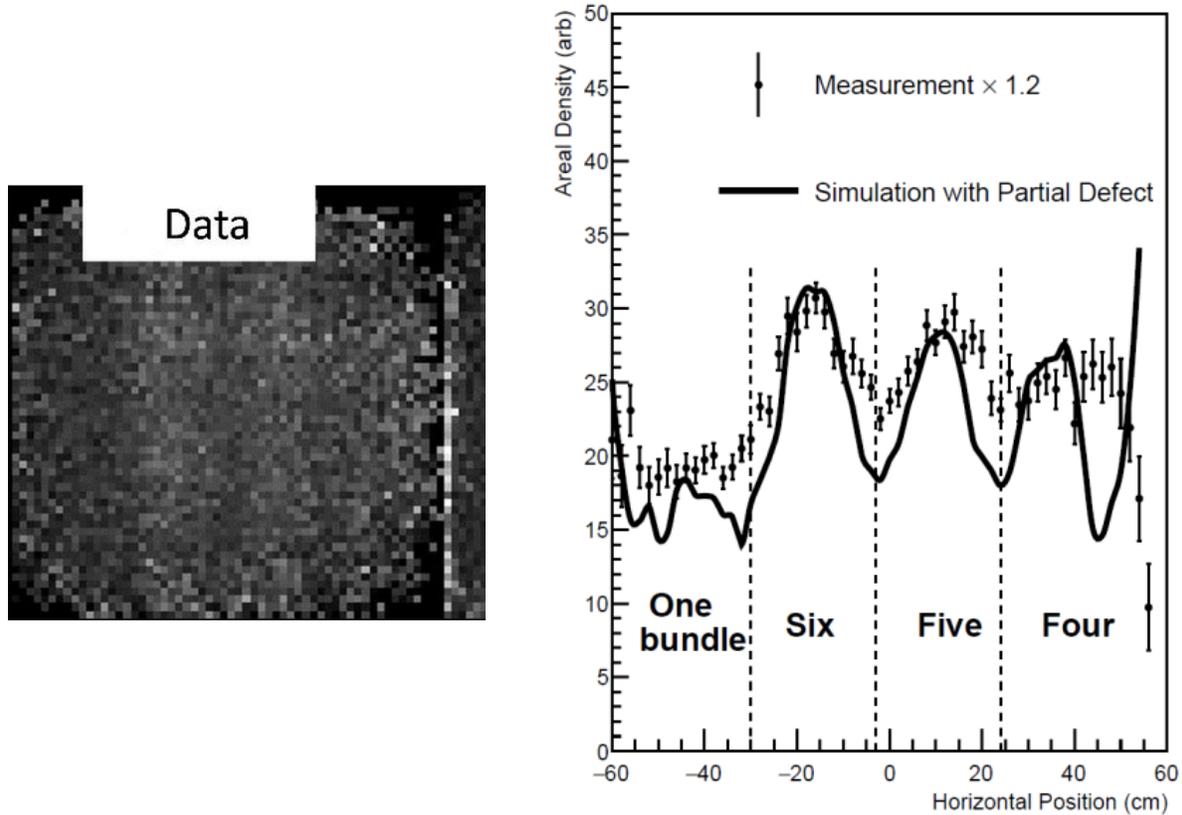

Figure 1.6        Image of cask in terms of areal density on the left and vertically integrated areal on the right (from [7], used with permission)

### 1.2.3    Imaging dry storage casks with muon using a simple PoCA algorithm

As a heavy charged particle, muons would experience multiple coulomb scattering when crossing an object, which leads to deflection from the incident direction. Point of Closest Approach (PoCA) assumes that muons only experience a single Coulomb scattering at the closest point between their incident and exiting trajectory. The variance of the scattering angle can reflect a certain amount of material information of the object crossed by these muons. In Purdue's work, it was proposed to image a spent nuclear fuel dry storage cask with cosmic ray muons using simple PoCA method. Both vertically and horizontally loaded storage casks were simulated. The experimental configurations are shown in Figure 1.7. The relative position between radiation detectors and the dry storage cask was fixed letting muons pass through the dry cask.

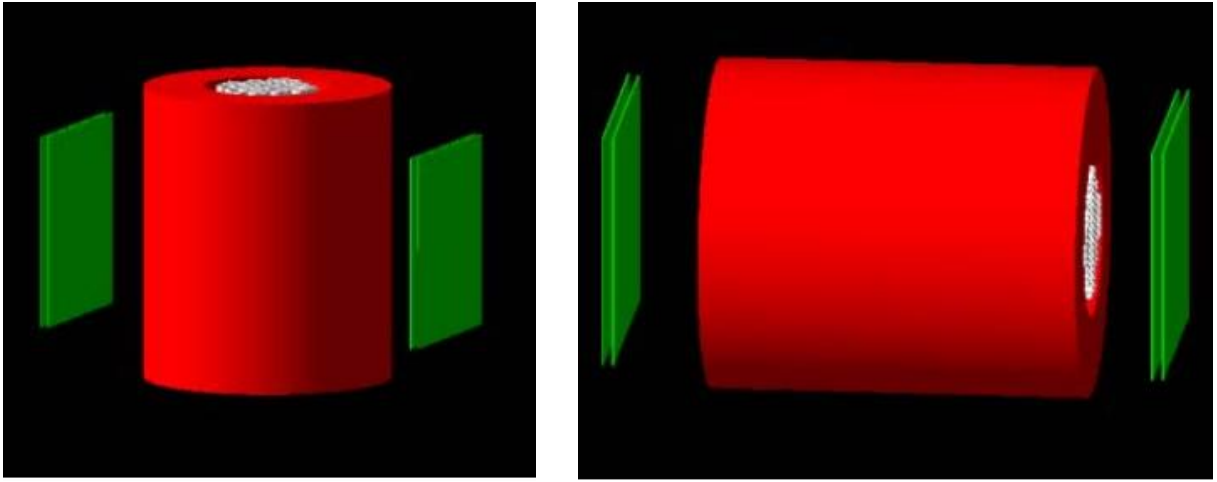

Figure 1.7    A GEANT4 model of a vertical (left) and horizontal (right) fully loaded dry cask with associated detector (from [6], used with permission)

Two pairs of muon detectors were used to register incident and exiting positions and directions of the muons. Only the muons which crossed these two pairs of muon detectors were used for image reconstruction using simple the algorithm.

The image reconstruction volume enveloped by two pairs of muon detectors was discretized into voxels. The scattering angle of each muon was stored in the voxel where its PoCA point was located. After storing all muon scattering angles in the image reconstruction volume, the voxel value is calculated using

$$s_i = \frac{1}{2}[(\theta_x{}^{out} - \theta_x{}^{in})^2 + (\theta_y{}^{out} - \theta_y{}^{in})^2] \qquad (1.1)$$

$$VoxelValue = \frac{1}{NL}\sum_{i=1}^{N} s_i \qquad (1.2)$$

where $\theta_x{}^{out}$, $\theta_x{}^{in}$, $\theta_y{}^{out}$ and $\theta_y{}^{in}$ are muon outgoing and incident direction angles on the XOZ and YOZ planes, N is the number of muons, and L is the side width of a voxel. Six different spent nuclear fuel missing patterns in dry storage casks were simulated: fully loaded, half loaded, one row of spent nuclear fuel assembly missing, one spent nuclear fuel assembly missing, and empty, as shown in Figure 1.8 from left to right. About $10^5$ muons were registered for each situation. The simple PoCA algorithm was used to reconstruct these dry storage casks with the different missing patterns from the simulated data. The reconstructed images of a vertical loaded (upper row) and horizontal loaded (lower row) dry storage cask are shown in Figure 1.9. The difference between the fully loaded cask, the half loaded cask, the cask with one row of spent nuclear fuel missing, and the empty cask can be easily identified with the exception of the missing pattern where there is only one spent nuclear fuel assembly missing. This may be attributed to the intrinsic accuracy of PoCA assumption.

Although the exposure time has been significantly reduced to about one day and no manmade radiative source is needed, the reconstructed image resolution is poor and may not be able to detect one spent nuclear fuel assembly missing scenario.

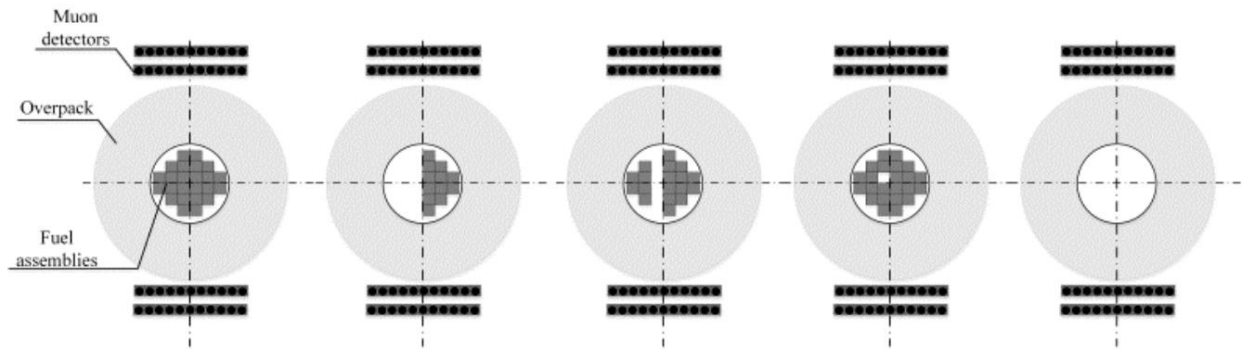

Figure 1.8        Simulated missing assemblies patterns (from [6], used with permission)

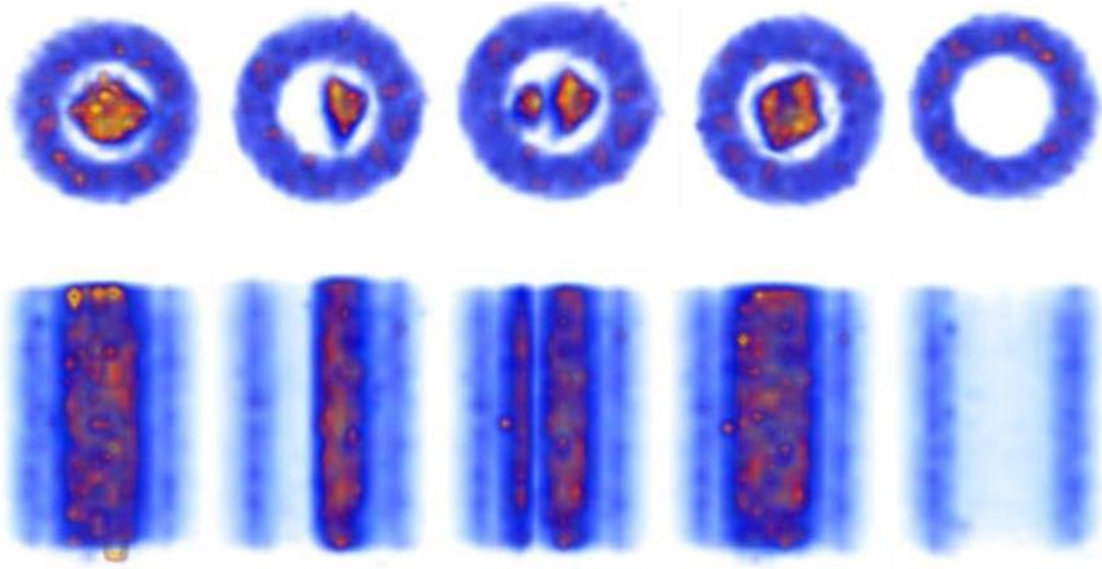

Figure 1.9      Reconstructed images of horizontal (upper row) and vertical (lower row) spent nuclear fuel dry storage casks with different fuel missing patterns. First column: full loaded, Second Column: half loaded, third column: a row of fuel assemblies missing, fourth column: one assembly missing, fifth column: empty. (from [6], used with permission)

### 1.2.4    Imaging dry storage casks with cosmic ray muon computed tomography

Similar to X-ray computed tomography, efforts were made by LANL to develop a cosmic ray computed tomography technique. In their simulation, two large area cylindrical detectors were used to register incoming and exiting muons from a $2\pi$ azimuthal angle as shown in Figure 1.10. Fuel missing pattern can also be seen in Figure 1.10. The incident muons were collected in one-degree wide bins of azimuthal angle and projected to a vertical plane in the center of the detector. The plane is discretized into 2 cm wide bins in horizontal direction. After muon scattering angles is stored into corresponding bins hit by its horizontal direction for all muons in one-degree wide bins, these scattering angles in each bin can be fit with the multigroup model [13] to extract the areal density. Because the large area cylindrical detectors wrap the entire dry storage cask, 360 views can be obtained with one degree wide per view. The extracted areal density from 360 views forms a complete sinogram of the dry storage cask. Filtered back projection was used to reconstruct dry storage cask from the sinogram.

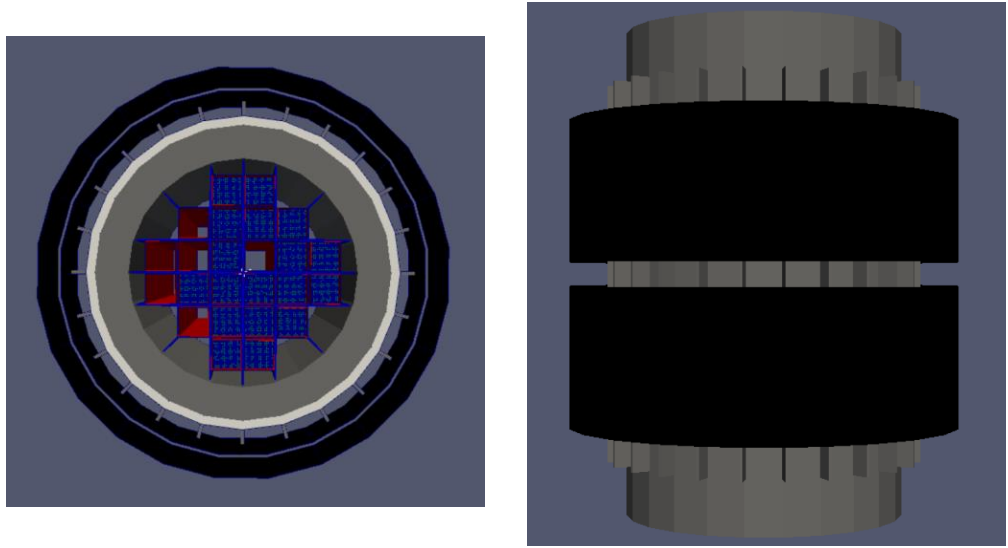

Figure 1.10    Simulated configuration of dry storage cask and two cylindrical detectors. Top down view on the left and side view on the right [9].

A quantity of $10^8$ muons were registered for reconstruction, which is equivalent to 1.6 days cosmic ray muon exposure. The reconstructed image is shown in Figure 1.11 on the left. The estimated values for spent nuclear assemblies and empty slot are 25.8 ±1.3 arb. and 2.5 ± 1.3 arb., which are separated by $18\sigma$. Thus, empty slots can be readily detected in a dry storage cask. However, the size of spent nuclear fuel assemblies in the reconstructed image is not correctly reconstructed and the gap between fuel assemblies is larger than it is in the simulated dry storage cask. Additionally, Figure 1.11 is a gray level image, the edge of dry storage cask is whiter than the spent nuclear fuel assemblies, which is misclassified. This may be caused by the polyenergetic nature of cosmic ray muon. Similarly, the transmission of muons through can form a complete sinogram of the dry storage cask. Then value in each bin become $-\ln(\frac{N_{out}}{N_{in}})$, where $N_{out}$ is the number of muons passing through upper and lower detectors and $N_{in}$ is the number of incoming muons registered by the upper detector. The reconstructed image is shown at right in Figure 1.11 [9].

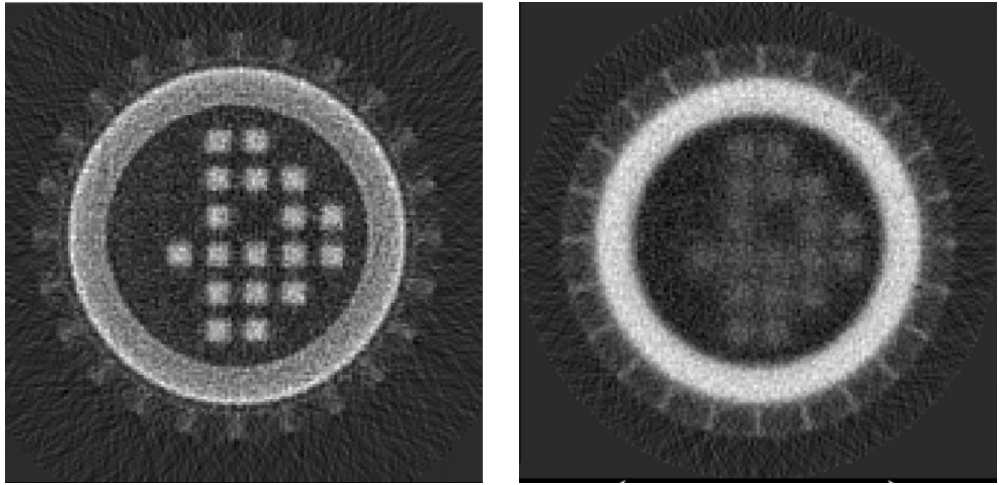

Figure 1.11    Reconstruct image of the dry storage cask with the partial fuel assemblies missing using scattering angle on the left and transmission rate on the right [9].

The reconstructed image of the dry storage cask using transmitted cosmic ray muons has a much worse image quality than the image reconstructed using scattering angle data. Similarly, the material of dry storage cask is also misclassified as higher Z material than spent nuclear fuel.

## 1.3   The author's role and original contributions

Cosmic ray muon tomography is a highly interdisciplinary subject, involving many different disciplines, like high energy physics, nuclear engineering, radiation detection and measurement, statistics, Geant4/MCNP simulation, image reconstruction techniques, digital signal processing and computer programming. Although this work was done as a part of a funded research collaboration, some parts were done independently by the author. Significant original contributions include:

- Comparing data from a GEANT4 simulation workspace against muon measurements on a MC-10 cask published by LANL

- Developing a VSC-24 dry storage cask simulation workspace in GEANT4

- Creating a methods development workspace including a muon computed tomography framework, ways of projecting muon interactions, and associated image reconstruction codes

- Investigating a new muon scattering angle projection method (later called method **b**)

- Investigating and quantifying the expected detection limits and imaging performance of developed methods for a dry nuclear fuel storage cask

- Investigating the expected effects of engineering restrictions including detector positional measurement uncertainty, detector size, and the ability to measure muon momentum on muon CT performance

The author also participated in: 1) the design and fabrication of a large area, position-sensitive detector for the muon imager system at Oregon State University and 2) the development of the most probable trajectory model for muon path modeling in collaboration with ORNL.

## 1.4   Dissertation outline

Chapter 1 briefly presents more background on this work. Due to the ban of the reprocessing of commercial spent nuclear fuel in the U.S. and the fat that no alterative storage method has been provided, more and more spent nuclear fuel has been transferred and stored in spent nuclear fuel dry storage casks after cooling down from water pool on nuclear power plant site. From the point of view of nuclear material protection and accountability, nondestructive assay techniques to probe into spent nuclear fuel dry storage casks are desperately needed given the prohibitively high cost of opening and resealing dry storage cask for visual check. Previous methods and their corresponding results are shortly covered in this chapter. New challenges have been put forward, like lowering detection limits, shortening measurement time, detecting the replacement of spent nuclear assemblies with dummy material and improving the reconstructed image quality.

In Chapter 2, basic knowledge of muon physics, the generation and spectrum of cosmic ray muons, detection and instrumentation, and muon path models in matter are provided.

In order to transfer traditional X-ray computed tomography techniques to a new probing particle, muons, Chapter 3 starts with an overview of traditional X-ray CT and its two common reconstruction methods: filtered back projection and algebraic reconstruction methods. Then a concrete analogy between traditional X-ray CT and the potential muon CT is drawn, wherein the muon CT frame is built. Unlike traditional X-ray CT, using of a straight path to approximate the muon path would not achieve the best result due to moderate deflection caused by multiple Coulomb scattering. Four muon path models and three different projection methods are embedded in the developed muon CT framework. Both FBP and algebraic reconstruction techniques can be used to reconstruct the objects using projection information.

In Chapter 4, after validating the simulation workspace against a physical experiment done by LANL at INL on a MC-10 cask with some spent nuclear assemblies missing, a VSC-24 dry storage cask is built in GEANT4 to generate simulation data for reconstruction. The six methods developed in Chapter 3 are then used to reconstruct a VSC-24 dry storage cask with one spent nuclear fuel assembly missing in the middle from the simulated data. To address the detection capability of our muon CT algorithms, half of an assembly and two a quarter of an assembly are simulated as missing and reconstructed. Results are also reported in this chapter.

Due to the low flux rate of cosmic ray generated muons, the measurement time is long from hours to days, meanwhile, an extended exposure time may not be worth the extra gain of reconstructed image quality. Thus, view sampling is investigated at the beginning of Chapter 5. When it comes to a field experiment, it is clearly different from the simulated ideal case, because of effects including finite detector position resolution, detection system temporal resolution, and background radiation. In this chapter, three different detector position measurement uncertainties are incorporated into the reconstruction process. Furthermore, imaging dry storage with small sized muon detectors is also simulated and reconstructed with economics in mid.

Finally, Chapter 6 summarizes the results using muon CT techniques with various path models and projection methods and describes other potential research topics that could add extra value for the bright future of muon CT.

# 2   THE MUON

A muon is an elementary particle similar to electron with an electric charge of −1 e and a spin of ½ and a rest mass of 105.7 MeV/c². Similar to all other elementary particles, muon has an antiparticle with one positive charge called antimuon μ⁺ [14]. Muon is a lepton, which is not thought to be comprised of any smaller particles. Muon is unstable. Its decay is governed by weak interaction only instead of more powerful strong interaction or electromagnetic interaction, thus muon has a much longer mean lifetime of 2.2 μs than many other subatomic particles. Due to its heavier mass, muon is much less likely to emit bremsstrahlung when interacting with matter, which makes it far more penetrating than electrons.

## 2.1   Cosmic ray muons

Muon is a naturally occurring particle created in the process of cosmic ray entering the earth upper atmosphere. Cosmic rays are high-energy radiation originating outside the solar system. About 90% of primary cosmic rays are simple protons (i.e., hydrogen nuclei), 9% are alpha particles and the rest 1% are nuclei of heavier elements [15]. When cosmic rays enter earth atmosphere, cosmic ray protons and other nucleus interact with air molecules in upper atmosphere about 45 km above the sea level creating pions ($\pi^-$, $\pi^0$ and $\pi^+$)  and kaons ($K^-$, $K^0$ and $K^+$). Charged pions ($\pi^-$ and $\pi^+$) and kaons ($K^-$, $K^0$  and $K^+$) soon decay into muons ( $\mu^-$ and $\mu^+$) with the same charge as illustrated in Figure 2.1.

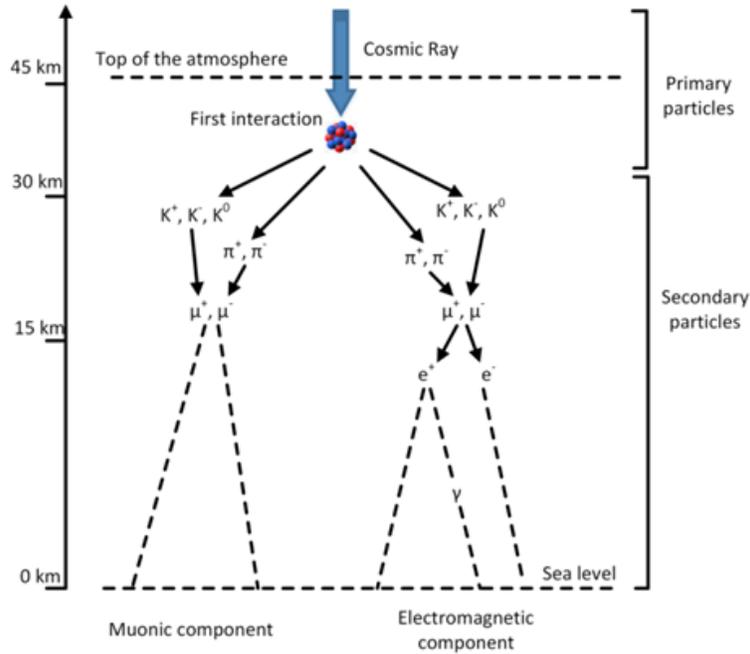

Figure 2.1　　　Illustration of the generation of cosmic ray muons (from [16], used with permission)

Neutral pions would decay into high energetic gamma rays (γ), whose energy is much above 1.022 MeV and subject to pair production effect. Muons would also decay into electrons, which are the dominant source of low energy electrons at the sea level. The ratio of $\mu^+$ and $\mu^-$ reflects the excess of $\pi^+$ over $\pi^-$ and $K^+$ over $K^-$.

　　　Except for electrons, muons are the most numerous charged particles at sea level. Typically, muons are created about 15 km above the ground as a result of the decay of pions and kaons. On average, muons lose 2 GeV before hitting the sea level. Their energy and angular spectrum at a specific altitude is a convolution of the production spectrum, thickness of air they have traversed and the decay. Based on measurements, muon intensity at vertical direction at the sea level is 70 $m^{-2}s^{-1}sr^{-1}$, which is equivalent to 10000 muons per $m^2$ per minute [17]. On the sea level, the overall angular distribution is roughly proportional to $\cos^2(\theta)$ with a mean energy of 3 ~4 GeV [18], where θ is the zenith angle. An experimentally measured muon differential intensity by Tsuji et al. [19], and Jokisch et al. [20] at 0°, 30°, 60° and 75° are shown in Figure 2.2.

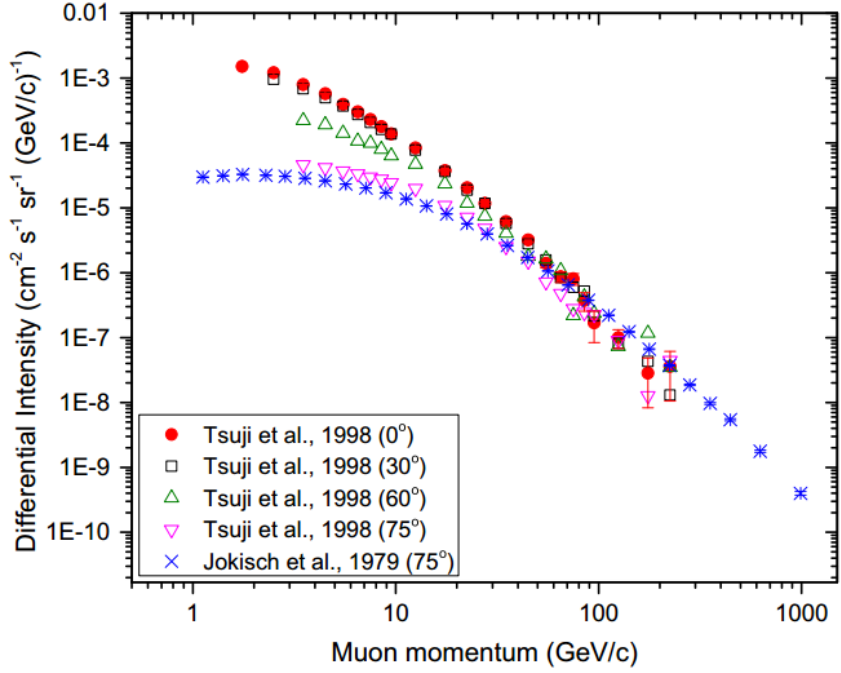

Figure 2.2    Measured muon differential intensities at 0°, 30°, 60° and 75° (from [21], used with permission)

Zenith angle has little influence on muon spectrum before it reaches 30°, which can be seen from the top two curves in Figure 2.2. For muons with an energy smaller than 1 GeV, the energy spectrum is quite flat due to range out of low energetic muons. For muons within the energy range of 10 to 100 GeV, the steep trends reflect that the primary spectrum is in this range. For energy above 100 GeV, the steeper trend is caused by the interaction of pion with atmospheric atoms before touching the ground. For muons within the extremely high energy range (bigger than 1 TeV), its energy spectrum is above one magnitude steeper than primary spectrum. At larger angles, the decay of low energy muon causes the drop of muon intensity because of limited lifetime and long-distance travel before hitting the ground. When earth curvature and muon decay are negligible, the following formula can be used to approximate muon spectrum [22]

$$-\frac{dN_u}{dE_u d\Omega} \approx \frac{0.14E_u^{-2.7}}{cm^2 \cdot s \cdot sr \cdot GeV} \times \left\{ \frac{1}{1 + \frac{1.1E_u cos^2(\theta)}{115 GeV}} + \frac{0.054}{1 + \frac{1.1E_u cos^2(\theta)}{850 GeV}} \right\} \quad (2.1)$$

## 2.2   Muon interaction mechanisms

### 2.2.1   Energy loss

When interacting with matter, muons lose energy via ionization and radioactive process: bremsstrahlung, direct production of e⁻ and e⁺ pairs, and photonuclear interactions. The total energy loss can be expressed by stopping power:

$$-\frac{dE_u}{dx} = a(E_u) + b(E_u)E_u \qquad (2.2)$$

where $a(E_u)$ is the ionization loss and $b(E_u)$ is the fractional energy loss by bremsstrahlung, pairs production effect and photonuclear interactions. Both are varying slowly with the change of muon energy. One example of muon stopping power in copper is shown in Figure 2.3 [23].

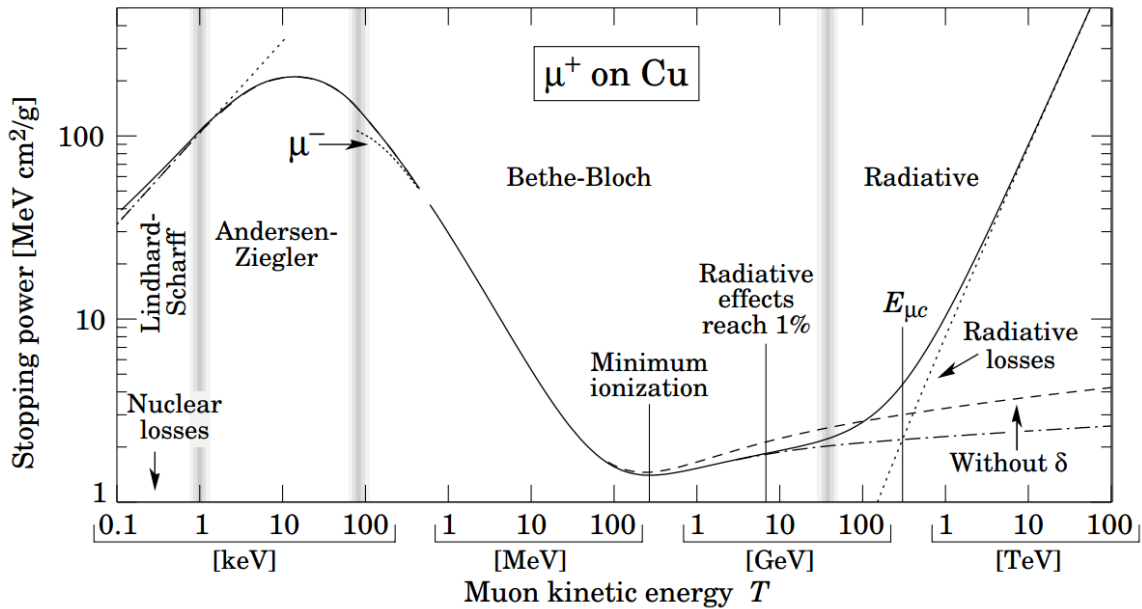

Figure 2.3          Muon stopping in Copper vs its kinetic energy [26].

It follows the same trend in other material as well. In the low energy region below 1KeV where the velocity of incident muon is smaller than that of valence electrons in the target, the stopping power is proportional to muon's incident speed, which can be described a semi-phenomenological model by Lindhard based on the work of research of Fermi and Teller [24]. With the increase of incident muon kinetic energy, electronic speed in the target can be neglected. In the interval between first two grey bands, a phenomenological fit was constructed by Andersen and Ziegler to bridge low energy region and high energy region. In high energy region, electronic energy loss is well depicted by widely known Bethe-Bloch formula [25]. Stopping power decreases as incident muon energy increase, however, this trend does not last long when it goes into very high energy region, because electronic stopping power decreases and radiative stopping power increases gradually. In the very high energy region, bremsstrahlung radiation gets more and more prominent and eventually dominants the energy loss mechanism. When $a(E_u)$ is equal to $b(E_u)$, muons reach its minimum ionization critical point, where $E_{uc}$ is the critical muon energy. Within the energy range of 0.1 to 1000 GeV, the minimum ionization is 2.2 MeV·cm$^2$/g, which varies with the change of matter it traversed.

### 2.2.2 Directional deflection

On top of the inelastic collisions with the atomic electrons leading to the energy loss of muons, meanwhile muons would experience elastic multiple Coulomb scattering [27] from nuclei with a smaller probability. Because of the greater mass of target nuclei than muons, the energy transferred from incoming muon to target nuclei caused by multiple Coulomb scattering is negligible. However, the scatter center continues adding a small scattering angle to the incoming muon causing it to deviate from a straight trajectory in the process of traversing objects. MCS can be deemed as integral effect of single scattering as shown in Figure 2.4. For a single scattering event that the target is extremely thin and no more two interactions happen, the process is shown in Figure 2.5. The cross section is well described by Rutherford formula [28]

$$\frac{d\sigma}{d\Omega} = \left(\frac{1}{4\pi\varepsilon_0}\right)^2 \frac{z^2 e^4}{M^2 c^4 \beta^4} \frac{1}{\sin^4(\theta_0/2)} \tag{2.3}$$

where z is the atomic number of incident particle, e is electron charge, M is the mass of target, $\beta$ is relativistic coefficient, $\varepsilon_0$ is the permittivity of free space and $\theta_0$ is the single scattering angle. When the thickness of target increases, and the number of single scattering events become significantly large, the MCS process has been formulated by several theories. Among these theories, Moliere's theory [29], which was analytical to the end and improved by Bethe [30] later, is in good agreement with most of experimental data [31-34].

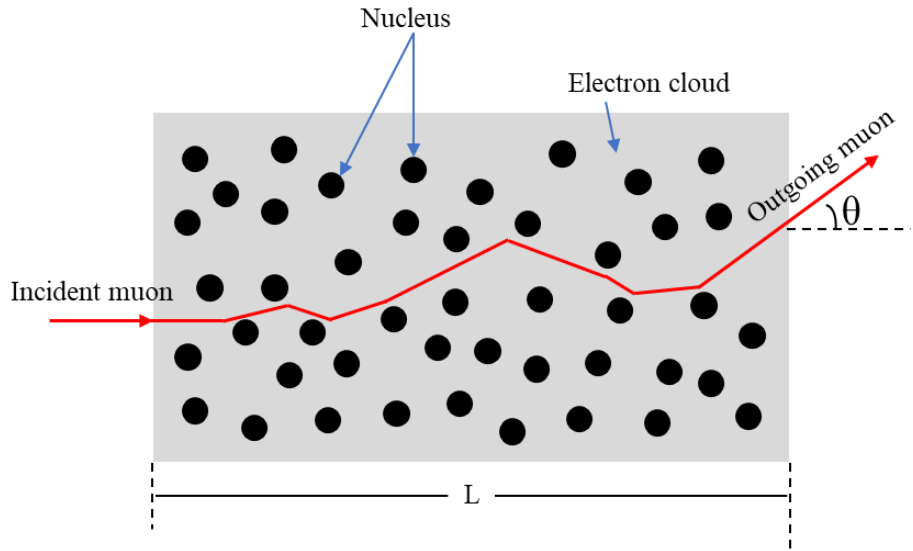

Figure 2.4          Scheme of multiple Coulomb scattering

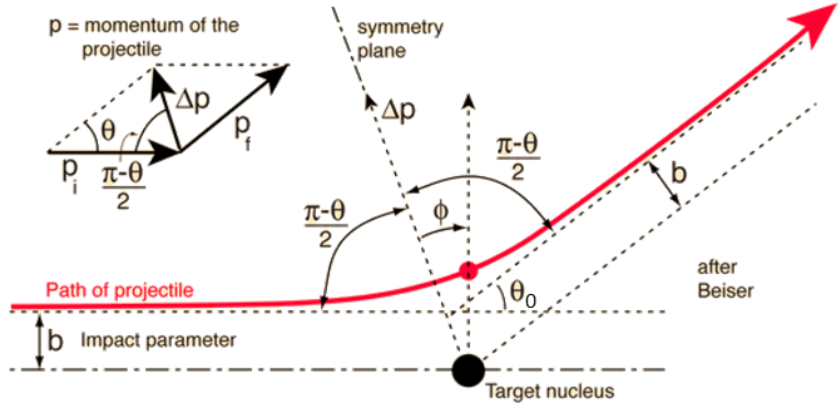

Figure 2.5    The scattering of incoming particle by central repulsive Coulomb force [28].

Although Moliere's derived scattering distribution is generally complex, a Gaussian approximation with zero mean value and variance $\sigma_\theta^2$ has been shown to adequately represent the central 95% of the scattering angle distribution [36] as:

$$f_\theta(\theta) \cong \frac{1}{\sqrt{2\pi}\sigma_\theta} e^{-\frac{\theta^2}{2\sigma_\theta^2}} \tag{2.4}$$

$$\sigma_\theta \cong \frac{15}{\beta c p}\sqrt{\frac{L}{L_{rad}}} \tag{2.5}$$

where $\beta c$ and $p$ are the velocity and momentum of incident muon, L is the thickness of object traversed by muons and $L_{rad}$ is radiation length of material.

Radiation length $L_{rad}$ is a characteristic amount of matter for radioactive process. It is the mean distance over which a highly energetic charged particle loses all but $1/e$ of its energy by bremsstrahlung, and $e^-$ and $e^+$ pair production. For simple substance, it can be approximated with

$$L_{rad} = \frac{716.4 \cdot A}{Z(Z+1)\ln\frac{287}{\sqrt{2}}} \, g \cdot cm^{-2} \qquad (2.6)$$

where Z is atomic number and A is mass number of the nucleus

More accurately it is

$$\frac{1}{L_{rad}} = \frac{4\alpha N_A Z(Z+1)r_e^2 \log(183Z^{-\frac{1}{3}})}{A} \qquad (2.7)$$

where $N_A$ is Avogadro's number, $\alpha$ is the fine structure constant and $r_e$ is the classical electron radius. For the estimation methods for radiation length of composite material and compound, please refer to APPENDIX B.

In 3-dimensional space, muon path, scattering angle can be decomposed into two orthogonal planes, XOZ plane and YOZ plane. Both $\theta_x$ and $\theta_y$ follow the same Gaussian approximation similar to Eq. (2.4).

$$f_{\theta_x}(\theta_x) \cong \frac{1}{\sqrt{2\pi}\sigma_\theta} \exp(-\frac{\theta_x^2}{2\sigma_\theta^2}) \qquad (2.8)$$

$$f_{\theta_y}(\theta_y) \cong \frac{1}{\sqrt{2\pi}\sigma_\theta} \exp(-\frac{\theta_y^2}{2\sigma_\theta^2}) \qquad (2.9)$$

where in 3D space $\theta$ may defined as

$$\theta = \sqrt{\frac{\theta_x^2 + \theta_y^2}{2}} \qquad (2.10)$$

To quantitively describe a characteristic information of material without the influence of its size, scattering density [36] as shown in Eq. (2.11) is introduced as the mean square scattering expected for muons with nominal momenta passing through a unit depth of a material with radiation length $L_{rad}$

$$\lambda_{L_{rad}} \equiv \left(\frac{15}{p_0}\right)^2 \frac{1}{L_{rad}} = \frac{\sigma_\theta^2}{L} \qquad (2.11)$$

By shining an object with numerous muons and measuring each scattering angle, the variance of these measured scattering angle can be calculated and the path length in a given object is roughly known, thus the atomic number can be estimated according to Eq. (2.5) and Eq. (2.6).

## 2.3   Muon detection and instrumentation

Muons are easy to detect with high accuracy and low backgrounds. For muons with an energy above a few GeV, their identification is based on low rate of interaction with matter: if charged particle penetrates a thick layer of absorber with minor energy losses and small angular displacement, such particle is considered as a muon. As a heavy charged particle with a huge amount of energy, muons deposit energy in all kinds of materials when passing through them. Thus, almost all kinds of radiation detectors can be used to detect muons. However, some special requirements are essential for muon imaging applications. The first consideration is coincidence timing in two or more detectors to register a muon event placed on each side of the item of interest. Secondly, in order to measure the muon flux or trajectories, which are the critical parameters for imaging purpose, detectors are required to be position sensitive. Third, due to low flux rate of muon, it shall be economically feasible to build the detector in large area. Various designs have been developed to determine the position information of muons. Arrays of drift tubes are very widely used as positive-sensitive detectors of muon [37][38]. Other types of common radiation detectors, like sodium-iodide, semiconductor and scintillation detectors, can be used for muon detection as well [39]. In muon imaging application, two most often chosen detectors for muon detection are gas wire detector and scintillation detector.

### 2.3.1    Drift tube detectors

Los Alamos National Laboratory used draft tube detectors to register muon passing through objects under investigation. One of muon tracker is shown in Figure 2.6. It is comprised of 88 drift tubes. Each tube is about 1.2 m long in a diameter of 5.08 cm and 0.89 mm thick filled with 1 bar of 47.5% Ar, 42.5% $CF_4$, 7.5% $C_2H_6$, and 2.5% He gas mixture with a gold-plated tungsten anode wire running down the center. At the time of operation, a potential of +2585 volts is applied between the central wire and draft tube wall. When muon crosses the draft tube, muon would ionize the filled gas atoms resulting free electrons. Electric filed accelerates the free ionized electrons toward central anode wire to ionize other gas atoms, which causes electron avalanche and makes the signal measurable.

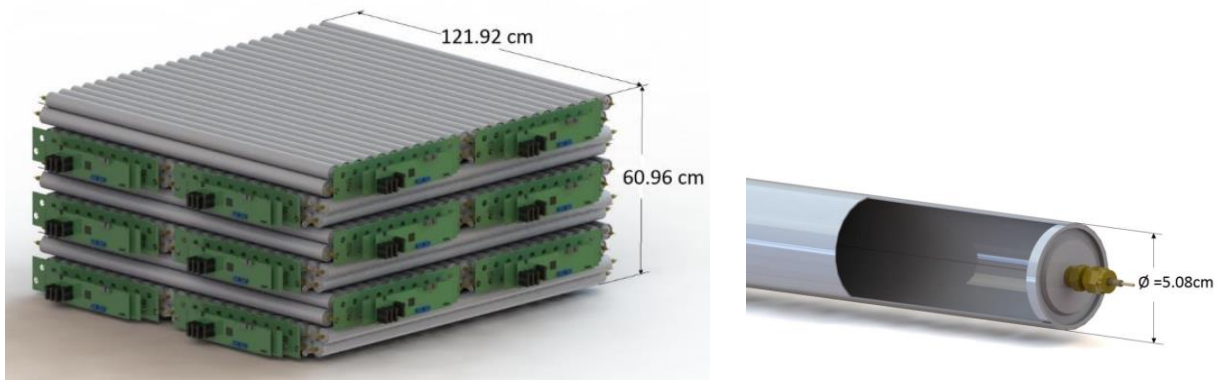

Figure 2.6        Draft tube muon tracker by LANL on the left, cut way view of one draft tube on the right. (from [7], used with permission)

## 2.3.2   Scintillation detectors with wavelength shifting fibers

Although gas detectors like drift tubes or GEM detector are extremely sensitive, which can reach a position resolution of submillimeter, due to its high cost especially for lager area detectors, there is a strong quest for alternatives.  In this work, we have designed and fabricated scintillator fiber position-sensitive detector for muon tomography use. It is based on a single layer plastic scintillator panel in a size of 32 cm × 32 cm × 2.5 cm. 32 parallel grooves with a pitch of 1 cm, 2 mm width and 4 mm depth are carved on the top and bottom sides of the scintillator panel. The directions of the grooves on the top and bottom sides are perpendicular to each other. Wave length shit (WLS) optical fibers are embedded in each groove for light transfer. WLS fibers are routed to individual pixels of a 64-pixel MAPMT. The WLS fibers absorb a portion of the scintillation light generated within the plastic scintillator by muon interactions and emit light with a different wavelength spectrum. Claddings surrounding the WLS fibers trap the light emission inside the fibers and transfer it to the MAPMT pixels for readout. The concept is shown on the top left in Figure 2.7 [40]. The position information in one direction can be obtained from the distribution of signal amplitudes in the corresponding 32 channels. As illustrated in Figure 2.8, when a scintillation event occurs at certain location, the light intensities, which are proportional to the signal amplitudes, of the fibers near the location follow a quasi-Gaussian distribution centered at the scintillation event. Therefore, an interpolation of the 32 signal amplitudes could reveal the origin of the scintillation light in each direction. By doing this, the position resolution around 1 cm is theoretically achievable. For muon imaging application, four of this detector are needed. One pair is used to register incident muon direction and position and another pair is used to register outgoing muon direction and position.

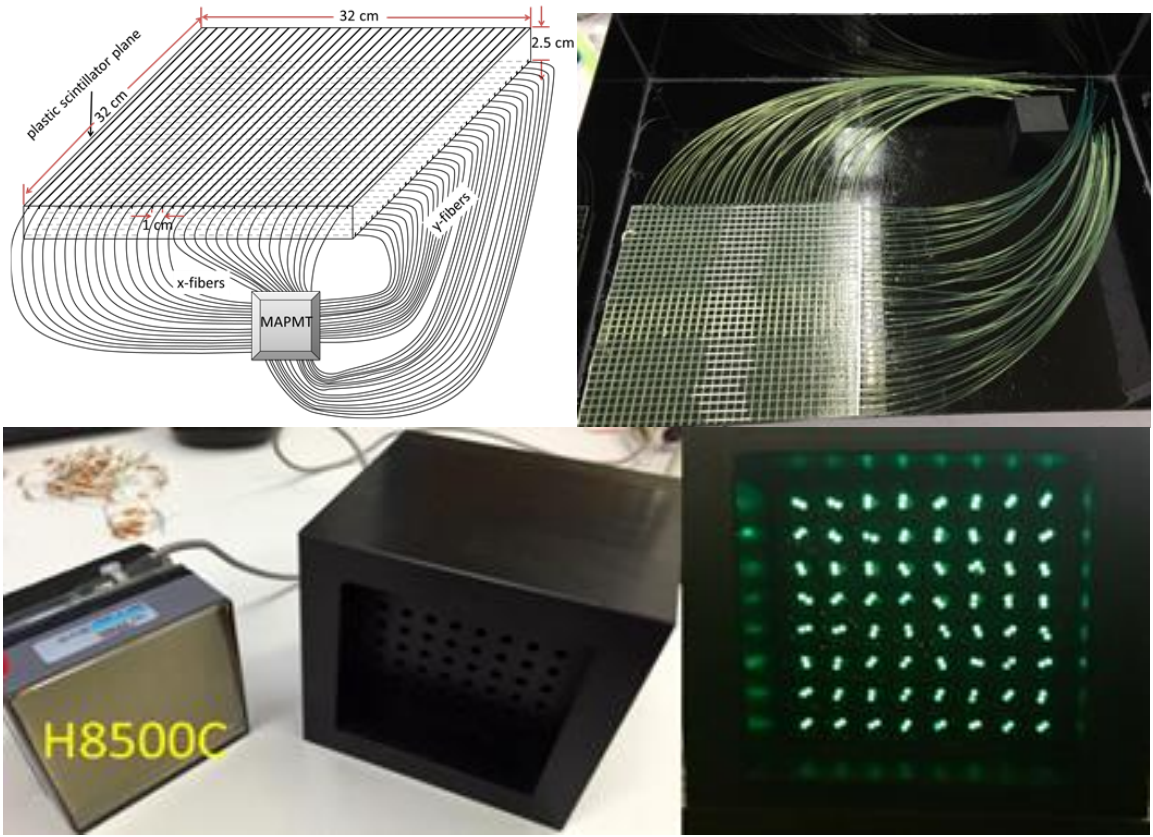

Figure 2.7 Scheme of the design of scintillator-fiber position-sensitive detector ( top left ), a picture of the design during assemblage (top right), PMT used in the system (bottom left), coupler used to connect WLS fiber and PMT(bottom middle) and a picture of coupler with WLS fiber installed (bottom right)

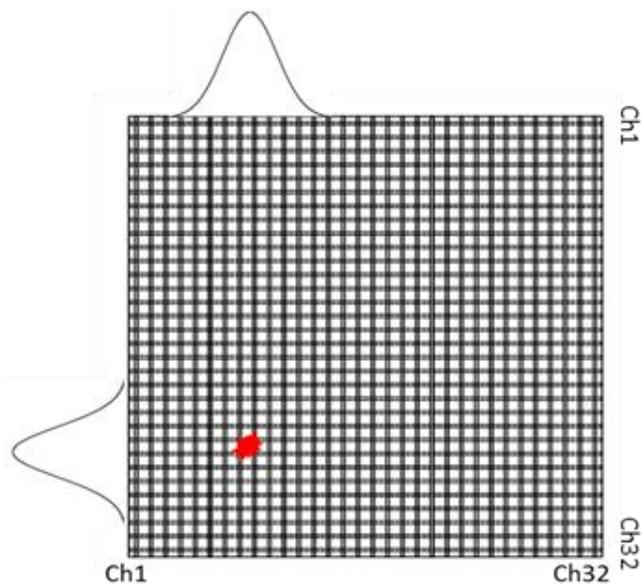

Figure 2.8 A scheme of a muon hitting scintillation detector plane and the light distribution from readout

The whole system contains four scintillation detectors imbedded in light tight boxes, four-fold coincidence module, 32 channel digitizer, high voltage supply and NIM crate as shown in Figure 2.9.

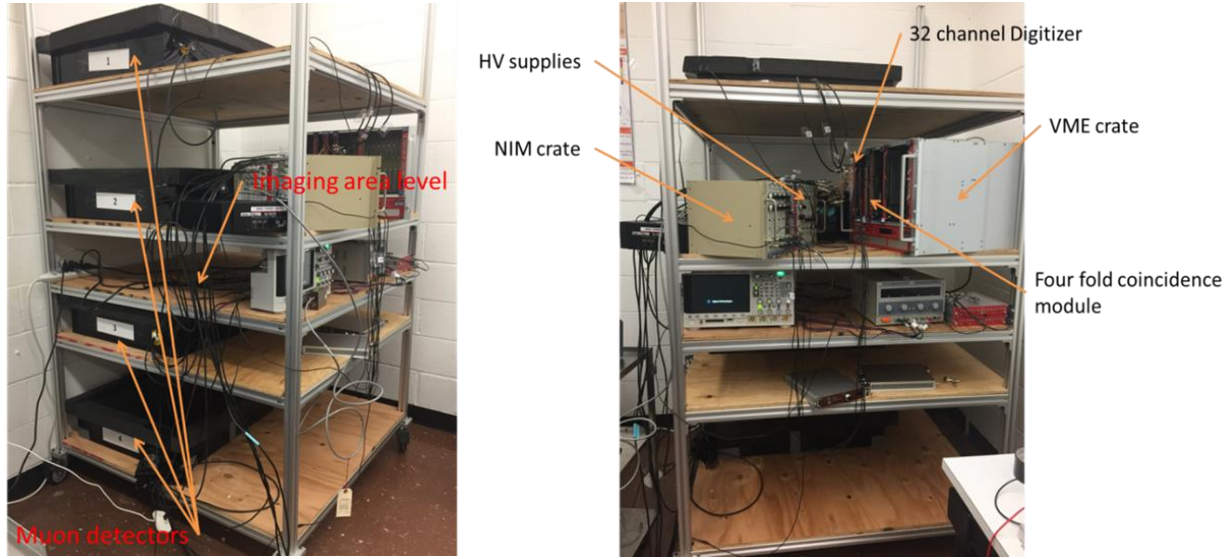

Figure 2.9        Pictures of system of scintillator-based WLS fiber detector for muon imaging application.

## 2.4   Muon path models

Muon path is subject to bending on account of multiple coulomb scattering with nuclei especially when crossing high Z atomic material. Thus, a curve may better describe muon path than a straight-line path (SLP) by connecting the incident point and exiting point. One of the muon curve path models is PoCA trajectory. It consists of two segments: (1) the segment connecting the point of muon incidence to the PoCA point and (2) the segment connecting the PoCA point and the muon exiting point. For simple geometry, PoCA trajectory model has been proven to work well as shown in  Figure 2.10 on the left, however, it has its own limitation in application when the PoCA point falls outside of the region of interest (ROI) as shown in Figure 2.10 on the right.  For muons with PoCA point falling outside the ROI, these muons must be rejected otherwise reconstructed result will be very noisy. With this said, most of effort in this section is to derive the development of a most likely path (MLP, or most probable trajectory, i.e., MPT, both are interchangeable) [41] published by the author and his colleagues and compare these four models with an GEANT4 simulated path.

Given the incident and exiting position and direction of muon, scattering angle and displacement can be calculated and projected onto two orthogonal planes XOZ and YOZ. The projection onto YOZ plane is shown in Figure 2.11. Similar projection onto XOZ plane can also drown, which is not shown here.

Scattering angle and displacement on two orthogonal planes follow a bivariate Gaussian distribution. Let $\mathbf{Y} = \begin{bmatrix} \theta_y \\ y \end{bmatrix}$, then

$$\mathcal{P}(\mathbf{Y}) \sim \mathcal{N}(0, \mathbf{\Sigma}) \tag{2.12}$$

where $\mathbf{\Sigma}$ is the covariance matrix of $\theta_y$ and $y$,

$$\mathbf{\Sigma} = \begin{bmatrix} \sigma_y^2 & \sigma_{\theta y}^2 \\ \sigma_{\theta y}^2 & \sigma_\theta^2 \end{bmatrix} \tag{2.13}$$

A maximum-a-posteriori algorithm (MAP) can used to maximizes the posterior distribution using to Bayes' theorem given the distribution of each point on the track govern by MCS processes. According to Bayesian theorem, the distribution of the scattering angle and displacement at an arbitrary point j on the trajectory given the incident and exiting information is:

$$P(\mathbf{y_j}|\mathbf{y_B}) = \frac{P(\mathbf{y_B}|\mathbf{y_j})P(\mathbf{y_j}|\mathbf{y_A})}{P(\mathbf{y_B})} \tag{2.14}$$

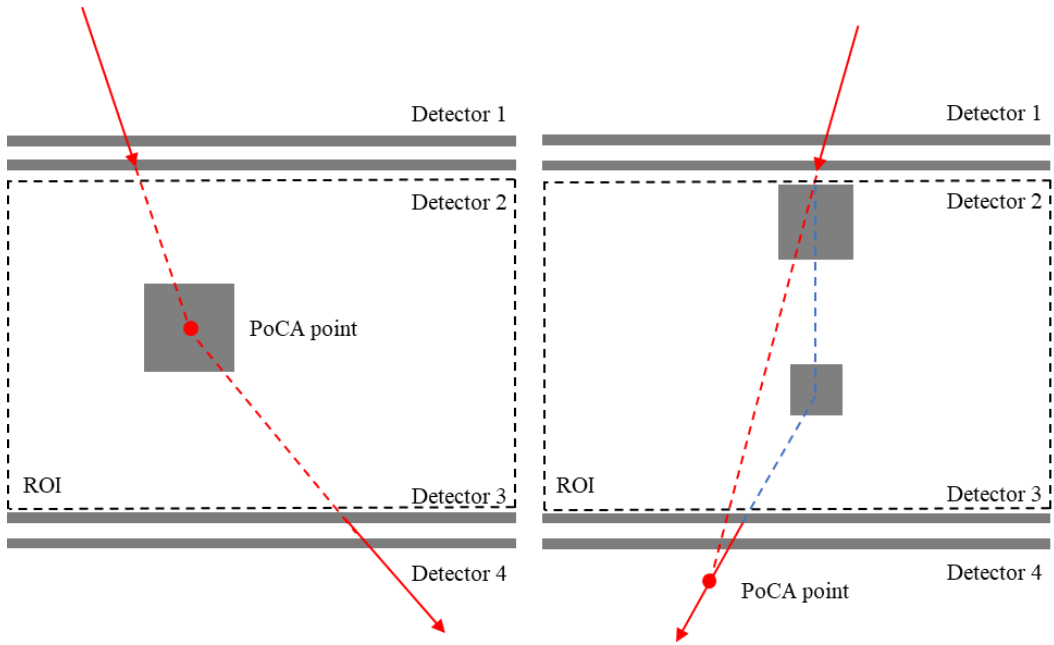

Figure 2.10    Illustration of a PoCA trajectory model. The PoCA trajectory is in red and the scenarios where the PoCA model works well on the left and where it does not work on the right.

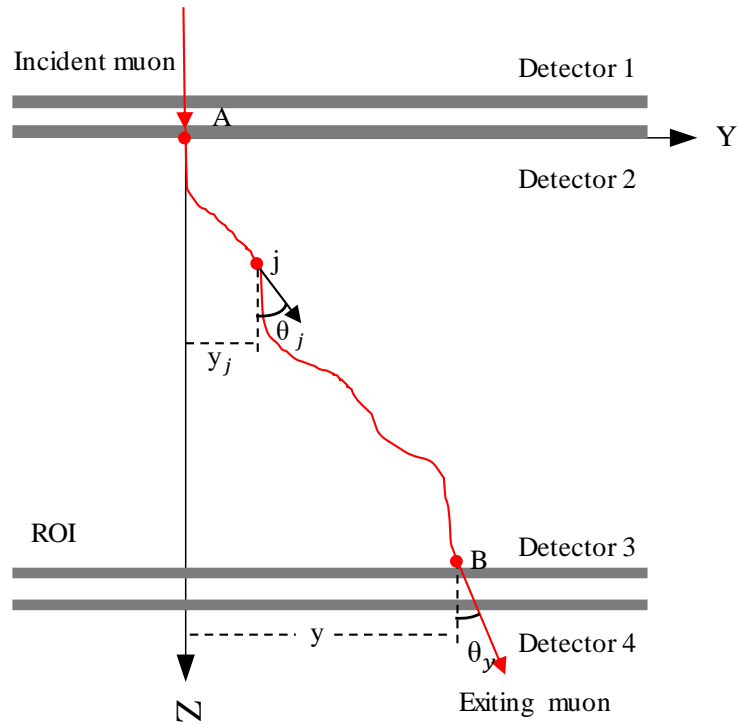

Figure 2.11 Projection of 3D muon track to YOZ plane

It can be considered as two processes: one is from the incident point A to point j and another is from point j to the exiting point B. The process from point A to point j follows the distribution

$$P(\mathbf{y_j}|\mathbf{y_A}) = \frac{1}{C_1}\exp\left(-\frac{1}{2}\left(\mathbf{y_j^T} - \mathbf{y_A^T R_A^T}\right)\boldsymbol{\Sigma_j^{-1}}\left(\mathbf{y_j} - \mathbf{R_A y_A}\right)\right) \tag{2.15}$$

where $C_1$ is a normalization constant and

$$\boldsymbol{\Sigma_j} = \begin{bmatrix} \sigma_{yj}^2 & \sigma_{\theta yj}^2 \\ \sigma_{\theta yj}^2 & \sigma_{\theta j}^2 \end{bmatrix} \tag{2.16}$$

$$\mathbf{R_A} = \begin{bmatrix} 1 & z_j - z_A \\ 0 & 1 \end{bmatrix} \tag{2.17}$$

Similarly, the process from point j to point B follows the distribution

$$P(\mathbf{y_B}|\mathbf{y_j}) = \frac{1}{C_2}\exp\left(-\frac{1}{2}\left(\mathbf{y_B^T} - \mathbf{y_j^T R_B^T}\right)\boldsymbol{\Sigma_B^{-1}}\left(\mathbf{y_B} - \mathbf{R_B y_j}\right)\right) \tag{2.18}$$

$$\mathbf{R_B} = \begin{bmatrix} 1 & z_B - z_j \\ 0 & 1 \end{bmatrix} \tag{2.19}$$

Plugging Eq. (2.15) and Eq. (2.18) into Eq. (2.14) and take natural logarithm of it, it becomes

$$\ln P(\mathbf{y_j}|\mathbf{y_B}) = -\frac{1}{2}\left(\mathbf{y_j^T} - \mathbf{y_A^T R_A^T}\right)\boldsymbol{\Sigma_j^{-1}}\left(\mathbf{y_j} - \mathbf{R_A y_A}\right)$$
$$-\frac{1}{2}\left(\mathbf{y_B^T} - \mathbf{y_j^T R_B^T}\right)\boldsymbol{\Sigma_B^{-1}}\left(\mathbf{y_B} - \mathbf{R_B y_j}\right) - \ln C_3 \tag{2.20}$$

where $C_3$ contains all other constants, which would be cancelled out in the following derivative step.

Taking derivative of Eq. (2.20) with respect to $\mathbf{y_j}$ and setting it zero, then most probable for arbitrary point j is obtained as:

$$\mathbf{y_{MLP}} = \left(\boldsymbol{\Sigma_j^{-1}} + \mathbf{R_A^T \Sigma_B^{-1} R_A}\right)^{-1}\left(\boldsymbol{\Sigma_j^{-1} R_A y_A} + \mathbf{R_A^T \Sigma_B^{-1} y_B}\right) \tag{2.21}$$

GEANT4 simulation package was used to simulate the process of 3 GeV muons crossing a uranium cube with a side width of 10 cm. It was found that about 30% muons have their PoCA point fall outside region of interest, which would reduce useful muon flux rate and lengthen measurement time if PoCA path model was adopted. For such an extreme case, a comparison between the calculated SLP, PoCA path, MLP and simulated path in GEANT4 is shown in Figure 2.12 on the left. It turned out MLP is in good agreement with muon real path. A 2σ and 3σ envelopes of MLP is shown in Figure 2.12 on the right.

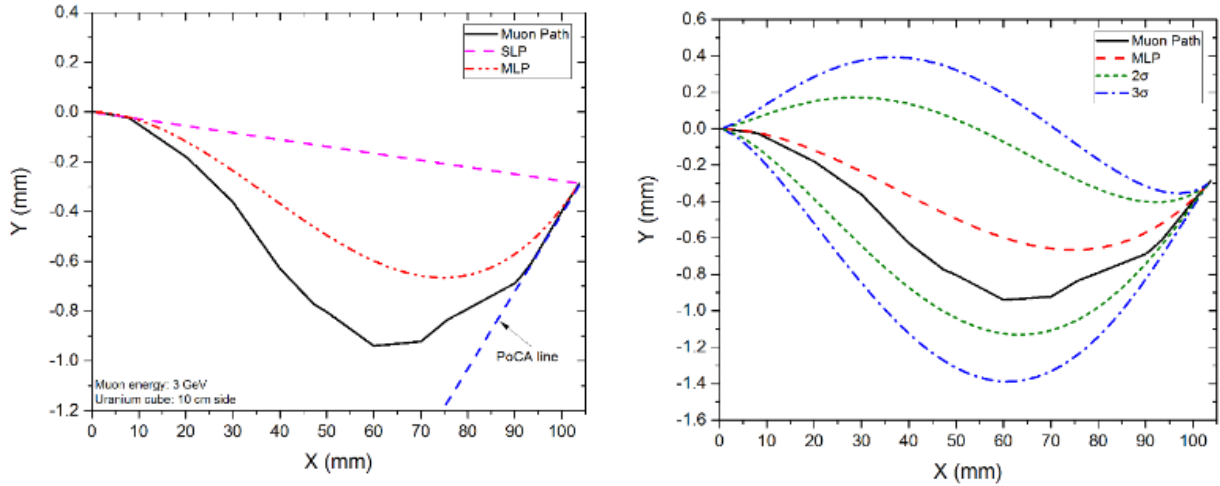

Figure 2.12    An example of the comparison between SLP, MLP and real muon path on the left and the associated 2σ and 3σ envelopes on the right.

## 2.5   Summary

This chapter briefly reviewed the generation of cosmic ray muons, the muon spectrum and basic muon physics. Muon interacts with material primarily via two interaction mechanisms: ionization and MCS, which leads to energy loss and deflection from its incident direction, separately. Scattering angles of a monoenergetic muon beam after crossing an object follow a Gaussian distribution with zero mean value. The variance of these scattering angle is directly related to the atomic number of the object, which serves as the fundamental rationale for using muon imaging to differentiate material. Muon detection and muon instrumentation were also addressed. Muon detection generally involves two major steps: muon identification and parameter measurement. If a charged particle penetrates a large amount of absorber with relatively minor energy losses and small angular deflection, then such a particle is considered to be a muon. Key muon parameters for imaging are the muon position, direction and its momentum. Due to MSC, a straight path normally cannot approximate the muon trajectory well. For simple geometry, the PoCA path can be used to roughly locate the object under interrogation; however, the PoCA point is likely to fall outside the reconstruction volume. For these muons, they are usually discarded to avoid causing extra noise to the reconstructed image. Thus, a most probable trajectory model based on Bayesian theory and MCS formula was put forward to better describe muon trajectory in objects.

# 3  MUON CT

Traditional X-ray computed tomography technique is utilizing the transmission rate of incident x-ray photons due to attenuation caused by photoelectric effect, Compton scattering, even pair production effect for high energy photons. Similarly, muon transmission rate can be used to reconstruct the object crossed by muon, however, it may not yield a good result. In this chapter, most of the efforts were made to develop a muon computed tomography technique using muon scattering angle as input information. Muon CT is primarily exploiting the fact that the scattering angles of a monoenergetic muon beam after traversing an object follow a Gaussian distribution with zero mean and a variance, which is related to the material they have traversed. Although X-ray computed tomography technique and muon computed tomography are based on different interaction mechanism with matter and use different information to reconstruct the image of object under investigation, transform and reconstruction techniques are quite similar. The first concrete mathematical description of muon computed tomography was published in [42] by the author. A quick review of traditional X-ray computed tomography technique and reconstruction methods is given below to be applied to muon computed tomography shortly.

## 3.1  Transmission X-ray CT

Radon transform and its inverse transform were developed by Johann Radon [43] in 1917 in two dimensional plane and later he extended these formulas into three dimension space, which has a very wide application in modern technologies, like barcode scanners, electron microscopy of macromolecular assemblies, computed tomography, reflection seismology and in the solution of hyperbolic partial differential equations. It takes the integral of an object f(x, y) defined on a plane along a line L at angle θ to a function p(θ, r), where r is the distance from the origin to the line L as shown in Figure 3.1 on the left. In space domain, mathematically it can be expressed as [44]

$$p(r, \theta) = \int_{-\infty}^{\infty} \int_{-\infty}^{\infty} f(x, y) \delta(x\cos\theta + y\sin\theta - r) dl \tag{3.1}$$

Projection data p(r, θ) represents the tomographic scan of the object f(x, y) at (r, θ), which can be used to reconstruct the original object f(x, y). It is usually called sinogram because all the points except the origin in object f(x, y) are a sinusoid in projection data p(r, θ). Thus, any objects are a number of sinusoids with different amplitudes and phases in (r, θ) domain.

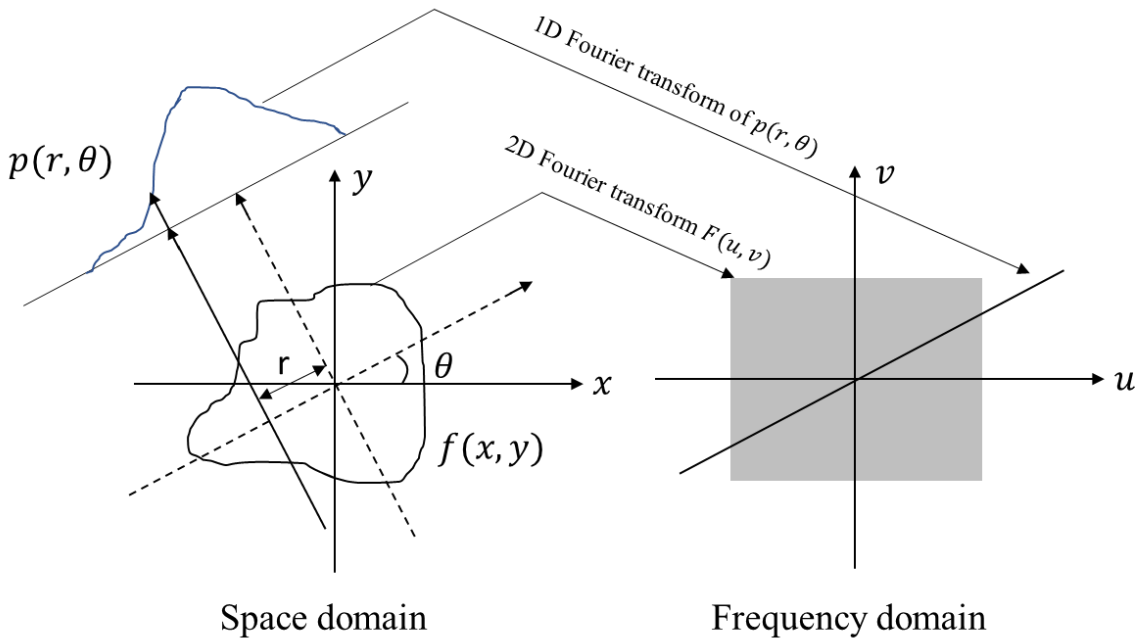

Space domain          Frequency domain

Figure 3.1        Illustration of Fourier-slice theorem

One dimensional Fourier transform of the projection at $(r, \theta)$ with respect to r is

$$G(\omega, \theta) = \int_{-\infty}^{\infty} p(r, \theta) e^{-j2\pi\omega r} dr \qquad (3.2)$$

where $\omega$ is the frequency variable.

Plug Eq. (3.1) into Eq. (3.2) and after simplification, $G(\omega, \theta)$ becomes

$$G(\omega, \theta) = \int_{-\infty}^{\infty} \int_{-\infty}^{\infty} f(x, y) e^{-j2\pi\omega(x\cos\theta + y\sin\theta)} dx dy \qquad (3.3)$$

Let $u = \omega\cos(\theta)$ and $v = \omega\sin(\theta)$, the above equation becomes

$$G(\omega, \theta) = \left[ \int_{-\infty}^{\infty} \int_{-\infty}^{\infty} f(x, y) e^{-j2\pi\omega(ux + vy)} dx dy \right]_{u = \omega\cos(\theta), v = \omega\sin(\theta)}$$

$$= [F(u, v)]_{u = \omega\cos(\theta), v = \omega\sin(\theta)} \qquad (3.4)$$

$$= F(\omega\cos(\theta), \omega\sin(\theta))$$

Above equation is the widely known Fourier-slice theorem, stating that the Fourier transform of a slice of projection $p(r, \theta)$ is a slice of the 2-D Fourier transform of the object at the same angle $\theta$ as shown in Figure 3.1 on the right.

### 3.1.1 Filtered back projection reconstruction technique

$f(x, y)$ can be calculated by taking the 2-dimensional inverse Fourier transform of $F(u, v)$ as indicated by Fourier-slice theorem

$$f(x, y) = \int_{-\infty}^{\infty} \int_{-\infty}^{\infty} F(u, v) e^{j2\pi\omega(ux + vy)} du dv \qquad (3.5)$$

Let $u = \omega\cos(\theta)$ and $v = \omega\sin(\theta)$. According to Jacobian in the change of variables, $du dv = \omega d\omega d\theta$, then the above equation becomes

$$f(x,y) = \int_0^{2\pi} \int_0^{\infty} F(\omega\cos(\theta), \omega\sin(\theta))\, e^{j2\pi\omega(x\cos\theta + y\sin\theta)}\omega d\omega d\theta \qquad (3.6)$$

Then plug Eq. (3.4) into the Eq. (3.6), it becomes

$$f(x,y) = \int_0^{2\pi} \int_0^{\infty} G(\omega,\theta)\, e^{j2\pi\omega(x\cos\theta + y\sin\theta)}\omega d\omega d\theta \qquad (3.7)$$

$2\pi$ in this equation can be split into two parts: $0$ to $\pi$ and $\pi$ to $2\pi$. Due to the circular symmetry property $G(\omega, \theta + \pi) = G(-\omega, \theta)$ and $x\cos\theta + y\sin\theta = r$, Eq. (3.7) can be further simplified as

$$f(x,y) = \int_0^{\pi} \left[ \int_0^{\infty} |\omega| G(\omega,\theta)\, e^{j2\pi\omega r} d\omega \right] d\theta \qquad (3.8)$$

The inner part between square parenthesis is an one dimensional inverse Fourier transform of the $G(\omega, \theta)$ multiplied with ramp filter $|\omega|$, which is one dimensional inverse Fourier transform of a slice of projection information $p(r, \theta)$. The outer part is the integration of the inner part over each view. Before the development of filtered back projection reconstruction technique, back projection reconstruction technique was used without filter $|\omega|$, which yielded a blurred reconstructed image.

Using filtered back projection to reconstruct image of objects under investigation, it can be carried out with following steps:

1. Measure the projections $p(r, \theta)$ of the object $f(x, y)$ from different angle $\theta$

2. Calculate one dimensional Fourier transform of each projection $p(r, \theta)$

3. Multiply Fourier transform of each projection $p(r, \theta)$ with filter $|\omega|$

4. Calculate one dimensional inverse Fourier transform of the result in step 3

5. Take the summation of all results in step 4

### 3.1.2 Algebraic-based reconstruction technique

Although algebraic reconstruction technique is using the same integral projection $p(r, \theta)$ as FBP does to reconstruct the image of object $f(x, y)$ under investigation, it is an entirely different way. It takes each beam crossing the object as an algebraic equation consisting of unknown $f(x, y)$ and the measured projection data $p(r, \theta)$. Superficially it looks much simpler than the filtered back projection reconstruction technique, however, it is subject to intensive computation and lack of accuracy. Algebraic based reconstruction techniques could get more useful when ray path is not straight, or more physics process needs to be considered, or it is under sampling. The framework of algebraic reconstruction technique is shown in Figure 3.2.

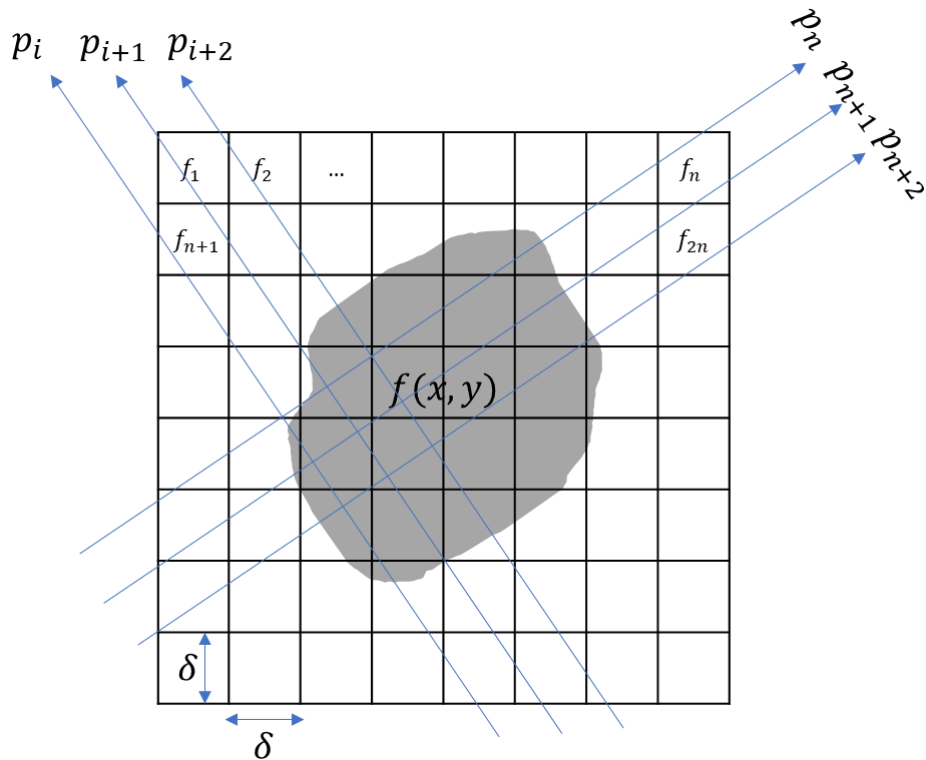

Figure 3.2       Scheme of algebraic method. The image value within each grid is assumed to be constant

The object is discretized into n×n cells with a side width of δ and assumed to be constant in each cell. Let $f_j$ be the constant value in the $j^{th}$ cell. Each ray is represented by an arrow line. Projection data $p_i$ is the summation of values in cells crossed by $i^{th}$ ray and expressed as:

$$\sum_{j=1}^{N} w_{ij} f_j = p_i \qquad i = 1, 2, \ldots, n^2 \qquad (3.9)$$

where $w_{ij}$ is a weighting factor representing the contribution of $j^{th}$ cell to the projection information $p_i$. If $j^{th}$ cell is not crossed by $i^{th}$ ray, then $w_{ij}$ is zero.

With M rays crossing the object, then this tomographic process can be described by M equations:

$$w_{11} f_1 + w_{12} f_2 + \cdots + w_{1N} f_N = p_1$$

$$w_{21} f_1 + w_{22} f_2 + \cdots + w_{2N} f_N = p_2$$

$$\vdots \qquad\qquad (3.10)$$

$$w_{M1} f_1 + w_{M2} f_2 + \cdots + w_{MN} f_N = p_M$$

Each equation can be deemed as a hyperplane in N-dimensional space represented by ($f_1$, $f_2$, …, $f_N$). If there is a unique solution to the above M equations, it would be at the intersection of all hyperplanes. One way to solve the M equations is to pick up an initial guess, denoted by ($f_1^{(0)}$, $f_2^{(0)}$, …, $f_N^{(0)}$), which are usually zeros and project this point to the first hyperplane to get the projection point ($f_1^{(1)}$, $f_2^{(1)}$, …, $f_N^{(1)}$). Then project the resulting projection point to the following hyperplane and continue doing so. This process can be described by Eq. (3.11) [45]

$$\mathbf{f}^{(i)} = \mathbf{f}^{(i-1)} - \frac{(\mathbf{f}^{(i-1)} \cdot \mathbf{w_i} - p_i)}{\mathbf{w_i} \cdot \mathbf{w_i}} \mathbf{w_i} \qquad (3.11)$$

where $\mathbf{w_i} = (w_{i1}, w_{i2}, \ldots, w_{in})$ is the coefficients of the $i^{th}$ equation in Eq. (3.10)

### 3.1.2.1 Algebraic reconstruction techniques (ART)

At most of the time $w_{ij}$ is simply replaced with 1 if $i^{th}$ ray crosses $j^{th}$ cell or 0 if $i^{th}$ ray does not cross $j^{th}$ cell. This simple change could make the whole process much easier to carry out during the run time. The physical meaning of Eq. (3.11) is smearing the difference between estimated value $\mathbf{f}^{(i-1)} \cdot \mathbf{w}_i$ and projection date $p_i$ over the cells crossed by $i^{th}$ ray. The resulting image by solving Eq. (3.11) is usually a poor approximation of the object $\mathbf{f}$ and suffers salt and paper noise. One way to alleviate these effects is by applying a relaxation $\alpha$ less than 1 to the difference between estimated value $\mathbf{f}^{(i-1)} \cdot \mathbf{w}_i$ and projection date $p_i$. $\alpha$ can be a constant for average use or get progressively smaller with the increase of iteration number for more sophisticated application.

### 3.1.2.2 Simultaneous algebraic reconstruction techniques (SART)

To further reduce noise caused by unavoidable small inconsistency with projection data $p_i$, Simultaneous algebraic reconstruction technique applies the corrections for all ray in one projection, which is capable of generating reconstructed image with good quality and numerical accuracy in one iteration.

$$f_j^{(K+1)} = f_j^{(K)} + \frac{\alpha \sum_i [w_{ij} \frac{p_i - \mathbf{w}_i^{\mathsf{T}} \mathbf{f}^{(K)}}{\sum_{j=1}^{N} w_{ij}}]}{\sum_i w_{ij}} \qquad (3.12)$$

## 3.2   Analogy between X-ray CT and muon CT

In transmission-based medical imaging, the incident beam is usually made of x-rays, which are neutrally charged particles. This beam typically undergoes photoelectric effect, Compton scattering, pair production (if E > 1.022 MeV), or no interaction, as it traverses through a patient or, most generally, an object. The incident beam often has a significant probability of experiencing Compton scattering in an object, which can scatter x-rays at large angles. Thus, the detected beam flux is typically the uncollided beam. As illustrated in Figure 3.3, let $I_0$ and $I$ be the incident beam and outgoing beam intensities, respectively. The object under investigation is represented by its attenuation coefficients and it is assumed that material in each voxel is homogeneous. The ratio $I/I_0$ is often used to reconstruct the investigated object using filtered back projection or algebraic based reconstruction technique. In Figure 3.3, the attenuated intensity $I$ can be described by:

$$I = I_0 e^{-d \sum_{i=1}^{n} \mu_i} \tag{3.13}$$

where $d$ is a selected discretized length in cm and $\mu_i$ is the attenuation coefficient of the $i^{\text{th}}$ pixel in cm$^{-1}$. After rearrangement, Eq. (3.13) becomes

$$\ln\left(\frac{I_0}{I}\right) = d \sum_{i=1}^{n} \mu_i \tag{3.14}$$

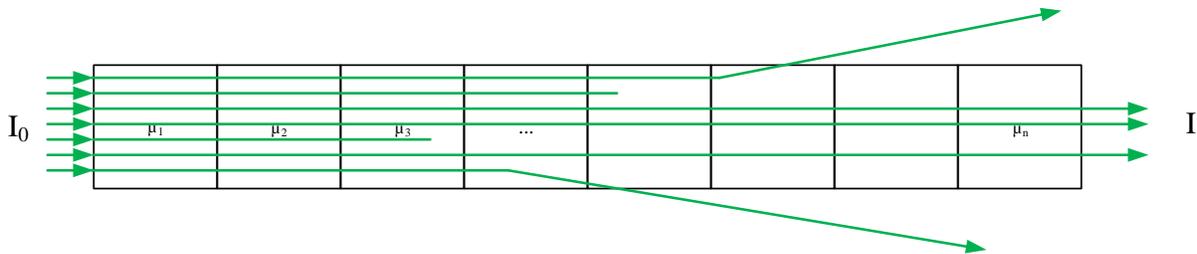

Figure 3.3        Illustration of a neutral beam crossing a discretized object.

The signal obtained from one projection or view is not enough to reconstruct an image. One typically rotates the radiation source and the detectors, while the object remains stationary, to obtain additional information from different angles.

In contrast to X-ray beam, muons are charged particle. They will not only experience ionization effect leading to energy loss and complete stop of part of muons in the material indicated by green lines without arrow but also undergo multiple coulomb scattering resulting in deflection from its incident direction indicated by red lines and green lines with arrow as shown in Figure 3.4. Thus, two potential image modalities can be used in muon computed tomography: using muon transmission rate $I/I_0$ to reconstruct the object in muon attenuation coefficients $\mu$ or using scattering angles to reconstruct the object in scattering density $\lambda$. Muon computed tomography using transmission rate is very similar to traditional X-ray computed tomography, which has been address at length above.

When muons traverse an object, many different scattering angles can be registered, which follow a Gaussian distribution with zero mean value and variance given by:

$$\sigma_\theta \cong \frac{15\text{MeV}}{\beta cp} \sqrt{\frac{L}{L_{rad}}} \qquad (315)$$

where p is the muon's momentum in MeV/c, L is the length of the object, and $L_{rad}$ is the radiation length of the material. The concept of a muon traversing an object is shown in Figure 3.4.

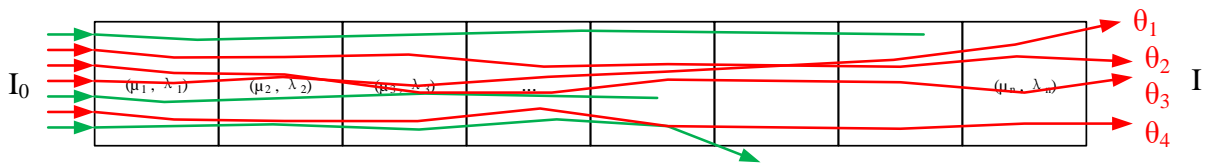

Figure 3.4    Illustration of a muon beam traversing a discretized object. The magnitude of the scattering angle is exaggerated in the figure for illustration purposes.

The variance of scattering angle of a monoenergetic muon beam caused by the $i^{th}$ voxel can be written as [46]:

$$\sigma_{\theta_i}{}^2 = d\lambda_i \tag{3.16}$$

where $\lambda_i$ is the scattering density of the $i^{th}$ pixel. The scattering density is defined as:

$$\lambda(L_{rad}) \equiv \left(\frac{15}{p_0}\right)^2 \frac{1}{L_{rad}} \tag{3.17}$$

where $p_0$ is the nominal momentum. Since MCS in individual voxels can be treated as independent, the variance of the scattering angle of a muon beam after traversing the entire object may be written as:

$$\sigma_{\theta}{}^2 = d\sum_{i=1}^{n}\lambda_i \tag{3.18}$$

Note that Eq. (3.14) and Eq. (3.18) share the same form, i.e., the right side of these two equations is a linear integration of a characteristic parameter over the particle's path. Thus, the scattering density $\lambda$ may be treated similar to the attenuation coefficient $\mu$ used in the x-ray computed tomography image reconstruction process.

Before jumping to image 3D objects, a simple simulation experiment was conducted to verify this initial thought on building muon computed tomography technique using muon scattering angle. Six cubes made of Al, Fe, Cu, Pb, W, and U were simulated in GEANT4 to capture the process of muon traversing different materials. Each cube has a side width of 21.4 cm. These cubes were placed between two pairs of position sensitive detectors. Parallel monoenergetic muon beams were generated to shine the objects. The muon source and detectors were rotated simultaneously for 90 times at an azimuthal angle increment of 2 degrees to generate 90 views. All things were on the same horizontal plane as shown in Figure 3.5. The scattering angles were calculated for each registered muon and FBP was used for image reconstruction. The reconstructed image is shown in Figure 3.6. The simulated and reconstructed image are in very good agreement. Next, the scattering density of each cube was estimated with the reconstructed image using two pixel sizes, 1 cm × 1 cm and 2 cm × 2 cm, and was compared with the theoretical values. The theoretical and estimated scattering densities are shown in Table 3.1. Although slightly different values are observed between the actual and reconstructed scattering density, the values are within 3σ, with the exception of tungsten. This result can be attributed to shadowing effect from the surrounding materials. The results demonstrated that the proposed muon CT framework can accurately reconstruct geometrical and material information.

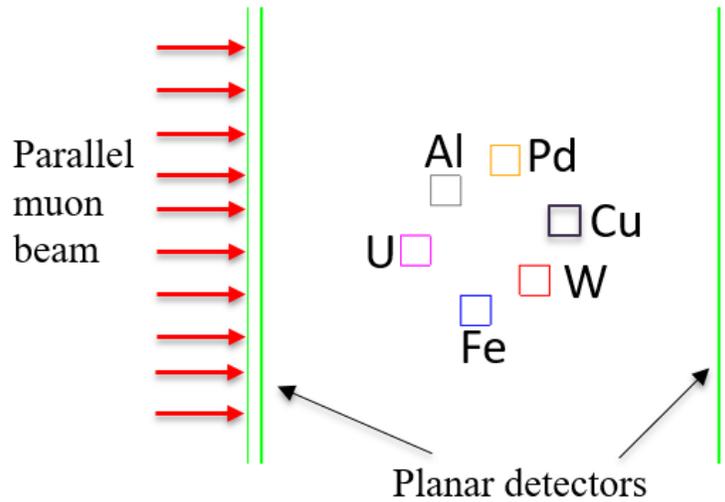

Figure 3.5        Top-down view of configuration of six cubes, detectors and muon beam

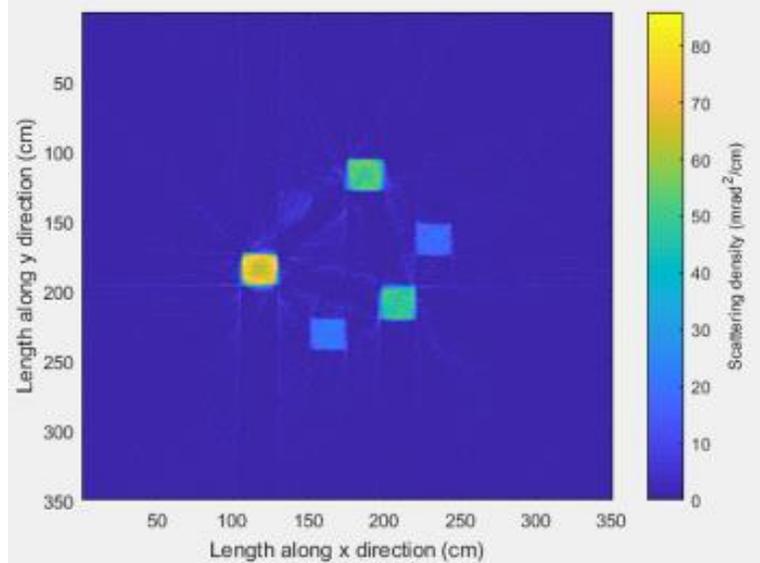

Figure 3.6        Reconstructed image of six cubes using FBP [42]

With the successful realization of muon computed tomography using scattering angle created by parallel monoenergetic muon beam crossing objects on a 2-dimensional plane demonstrated above, efforts to create muon computed tomography in 3 dimensions using cosmic is shown below. There are two possible setups for μCT as shown in Figure 3.7: setup 1 is letting muon detectors be horizontally placed and rotate around y axis, and setup 2 is letting muon detectors be vertically placed and rotate around z axis. From the point of view of shortening measurement time, setup 1 is better than setup 2 because cosmic ray muons are preferentially showering down from vertical direction. This setup can be used to screen cargos and trucks for special nuclear material, where it is possible to put a pair of muon detector underneath the objects that are under interrogation. However, in this work the object under interrogation is dry storage cask containing 24 or even more spent nuclear fuel assemblies, which makes it impossible to put a pair of muon detectors underneath, thus setup 2 was adopted in the algorithm development and reconstruction process. In order to increase muon flux rate crossing the detectors, a vertical offset is applied to the setup as shown in Figure 3.7 on the right.

Table 3.1         Actual and reconstructed scattering density of six materials

| Material | Scattering density (mrad²/cm) | | |
|---|---|---|---|
| | Theoretical | Estimated (pixel=1 cm) | Estimated (pixel=2 cm) |
| Al | 2.81 | 4.5±2.9 | 4.4±2.5 |
| Fe | 14.22 | 17.8±3 | 17.8±1.5 |
| Cu | 17.41 | 21.0±4 | 21.0±1.4 |
| W | 71.35 | 54.5±4 | 52.3±2.6 |
| Pb | 44.55 | 47.6±5 | 47.5±2 |
| U | 78.96 | 72.2±7 | 71.3±4 |

Unlike the above demonstrative simulation experiment where monoenergetic and parallel muons can be created in GEANT4 workspace, in reality muons are created in high atmosphere and shower down to the ground from 0 degree to almost 90 degree zenith angle and isotopically in $2\pi$ azimuthal angle. Two steps are used here to collect quasi parallel muon beams. Let the position of $i^{th}$ muon be $(x_{1i}, y_{1i}, z_{1i})$, $(x_{2i}, y_{2i}, z_{2i})$, $(x_{3i}, y_{3i}, z_{3i})$, and $(x_{4i}, y_{4i}, z_{4i})$ on four detectors separately as labeled in Figure 3.7, two before the object and two after. The incident azimuth angle of the $i^{th}$ muon $\varphi_i$ is

$$\varphi_i = \arctan\left(\frac{y_{2i} - y_{1i}}{x_{2i} - x_{1i}}\right) \tag{3.19}$$

The scattering angles $\theta_i$ is calculated with:

$$\theta_{ix} = \operatorname{atan}\left(\frac{x_{4i} - x_{3i}}{z_{4i} - z_{3i}}\right) - \operatorname{atan}\left(\frac{x_{2i} - x_{1i}}{z_{2i} - z_{1i}}\right) \tag{3.20}$$

$$\theta_{iy} = \operatorname{atan}\left(\frac{y_{4i} - y_{3i}}{z_{4i} - z_{3i}}\right) - \operatorname{atan}\left(\frac{y_{2i} - y_{1i}}{z_{2i} - z_{1i}}\right) \tag{3.21}$$

$$\theta_i = \sqrt{\frac{\theta_{ix}^2 + \theta_{iy}^2}{2}} \tag{3.22}$$

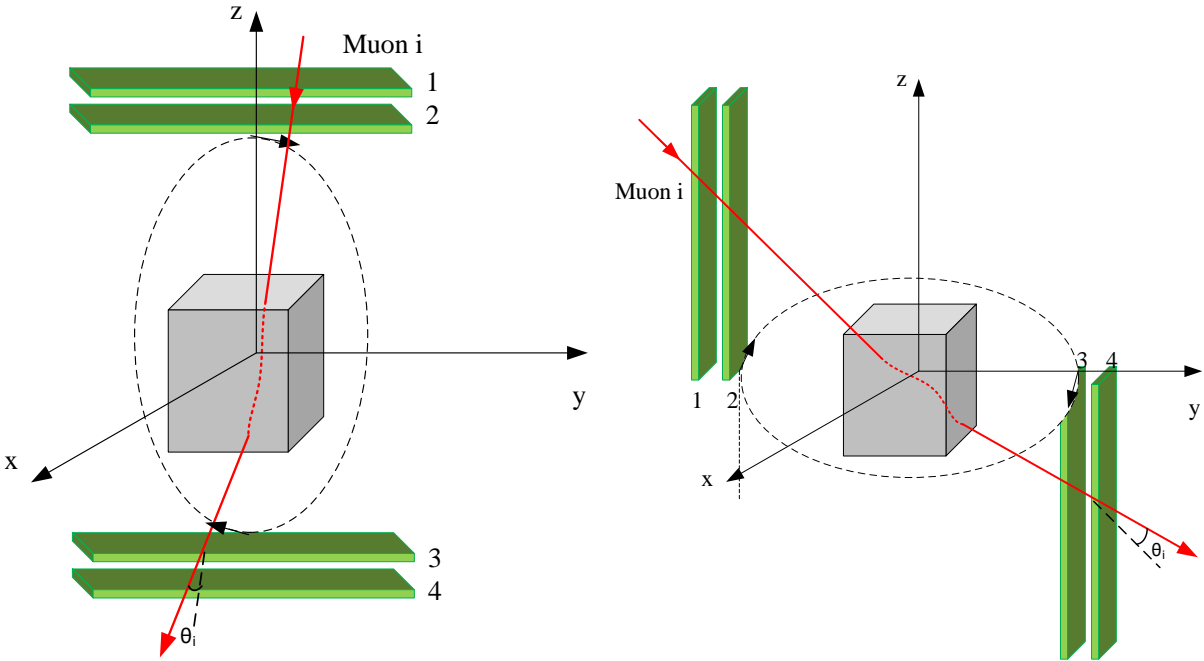

Figure 3.7        Two possible μCT setups: setup 1 is on the left and setup 2 is on the right.

In order to obtain parallel or quasi parallel muon beams, the detectors are rotated around the object under interrogation for π degree with an angle increment of π/(n-1), where n is number of views intended to be used.

For a monoenergetic muon beam crossing a homogeneous object, its scattering angles follow a gaussian distribution $(0, \sigma_\theta^2)$ described in Eq. (3.15). For a polyenergetic muon beam crossing a homogeneous object, these muons with larger kinetic energy tend to scatter less than those who has smaller kinetic energy. In order to make a single gaussian model better describe scattering angle distribution of a polyenergetic muon beam crossing an object, its momentum can be used to correct the influence of muon kinetic energy on its scattering angle. One way is to normalize the scattering angle relative to a nominal momentum as in Eq. (3.23)

$$\theta_i{}' = \frac{p_i}{p_0}\theta_i \qquad (3.23)$$

where in this work the nominal momentum $p_0$ is chosen to be 3 GeV/c and the quantity $p_i$ is chosen to be muon initial momentum. It is assumed that there is no energy loss in the process of muon crossing objects. There may be other better ways to correct the influence of muon momentum on its scattering angle from theoretical stand or from the stand of fitting of simulated or measured scattering angles of muons crossing a known simple substance, which will be discussed in section 6.2.1 the future work and APPDIX B. If momentum information is not available, it will be assumed that all muons are of the same momentum $p_0$ and no momentum correction for scattering angle will be applied.

Due to the nonuniform incident position and direction of muons, it would be hard to construct 3D muon computed tomography using cosmic ray muons, which will be addressed in section 6.23 the future work. Fortunately, because of the uniformity along Z axis direction of our investigation subject spent nuclear fuel dry storage cask, it is possible to collapse the registered muon data down to a two-dimensional plane without losing the fidelity to object under interrogation, which could significantly simplify the reconstruction process.  In the following, 3D muon information will be collapsed down to a horizontal plane by applying length correction. Besides muon momentum, the magnitude of muon scattering angles also has a dependence on the depth it traversed in the object. The longer distance muon travels in an object, the bigger the scattering angle will be. The incident muons from different zenith angle have different path length in image reconstruction volume enclosed by two pair of muon detectors. However, we were trying to project the registered muons down to a horizontal plane, thus difference of muon scattering angles caused by different path length needs to be normalized. One way to use path length as a correction for the influence of different path length along vertical direction is describe in Eq. (3.24), after path length correction the normalized scattering angle of a muon becomes

$$\theta_i{''} = \sqrt{\frac{L_{ih}}{L_i}} * \theta_i{'} \qquad (3.24)$$

where $L_i$ is the distance between point $(x_{2i}, y_{2i}, z_{2i})$ and point $(x_{3i}, y_{3i}, z_{3i})$, and $L_{ih}$ is the horizontal projection of $L_i$. Finally, the registered incident muon spectrum can be divided into $\pi/N$ -degree-wide bins according to their incident azimuth angle $\varphi$, re-sorting the incident muons into N quasi-parallel groups and projecting registered muon information down to a horizontal plane.

## 3.3   Muon path models and projection methods

Three different muon tracing algorithms are proposed and investigated. These are: (1) use of a straight path along a muon's incident trajectory, (2) use of a straight path along a muon's incident direction crossing its PoCA point, (3) use of a muon's POCA trajectory. For trajectory model 1, path length in each pixel can be calculated in a 2-dimensional plane directly using muons' incident horizontal direction. For trajectory 2 and 3 which use PoCA point or something similar to PoCA point, their trajectories need to be calculated in 3D space first then project down to a horizontal plane, otherwise it is very likely to misrepresent muon trajectory in the process of collapsing 3D muon information down to 2D plane especially when incident trajectory and exiting trajectory does not insect.

For each muon tracing algorithm, three different scattering angle projections methods were tried: projection method **a**, each scattering angle is used only once and stored directly into the corresponding detector bins to generate variance of scattering angle; projection method **b,** each scattering angle is back projected into the image space first to calculate the variance of the scattering angle in each pixel, then the summation of the variances weighted by average path in corresponding pixel is forward projected to the corresponding detector bins; or projection method **c,** similar to projection method **b,** each scattering angle is back projected into the image space and weighed by its path length in corresponding pixel first then to calculate the variance of the weighted scattering angle in each pixel, finally the summation of the variances in corresponding pixel is forward projected to the corresponding detector bins. Trajectory method **1** along with projection method **a** is widely adopted, however, it has a remarkable drawback to yield low quality result of reconstructed image especially when muon momentum is not available, which is close to the real case due to difficulty to measure the muon's momentum. Projection method **b** and **c** was put forward by the author to reduce or get around the reliance of reconstructed image quality on muon momentum. Three muon tracing algorithms and three scattering angle projection methods result in nine combinations: 1**a**, 1**b**, 1**c**, 2**a**, 2**b**, 2**c**, 3**a**, 3**b** and 3**c.** These methods were used to generate projection information and corresponding system matrices to investigate how muon ray tracing algorithms and scattering angle projection methods affect muon CT reconstruction image quality and detection capability. Meanwhile it was assumed that muon beam has the same width as the detector bin size.

Tracing algorithm 1: Use of a straight path along the incident muon trajectory (straight path tracing)

   This algorithm assumes that a muon experiences no scattering events or that scattering is negligible. It results in a straight path crossing the object along the muon incident trajectory, and completely ignoring the exiting position. The scattering angle can be back projected in three different ways:  method 1**a** is used to directly store the scattering angles for each muon from the same quasi-parallel beam subset into the corresponding detector bins hit by its path, then the variance is calculated from the scattering angles in each bin and used as projection information P. Method 1**b** is used to back project each scattering angle into the pixels crossed by this straight path for all muons in the same quasi-parallel beam subset, and then to calculate the variance of the scattering angle in each pixel, and, finally, it takes the summation of the variances weighed by the average path length in corresponding pixels along this path and stores it into the corresponding detector bin as projection information P. Method 1**c** is used to back project each scattering angle into the pixels crossed by this straight path and weighted by $L_{ij}/L_i$ for all muons in the same quasi-parallel beam subset, where $L_i$ is the total path length of $i^{th}$ muon and $L_{ij}$ is the path length of $i^{th}$ muon in the $j^{th}$ pixel, and then to calculate the variance of the weighted scattering angle in each pixel, and, finally it takes the summation of the variances along this path and stores it into the corresponding detector bin as projection information P.

Both filtered back projection (FBP) and simultaneous algebraic reconstruction technique (SART) were used to reconstruct the image using projection information P. For FBP, one simply applies a high pass filter to the projection information P stored in the detector bins before back projecting it into the image domain. For SART, the average path length of the muons in each pixel in the same quasi-parallel beam subset is used to build a system matrix W.

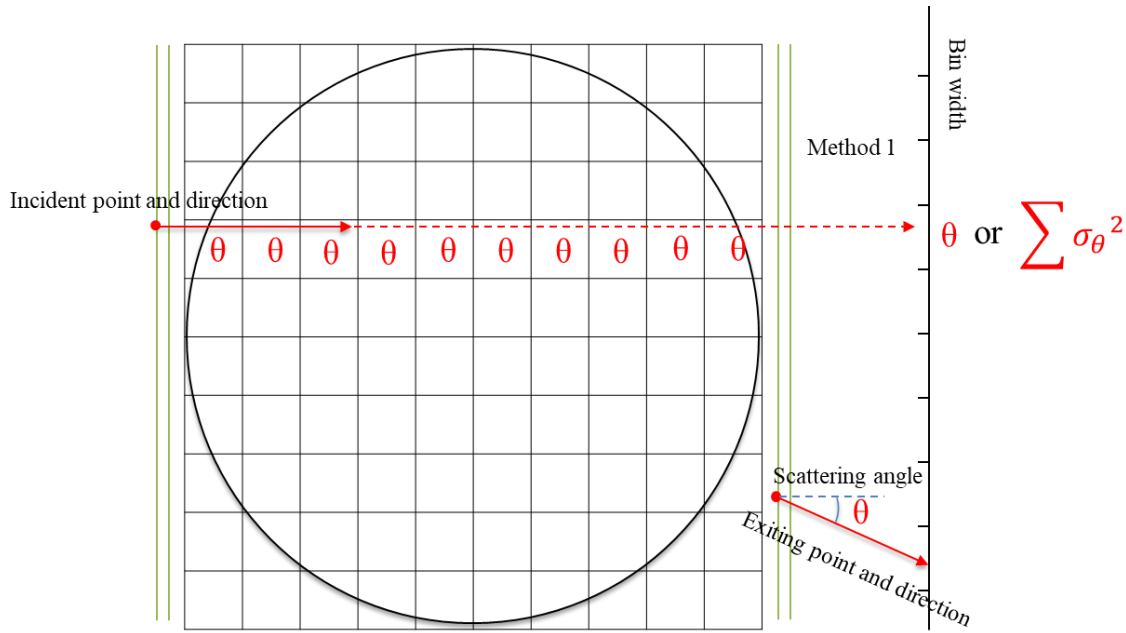

Figure 3.8    Top-down view illustration of tracing algorithm 1 (straight path tracing). Approach 1**a** projects the scattering angle within a defined volume (shown as a large square, the black circle indicates dry storage cask) along a straight line to a segmented detector (shown at far right). In approach 1**b**, $\theta_i$ is back projected into the pixels crossed by this straight path indicated by the red dashed line. In approach 1**c**, $\theta_i$ is back projected into the pixels crossed by this straight path indicated by the red dashed line and weight $L_{ij}/L_i$, which is not shown in the above figure.

Tracing algorithm 2: Use of a straight path along the muon incident direction that crosses the PoCA point (PoCA point tracing)

This algorithm assumes that a muon experiences a single Coulomb scattering event within a defined volume. The scattering angle is taken to occur at the closest point distance between the incident and exiting trajectories. This point is also known as the PoCA point. Instead of completely ignoring the exiting position, tracing algorithm 2 makes a compromise between a muon's incident and exiting positions by assuming the muon crossed the object along incident direction crossing the PoCA point, as shown in Figure 3.9. The rest of the steps are the same with tracing algorithm 1.

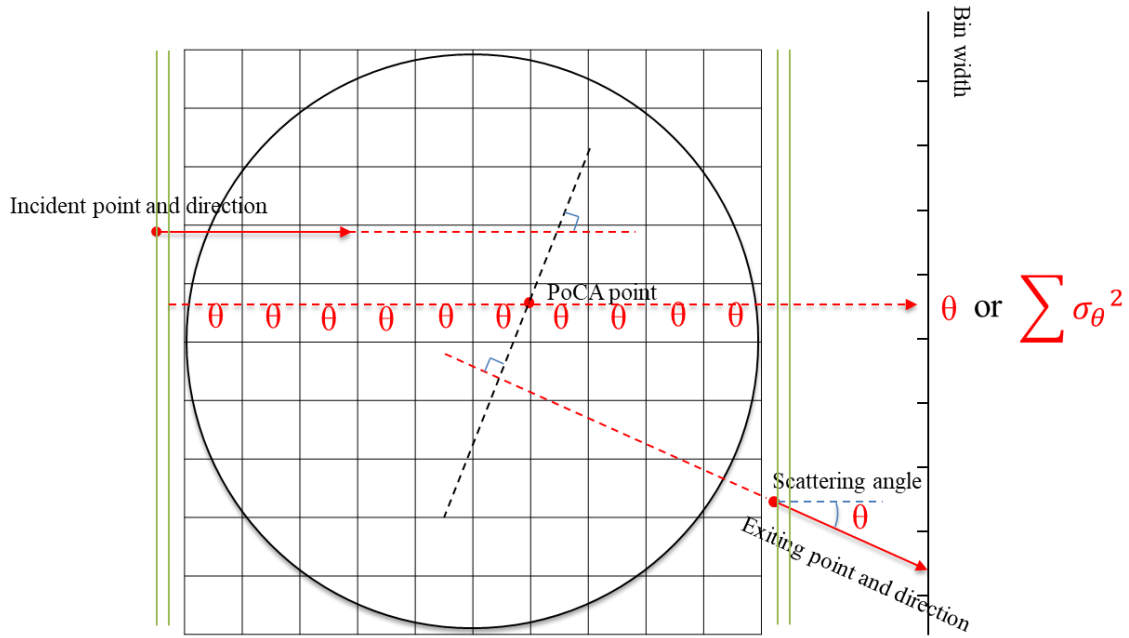

Figure 3.9    Top-down view illustration of tracing algorithm **2** (PoCA point tracing). Approach 2**a** projects the scattering angle within a defined volume (shown as the large square, the black circle indicates dry storage cask) along a straight line to a segmented detector (shown at right). In approach 2**b**, $\theta_i$ is back projected into the pixels crossed by this straight path indicated by the red dashed line. In approach 2**c**, $\theta_i$ is back projected into the pixels crossed by this straight path indicated by the red dashed line and weight $L_{ij}/L_i$, which is not shown in the above figure.

Tracing algorithm 3: Use of PoCA trajectory (PoCA trajectory tracing)

   As a heavy charged particle, muons experience multiple Coulomb scattering when they traverse objects, causing it to deviate from a straight path. Thus, a curve path may better approach muons trajectory crossing objects than a simple straight line. This tracing algorithm assumes that a muon travels along the so-called PoCA trajectory within our defined volume. The PoCA trajectory consists of two segments: (1) the segment connecting the point of muon incidence to the PoCA point and (2) the segment connecting the PoCA point and the point at which the muon exits said volume, as descried in Figure 3.10. The rest of the steps are the same as those described in tracing algorithm 1.

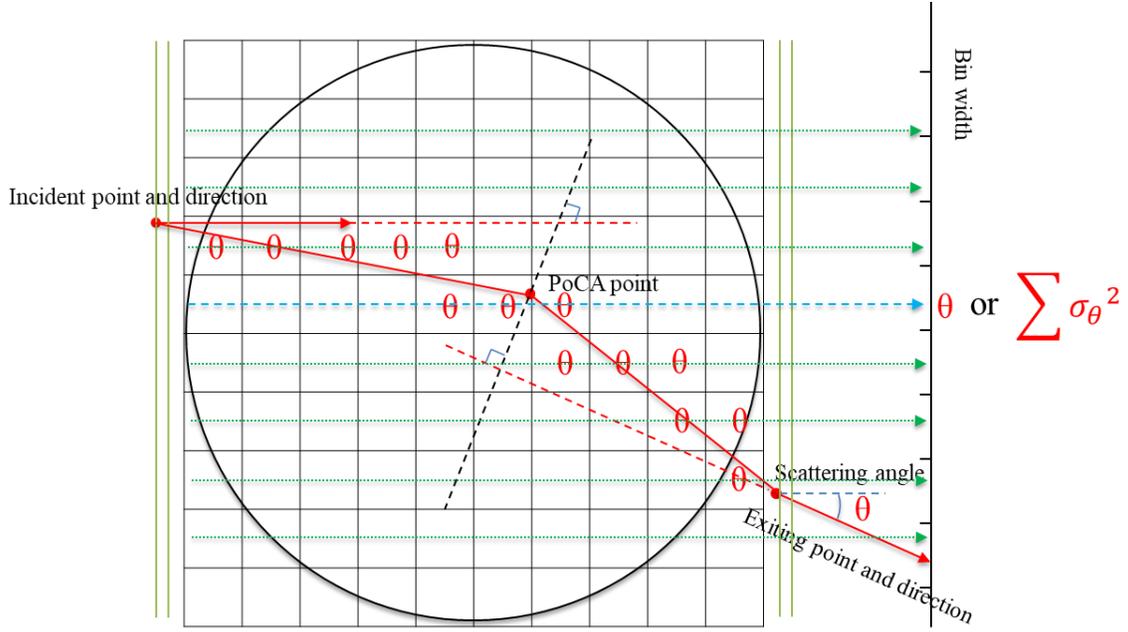

Figure 3.10    Top-down view illustration of tracing algorithm 3 (PoCA trajectory tracing). Approach 3**a** projects the scattering angle along the blue dashed line to a segmented detector (shown at far right), and the system matrix is calculated with the PoCA path indicated by the red segments. According to approach 3**b**, the scattering angle is projected back into the pixels crossed by the PoCA trajectory for all muons in a quasi-parallel beam subset, the variance in each pixel is calculated, and the sum of the variances of scattering angle along its incident horizontal direction indicated by green line is stored in the corresponding detector bins. Approach 3**c** is similar to approach 3**b** except that the scattering angle is weighted by $L_{ij}/L_i$.

Due to the poor result of using projection method **c**, this project method was not reported in the section of results and analysis in Chapter 4. It is only shortly mentioned here to let readers know that there are many other ways to project scattering angles. There could be an even better one.

## 3.4   Image reconstruction

Once the system matrix **W** and projection information **P** are obtained, the image reconstruction process is the same with X-ray CT using transmission rate. In this dissertation FBP is used for comparison purpose. Due to under sampling and muon curve path, SART is expected to achieve a much better result.

A certainty number of views were used to cover 180 degrees. The object to be reconstructed was discretized into N×N pixels, and the scattering density was expressed as a $N^2 \times 1$ vector X. The iterative reconstruction process was stopped when the maximum iteration number was reached or when the difference in successive iterations was below a car certain threshold.

A pseudocode is detailed below:

1) Collect incident and exiting muon position $(x_{1i}, y_{1i}, z_{1i})$, $(x_{2i}, y_{2i}, z_{2i})$, $(x_{3i}, y_{3i}, z_{3i})$, and $(x_{4i}, y_{4i}, z_{4i})$ for M views

2) Calculate scattering angle θ, PoCA point and azimuth angle for each muon.

3) Correct scattering angle $θ_i$ with path length $L_i$ and muon momentum $p_i$ (if available).

4) Re-sort muons into quasi-parallel subsets according to their azimuth angle φ.

5) Discretize the reconstruction volume into N×N pixels.

6) For each muon quasi-parallel subset,

   if use projection method **a**, do:

   a) Calculate the path length in pixels crossed by each muon in this subset using of the three trajectory models. At the end of the last muon in this group, take the average path length in each pixel and store them in system matrix W.

   b) Project scattering angle $θ''$ into corresponding detector bins along it incident direction. At the end, calculate the variance $σ_θ^2$ of scattering angle in each bin and store them in P.

   if use projection method **b**, do:

   c) Similar to 6a.

d) Back project the scattering angle $\theta''$ of each muon in this subset into pixels crossed by its trajectory. At the end of last muon in this group calculate the variance of scattering angle in each pixel

e) take summation of variance of scattering in pixels along incident horizontal direction of beam and store it in P

7) Solve equation WX = P using SART or SIRT with or without regularization.

## 3.5 Summary

This chapter started with a quick review of traditional X-ray computed tomography technique along with its analytical reconstruction method FBP and a couple of algebraic based reconstruction methods. A detailed analogy between X-ray CT and a potential muon CT was drawn. The essence of X-ray transmission tomography is utilizing the fact that the negative natural logarithm of X-ray transmission rate is equal to the linear integral of X-ray attenuation coefficients of the object along its incident direction. Similarly, for muons who made it through the object, the variance of scattering angles of these muons is also a linear integral of scattering density of the crossed object along muon's trajectory. Due to the same form of Eq. (3.14) and Eq. (3.18), muon computed tomography using muon scattering angle was once born. To obtain muon beams, detectors were rotated around the object to cover 180 degrees. Then the registered data was collapsed down to a horizontal plane with a length correction along with momentum correction if muon momentum is available. Finally, these muons were resorted into quasi-parallel muon beams according to their incident horizontal direction. Because of multiple Coulomb scattering, muon's path is subject to bending. Three different path models were introduced to approximate real muon trajectory. Two different projection methods were put forward to utilize muon scattering angle. Three path models and two projection methods result in 6 methods in combination. Both FBP and SART can be used for image reconstruction.

# 4 DRY STORAGE CASK SIMULATIONS AND RECONSTRUCTION

## 4.1 Dry storage casks

So far, a variety of dry storage technologies have been approved by NRC for storage purpose or storage and transport dual purposes. Many of them are still being used for storing spent nuclear fuel at nuclear power plant or at storage facility site. According to how dry storage cask is placed, they can be classified into vertical cask and horizontal cask. Per material of cask itself, they can be categorized into metal cask and concrete cask. Some common casks are VSC-24, CASTOR V/21, CASTOR X, HI-STAR 100, NAC MPC, NAC UMS, NAC MAGNASTOR, TN Metal Casks, Westinghouse MC-10 [47].

### 4.1.1 VSC-24

VSC-24 cask is vertical concrete cask holding 24 spent nuclear fuel assemblies. It is composited of a multi-assembly sealed basket (MSB), a ventilated concrete cask (VCC) and an MSB transfer cask (MTC). The confinement and criticality of cask for storage and transfer are monitored by the welded MSB. ventilated concrete is used to shield radiation emitted by the spent nuclear assemblies inside the cask and used to cool down the fuel by allowing natural convention. MSB is made up of a steel cylindrical shell and two steel cover plates welded to both end of the cylindrical shell. The steel cylinder is about 420 cm to 460 cm tall in a diameter of 161.65 cm and 2.54 cm thick. There is an internal basket in inside cylindrical shell to fix 24 PWR spent nuclear fuel assemblies. This internal basket has 24 square storage slots. In case that the basket would get eroded by the spent nuclear fuel and water, the basket is plated with Carbon Zinc. VCC is used to reinforce the concrete cask outside of the steel cask. It has four air inlets and four outlets at the top and bottom of the concrete cask. During the fuel loading, MTC plays the role to shield and protect VCC.

### 4.1.2   CASTOR V/21

CASTOR V/21 was developed by General Nuclear Systems, Inc and received its certificate of compliance for use under general license in 1990. CASTOR V/21 is metal cask designed to store 21 PWR assemblies in vertical direction. It is 192 inches tall in a diameter of 2.44 m. When CASTOR V/21 cask is fully loaded, it weighs up to 117 tons. CASTOR V/21 cask body is made up of thick nodular iron casting with two lids bolted to cask body. The leak tightness between the cask main body and lids are maintained by metal seal. Thick steel cask wall not only provide structural integrity but also shields the Gamma ray and neutron particle from the spent nuclear fuel inside. Different from VSC-24 concrete cask using thick concrete overpack to shield neutrons, polyethylene rods are incorporated into the CASTOR V/21 cask wall to shield neutrons because steel relative low neutron shielding capability. Outside of metal cask is covered with fins to increase contact area of air and cask to transfer hear from away the cask, which may cask the overheat of cask and damage cask structure.  Inside of the cask there is a basket with 21 square tubes made of borate plated stainless steel. The inner surface of the cask is coated with nickel to protect it from corrosion. Pressure sensor is used to monitor the pressure in the interspace between primary and secondary lids to make sure of the leak tightness of cask.

### 4.1.3    CASTOR X

CASTOR X/33 is a metal storage cask with the capacity to hold 33 PWR spent nuclear fuel assemblies in vertical direction. This cask is about 480 cm high and 238 cm in diameter. When it is fully loaded, it weighs up to 118 tons. It was approved by NRC for storage in April 1994. CASTOR X/33S' cask body is made of 12 inches thick ductile cast iron. On both top and bottom sides, it is sealed with stainless steel lid along with metallic and elastomeric O-ring seals. Cask wall can be used to shield the gamma ray emitted by spent nuclear fuel. Similar to metal cask, polyethylene rods are incorporated into the cask wall to shield neutrons also coming from the spent nuclear fuel. A stainless steel basket with 33 square fuel tubes is used to hold spent nuclear fuel assemblies.  Borate is coated to the basket for criticality control. nickel coating is applied on the cask inner surface and epoxy resin is coated on the external surface of the cask for corrosion protection.  CASTOR X also has a pressure monitoring system to make sure of the leak tightness integrity.  Under a site-specific license the CASTOR X/33 is currently being used at Virginia Power's Surry ISFSI.

### 4.1.4    MC-10

MC-10 is metal cask designed by Westinghouse to store 24 PWR spent nuclear assemblies in vertical direction. The main body of MC-10 cask is a thick forged steel cylinder. Inside the steel cylinder is a cavity used to hold fuel basket. The whole MC-10 cask including fins is 188.4 inches high and 106.8 inches in diameter. The bottom is used for gamma ray shielding and structural integrity.  Primary containment seal is made possible by a carbon steel cover and a metallic O-ring seal. Also, on top of the primary containment seal, a second seal lid is also applied to maintain leak tightness.  Instead of using polyethylene rods, an additional cover contains BISCO NS-3 is used for neutron shielding. Inside of internal surface of MC-10 cask, there is thermally sprayed aluminum used to protect it from corrosion. Outside of external surface of MC-10 cask, 24 cooling fins are welded to cask body to transfer the heat generated by spent nuclear fuel from the cask in case of overheat. Fins are also connected by carbon steel plates.

## 4.2    Geant4 model validation

GEANT4 is the abbreviation of GEometry ANd Tracking code package using Monte Carlo methods [48][49][50]. It is the successor of GEANT series software developed by CERN for simulation of the passage of particle through matter. GEAT4 collaboration is taking care of its development and maintenance. It has very wide application areas, including high energy physics, nuclear engineering and accelerator physics, as well as studies in medical and space science. In this dissertation, most of the data used for image reconstruction is from GEANT4 simulation due to our limited resource (large area position sensitive detector) and access to a dry storage cask. However, there is a long history of doubt about the accuracy of GEAN4 simulation. Thus, in order to make our following reconstruction process more trustworthy and convincing, a GEANT4 model validation against physical experimental is presented below.

So far there are only two physical experiments [7][8] done on imaging a dry storage cask. Both measurements were carried out on a MC-10 metal dry storage cask with partial loading at Idaho National Laboratory by muon group of Los Alamos National Laboratory. Two experiments are quite similar to each other, except that in second experiment muon detectors on both sides of dry storage cask were shifted for 3 times to increase the field of view in order to create a full projection of the MC-10 cask. Because the second field experiment provided more information about spent nuclear assemblies configuration in the dry storage cask, we chose to validate our GEANT4 model against this experiment. In the physical experiment, the MC-10 dry storage is partially loaded for experimental purpose with 0, 1, 6, 5, 4 and 2 fuel assemblies in column 1 to column 6 separately. Two identical muon tracking detectors were placed on opposite sides of this MC-10 cask. Each muon tracker is about 1.2 m wide and wide and 61 cm thick. Because muons preferentially shower down to ground form vertical direction, in order to increase the useful muon flux rate crossing these muon detectors, one muon detector was elevated by 1.2 m. The diameter of MC-10 dry storage cask is about 2.7 m, which is more than twice longer than the width of muon tracking detector. Thus, in order to generate a full projection view of the dry storage cask, both the upper and lower muon tracking detectors were horizontally translated for three times in a combination of 9 different locations for this pair of muon tracking detectors as shown in Figure 4.1 on the left.

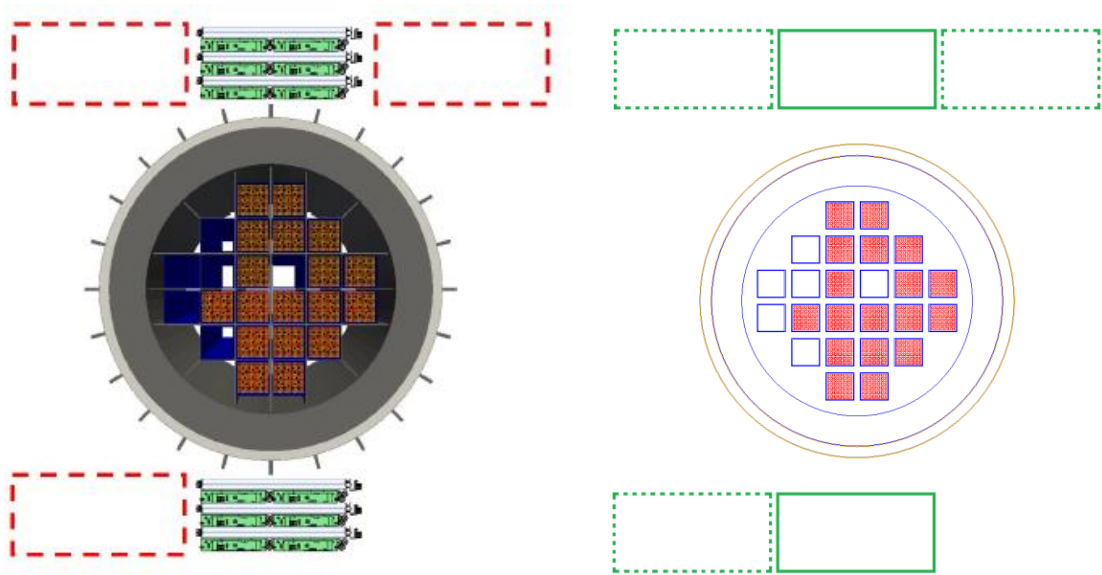

Figure 4.1      MC-10 cask configuration used in LANL experiment [15] (left) and GEANT4 simulated cask (right). The simulated detector positions are shown in green, the fuel assemblies in red, and the empty locations in blue.

As reported in [8], due to the strong wind during measurement when one muon tracking detector was placed at the lower right position (which is not shown in Figure 4.11 on the right), the detector suffered unexpected motion. Thus, data collected by this configuration was rendered unusable and discarded. In our simulation, this configuration was simply skipped, because the data collected with this configuration was not used in the image processing.

In despite of the heavy shielding made up of steel and BISCO NS-3, there is a strong radiation dose at the muon tracking detector surface emitted from the spent nuclear fuel: about 10 mrem/hour neutron dose and about 10 mrem/hour gamma rays dose. In order to reduce the false coincidence rate, a time window of 600 ns was introduced to consider particles falling in this time window as true muon event. With maximum drafting time of about 1 µs and by setting up this time window, muon tracking efficiency would be reduced by half. In our simulation workspace, the fins on cask surface and the basket used to fix spent nuclear were not simulated as shown in Figure 4.1, meanwhile decay process of spent nuclear fuel was not simulated either. The detectors were modelled to have 100% efficiency. The exact configurations of the muon tracking detectors and the MC-10 storage cask were adopted in an effort to reproduce the physical experiment. In the physical experiment, $4 \times 10^4$ to $9 \times 10^4$ muons crossed two muon tracking detectors in each configuration. In our simulation, the muon generator described in [51] was used to generate muons. For each configuration, about $6.5 \times 10^4$ muons crossed both the upper and lower muon tracking detectors. In data processing, the volume between two detector planes was discretized into voxels with dimensions of 2 cm in horizontal direction, 4 cm in the direction between two detector and 2.4 meter in vertical direction. The scattering angle of each muon was placed in the voxel where its PoCA point is located. Then the average value of scattering angle in each voxel is calculated. A comparison of experimental and simulated values of the average scattering angle in a slice crossing the center of the cask is shown in Figure 4.2. The results show that our simulation is reasonably close to the experimental measurements, except in the region from -20 cm to 40 cm where a maximum discrepancy of 3 mrad is observed. In particular, the cask wall and columns no. 2, 3, 4, and 5 are close to the experimental measurements—close enough so that one can argue that the Geant4 simulation should have some predictive value but not close enough to argue that they are in good agreement. One would expect column 1 to have lower experimental values than column 2 given that it contains no assemblies. This effect is captured by the simulation experiment but not by the physical experiment.

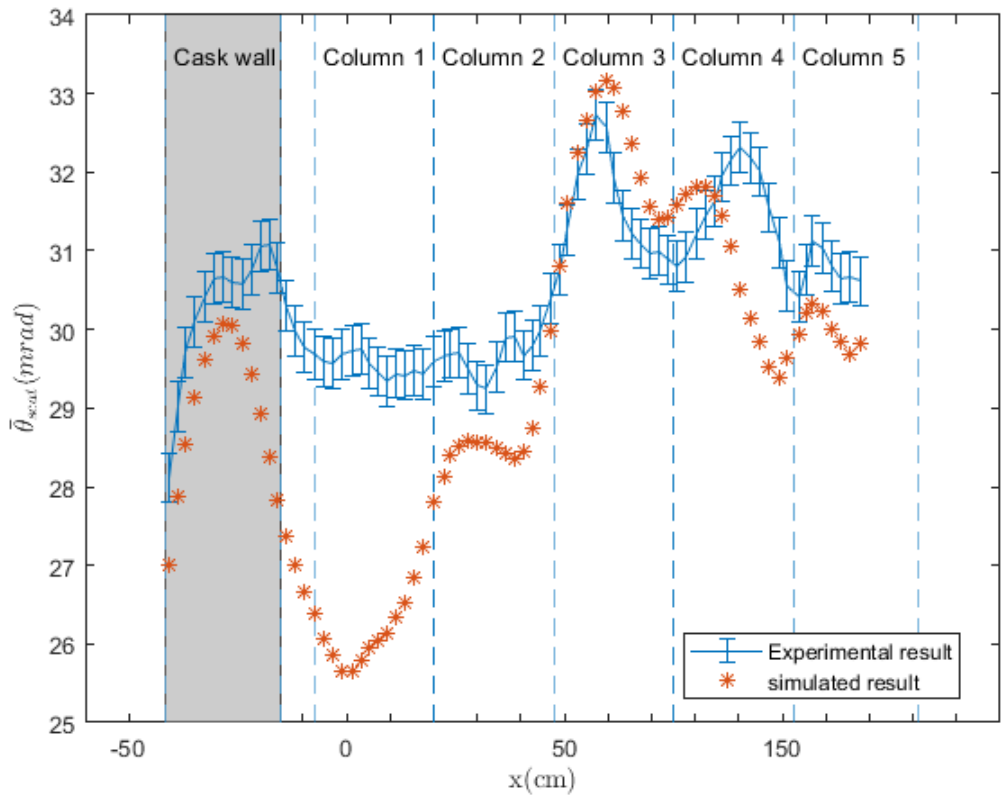

Figure 4.2     Experimentally measured [15] (blue line) and GEANT4 simulated (orange asterisk) average scattering angles for muons crossing a MC-10 dry storage cask.

Possible reasons that this was not captured in the experimental measurements are that strong winds shook detector at the time of the experiment as reported in [8] and that detection efficiency, especially geometrical efficiency was far away from perfect as in the simulation. We suggest that more measurements on dry nuclear fuel storage casks should be made by others so that the agreement between simulation and measurements can be better quantified.

## 4.3   Experiment configuration

After validating the GEANT4 simulation against one of two existed physical experiment on a dry storage cask, we started to illustrate and evaluate the proposed muon tracing algorithms and projection methods using filtered back projection and simultaneous algebraic reconstruction methods as discussed in Chapter 3. In order to better represent the population of dry storage cask in use, a concrete walled VSC-24 dry cask storing commercial spent nuclear fuel assemblies was chosen to be our object under investigation. This cask contains fuel assemblies in a thin steel canister that is shielded externally by a thick concrete overpack. Given the physical experiment result that multiple fuel assemblies missing could be detected by one projection view, thus one spent nuclear fuel assembly in the middle of dry storage cask was picked up as the initial spent nuclear fuel missing pattern. Two pairs of identical planar detectors with dimensions 350 cm×150 cm were simulated in GEANT4 to register crossing position and direction of incident and exiting muons. In order to shorten the measurement time, one pair of planar detectors were escalated by 100 cm in vertical direction to increase the muon flux rate crossing the dry storage cask. The separation between each planar detector was 10 cm. The magnitude of separation between a pair of muon detectors would affect the measured scattering angle when the muon detector positional uncertainty is considered, which will be discussed soon in Chapter 5. In this chapter, muon detectors are assumed to have perfect efficiency and position resolution. With this configuration as shown in Figure 4.3, the zenith angle of incident muons was ~50º, yielding a muon flux of ~20,000 muons/min. The details on how to calculate muon flux rate and the time needed to register a given number of muons will be addressed in Chapter 5 as well. The fuel assemblies in simulation was Westinghouse 15 x 15 pressurized water reactor fuel bundles with a burn up of 30,000 MWd. Each fuel assembly is about 21 cm long and wide and 3.66 m tall. Twenty out of 225 rod locations are used to hold control rods and the middlemost slot spared for instrumentation in nuclear reactor. The rest 204 slots are filled with cylindrical nuclear fuel rod in a diameter of 1.07 cm. These nuclear fuel rods are bundled together with stainless steel as shown in Figure 4.4.

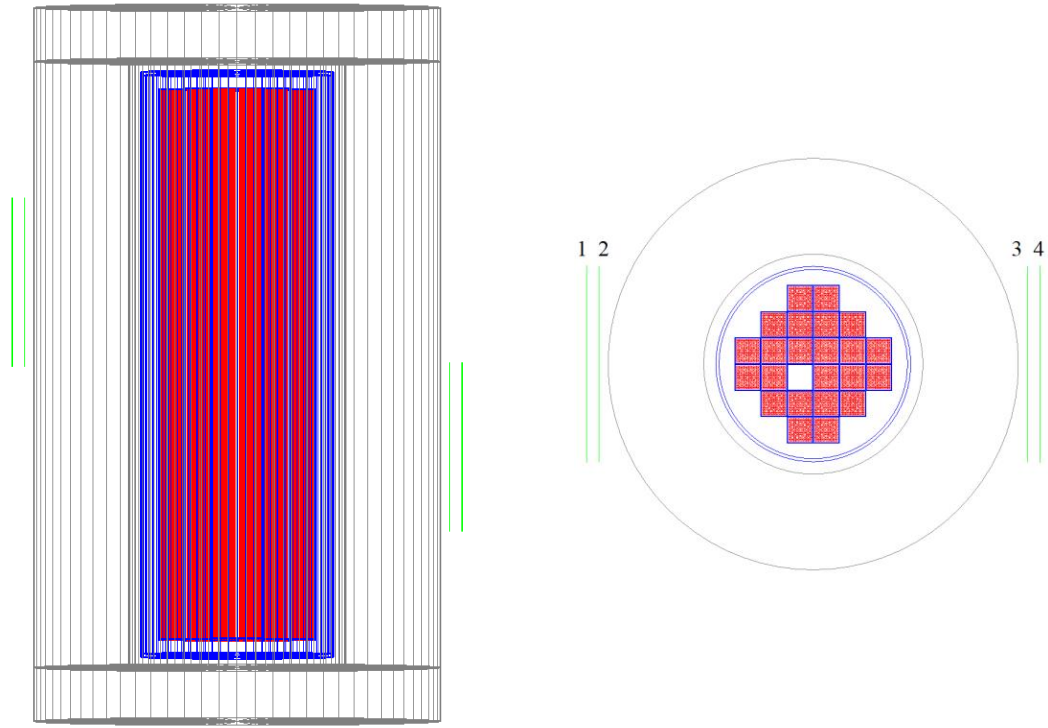

Figure 4.3        Side (left) and top-down (right) illustrations of the cask and detectors built in Geant4.

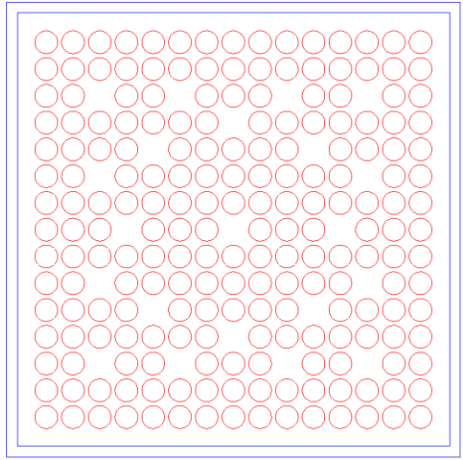

Figure 4.4          Westinghouse 15 x 15 pressurized water reactor nuclear fuel assembly

Similar to the simulation of GEANT4 model validation, neither the basket used to hold nuclear fuel assemblies nor the decay of spent nuclear fuel were simulated here in VSC-24 dry storage cask. Spent nuclear fuel is filled with GEANT4 built-in material G4_URANIUM_OXIDE. Muon source was simulated as vertical planar source as shown in Figure 4.5. In order to save running time in GEANT4, only the muons falling in the $[-\theta_2, \theta_1]$ in horizontal direction and $[\theta_3, \theta_4]$ in vertical direction were simulated.

$$\theta_1 = \theta_2 = \tan^{-1}\left(\frac{175}{350}\right) = 26.6 \tag{4.1}$$

$$\theta_3 = \tan^{-1}\left(\frac{350}{300}\right) = 49.39 \tag{4.2}$$

$$\theta_4 = \tan^{-1}\left(\frac{350}{150}\right) = 66.8 \tag{4.3}$$

During the simulation, the two pairs of muon detectors were rotated around the center of dry storage cask for 89 times in an increment of azimuthal angle of 2 degree, 90 views in total. Dynamic geometry was used to realize this in GEANT4, because during each run in GEANT4, the geometry can't be changed, however, it can be changed between two runs.

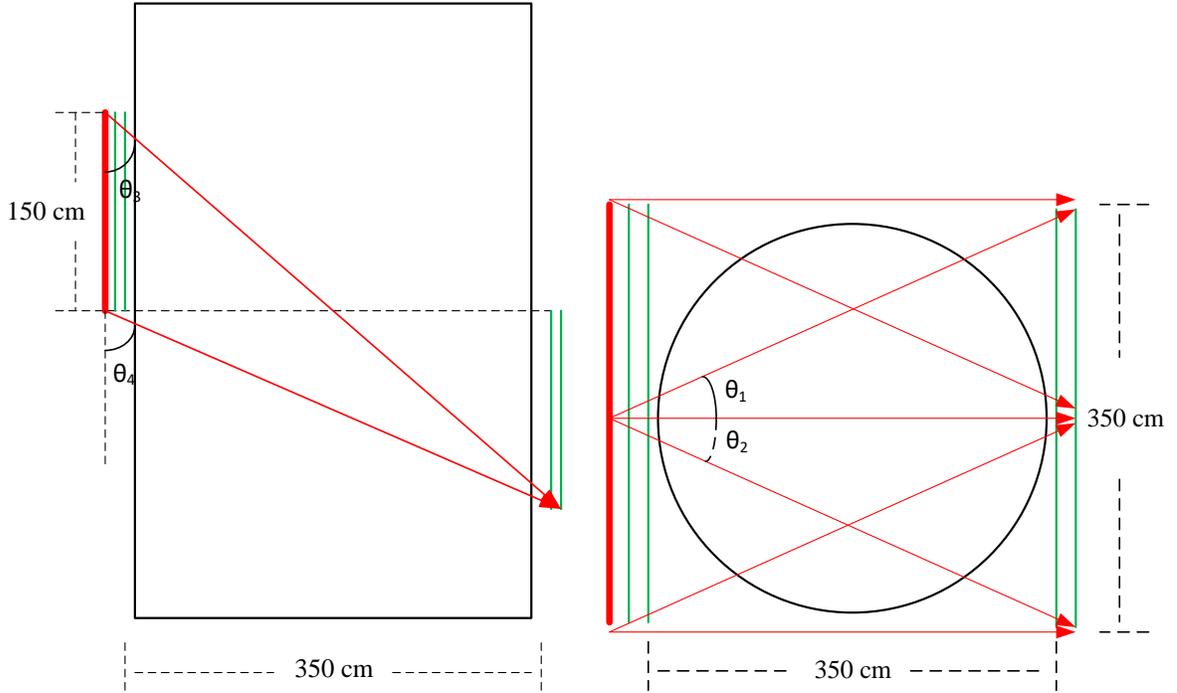

Figure 4.5          Side view of muon source on the left and top-down view of muon source on the left

For each view, 4 .gt4 files were used to record the position, momentum and event ID of muon when crossing the muon detectors 1, 2, 3 and 4. Event ID was used to pick up the muons crossed four detectors. Positions on four detectors were used to calculate the incident and exiting direction, and scattering angle of these muons as described in Eq. (3.19) and Eq. (3.20).

## 4.4   Results and analysis

For each view $1.5 \times 10^5$ muons were generated and about 2/3 of these muons made through four muon detectors 90 views in total, which led to $9 \times 10^6$ muons for reconstruction. Then re-sort these $9 \times 10^6$ useful muons into 180 quasi-parallel muon subsets according to their incident horizontal azimuthal angle and each quasi-parallel muon subset would have about 5000 muons. Method **1a** using a straight path along incident trajectory along with filtered back projection is used as benchmark. The reconstructed images of dry storage cask with fuel assembly missing in the middle using method 1**a** and 1**b** and FBP with and without muon momentum information is shown in Figure 4.6 and Figure 4.7. It can be seen when muon momentum information is available, large components, the concrete overpack and canister can be easily made out and it is obvious there is one spent nuclear assembly missing. But as to whether there is less than one spent nuclear assembly missing in other slots, i.e., half assembly missing or a quarter fuel assembly missing, it is hard to detect that with methods 1**a** or 1**b**. When no muon momentum information is incorporated in the reconstruction process to correct the scattering angle as suggested by Eq. (3.21), a significant degradation in image resolution is observed for both method 1**a** and 1**b** using FBP. One fuel assembly missing can barely be detected especially with method 1**a** using FBP. Overall, the reconstructed images using method 1**b** generated better results than 1**a** in both cases with and without muon momentum information, which can be told by comparing the 6 neighboring spent nuclear fuel assemblies with the empty fuel slot.

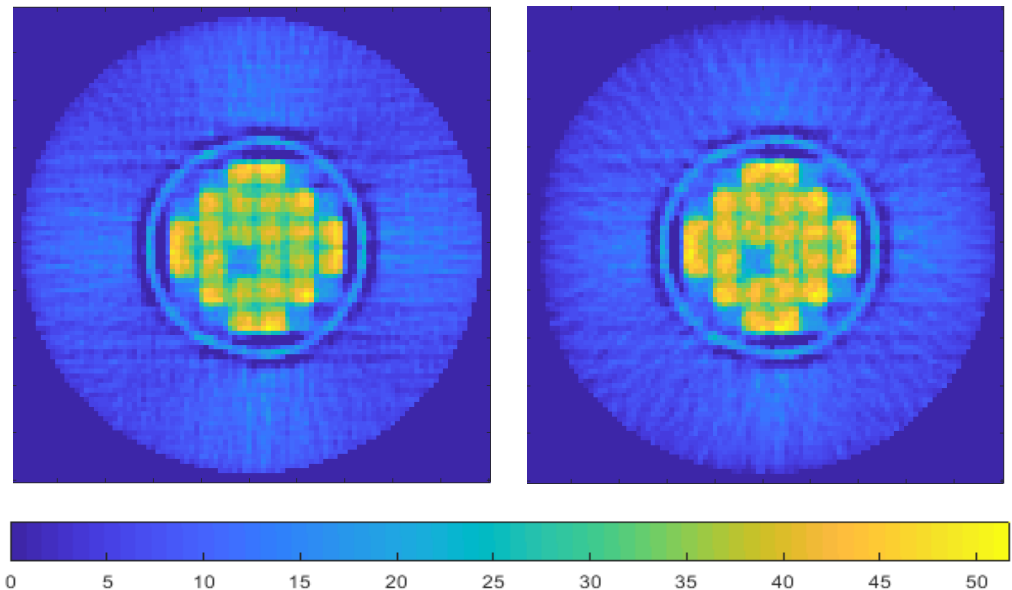

Figure 4.6    FBP reconstruction of a dry cask (refer to Figure 4.3) with perfect momentum measurement. Results are shown using straight path tracing 1a on the left and 1b on the right.

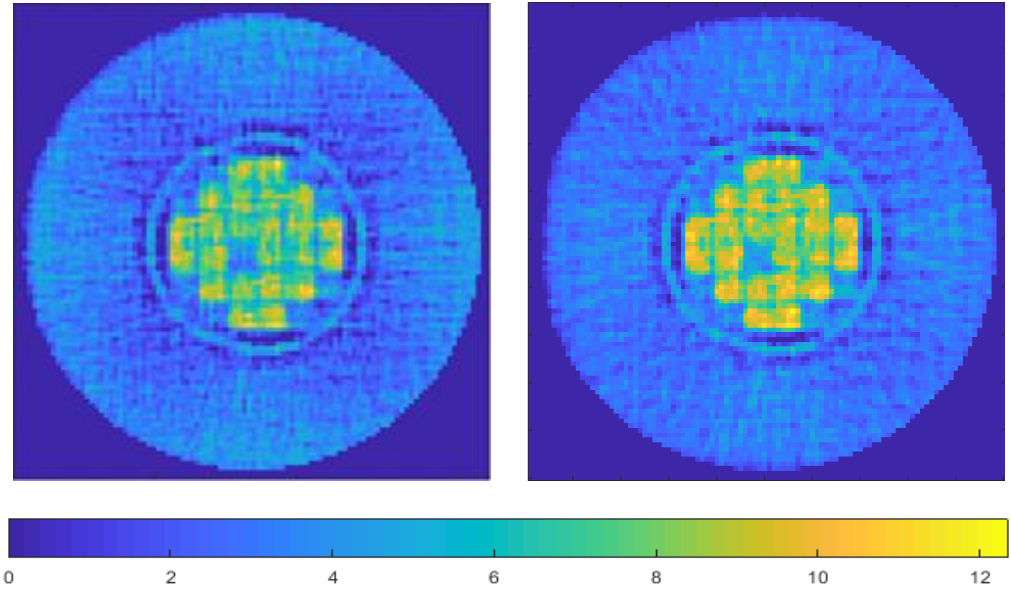

Figure 4.7    FBP reconstruction of a dry cask (refer to Figure 4.3) without momentum measurement. Results are shown using straight path tracing **1a** on the left and **1b** on the right.

The results reconstructed with SART, using tracing algorithms 1, 2 and 3 with and without muon momentum information are shown in Figure 4.8 and Figure 4.9. Similar to the result using FBP, the location of the missing fuel assembly is easily identifiable. Moreover, the image quality is remarkably improved when compared to FBP, even without muon momentum information. In addition, it appears that all tracing algorithms using SART have improved image quality when compared to FBP and algorithm **3b** achieves the best image quality. Signal to noise ratio (SNR), contrast to noise ratio (CNR) and detection power (DP) were used metrics to quantitatively assess how the muon tracing and scattering angle projection algorithms would be expected to affect the reconstructed image quality. SNR, CNR, and DP were calculated as [52]:

$$\text{SNR} = \frac{\text{mean}(8 \text{ assemblies surrounding missing one})}{\text{std}(8 \text{ assemblies surrounding missing one})}$$

$$\text{CNR} = \frac{\text{mean}(8 \text{ assemblies}) - \text{mean}(\text{missing one})}{\text{max}(\text{std}(8 \text{ assemblies}), \text{std}(\text{missing one}))}$$

$$\text{DP} = \text{SNR} * \text{CNR}$$

Figure 4.8    SART reconstruction of a dry cask (refer to Figure 4.3) with perfect momentum measurement. Results are shown using tracing algorithm 1**a** (top-left), 2**a** (middle-left), 3**a** (bottom-left), 1**b** (top-right), 2**b** (middle-right), and 3**b** (bottom-right).

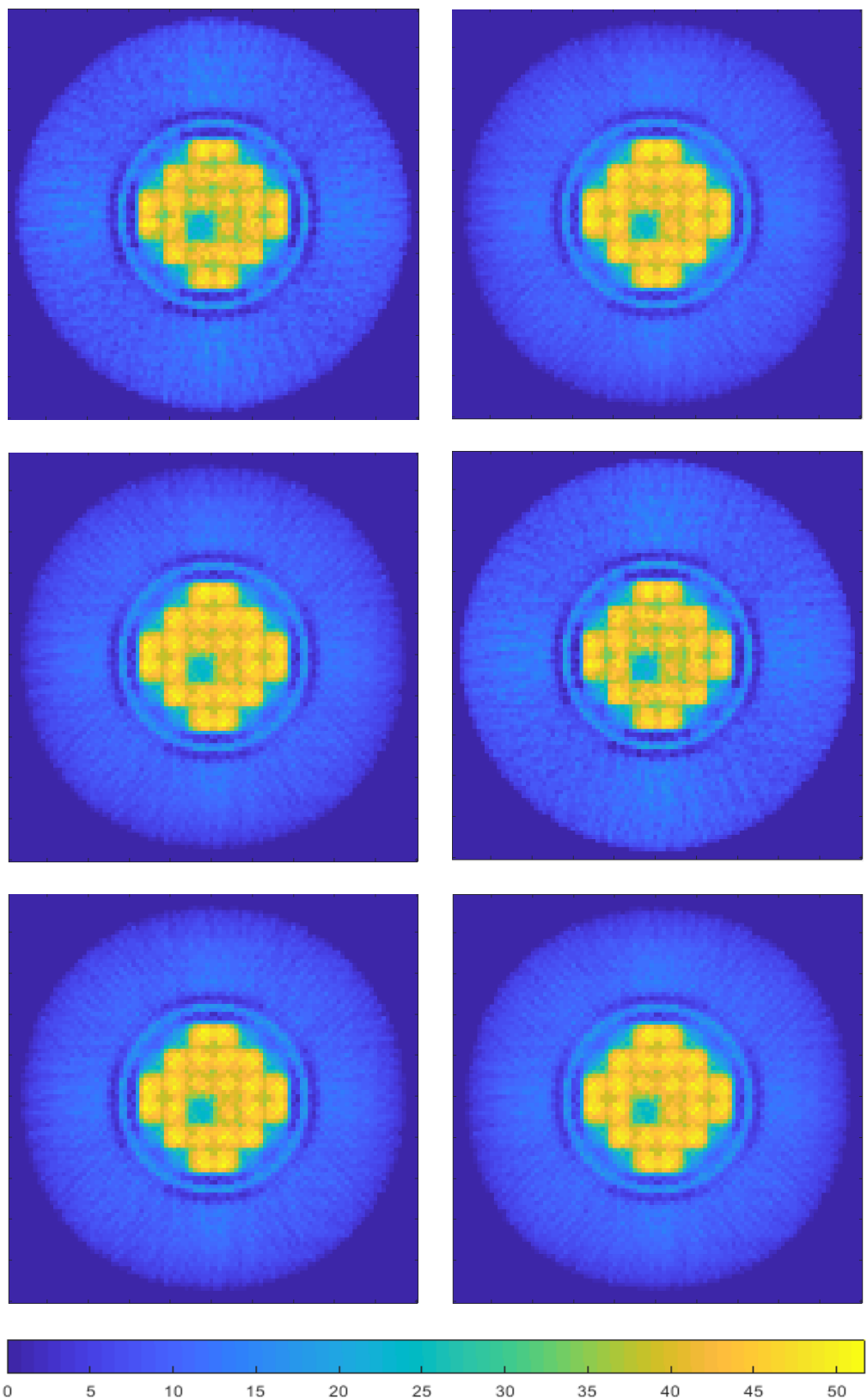

Figure 4.8 continued

Figure 4.9    SART reconstruction of a dry cask (refer to Figure 4.3) without momentum measurement. Results are shown using tracing algorithm 1a (top-left), 2a (middle-left), 3a (bottom-left), 1b (top-right), 2b (middle-right), and 3b (bottom-right).

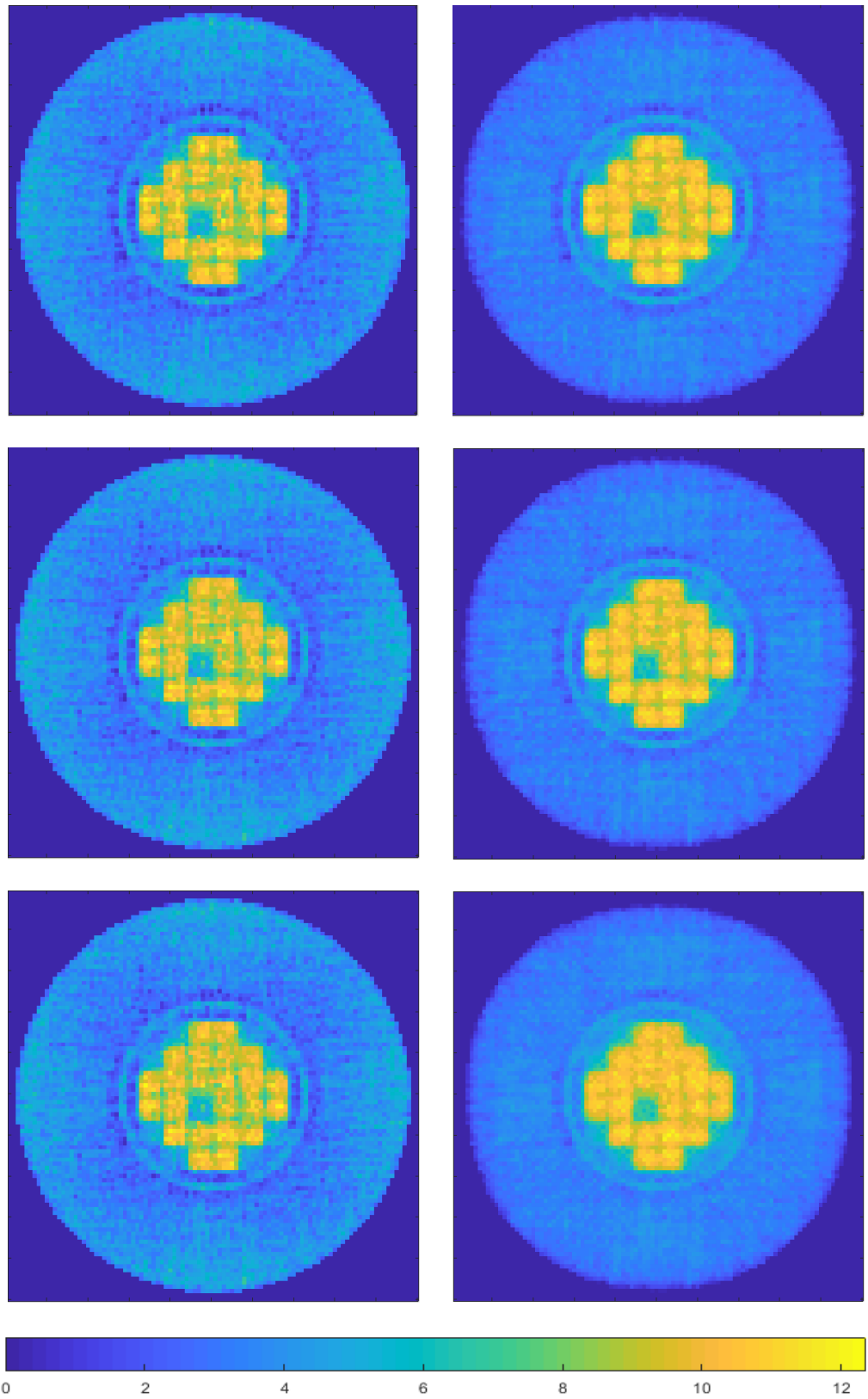

Figure 4.9 continued

where std denotes standard deviation. There are two different regions of interest in the model: (i) the empty slot and (ii) the surrounding 8 spent nuclear fuel assemblies. The SNR is used to quantify signal strength and how uniform the estimated scattering density is within the assemblies. CNR is used to quantify how different these two regions are. A large CNR means the reconstruction method is more capable to differentiate between the two regions. Since signal strength, uniformity and contrast are all important in detection of anomalies, the multiplication of SNR and CNR is defined to be detection power (DP) of the reconstruction method. The SNR, CNR and DP of the reconstructed images shown in Figure 4.6 to Figure 4.9 are given in Table 4.1. A comparison of SNR, CNR and DP for projection method **a** and **b** using 3 the different path models with momentum is shown in Figure 4.10 to Figure 4.12. When perfect muon momentum information is available, from use of the straight path along the muon incident trajectory (path type 1) to use of the straight path along the muon incident direction crossing the PoCA point (path type 2) to use of the PoCA trajectory (path type 3), the reconstructed image quality (SNR and CNR) is expected to be quite similar. But using projection methods **b** can boost CNR, SNR and DP in all three path types by an average of 17.2%, 8.5% and 27.1% relative to the use of projection method **a**.

Table 4.1          Expected image characteristics for the described methods

| Tracing algorithm | With momentum | | | Without momentum | | |
|---|---|---|---|---|---|---|
| | SNR | CNR | DP | SNR | CNR | DP |
| FBP | | | | | | |
| 1a | 6.30 | 2.73 | 17.20 | 5.13 | 1.98 | 10.17 |
| 1b | 7.00 | 2.60 | 18.25 | 6.36 | 1.94 | 12.35 |
| SART | | | | | | |
| 1a | 11.94 | 5.06 | 60.46 | 9.27 | 3.82 | 35.39 |
| 2a | 12.22 | 5.29 | 64.73 | 9.43 | 3.89 | 36.68 |
| 3a | 12.19 | 5.28 | 64.32 | 9.41 | 3.88 | 36.50 |
| 1b | 14.06 | 5.68 | 80.04 | 12.96 | 4.94 | 64.10 |
| 2b | 13.97 | 5.68 | 79.22 | 12.40 | 4.64 | 57.59 |
| 3b | 14.54 | 5.60 | 81.45 | 15.64 | 4.96 | 77.61 |

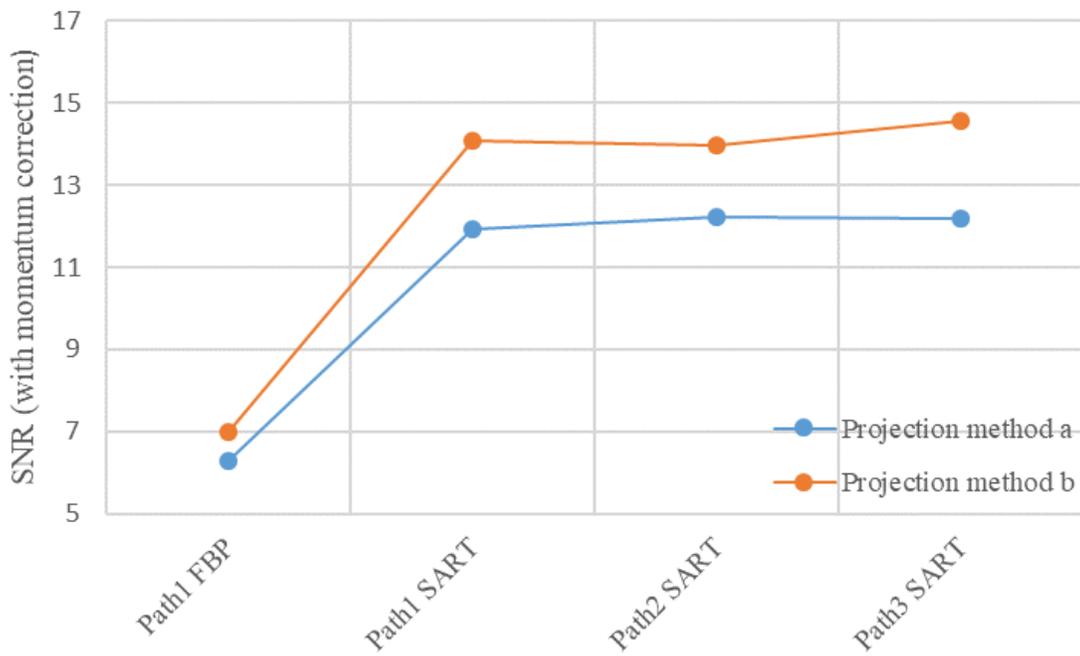

Figure 4.10          Sign to noise ratio of methods **1a** and **1b** using FBP as a benchmark, and methods **1a**, **1b**, **2a**, **2b**, **3a**, **3b** using SART with momentum information.

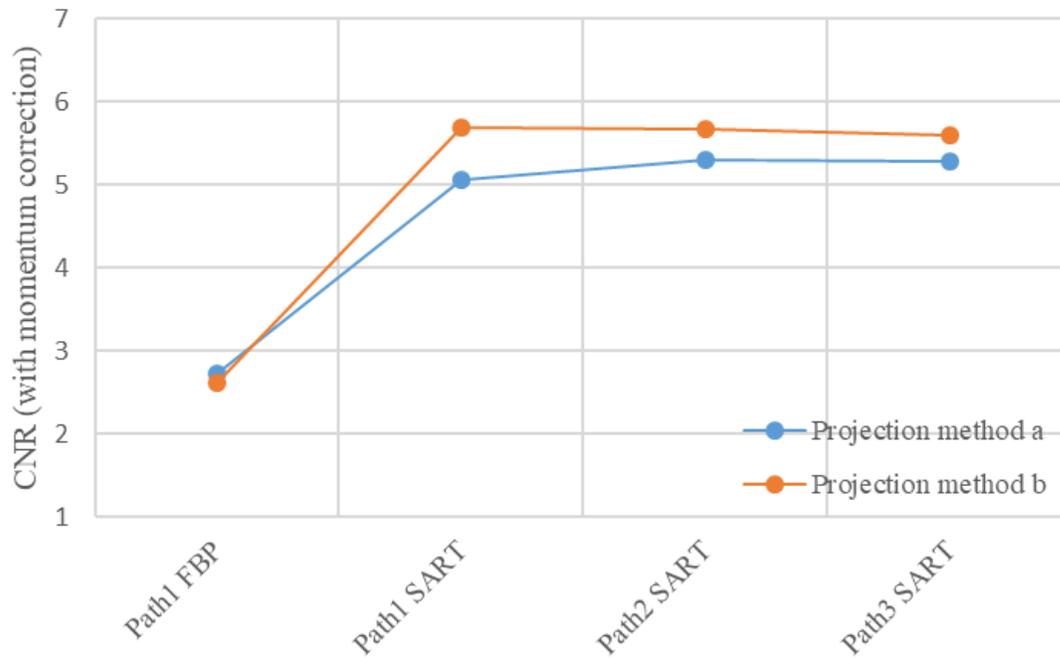

Figure 4.11    Contrast to noise ratio of methods 1a and 1b using FBP as a benchmark, and methods 1**a**, 1**b**, 2**a**, 2**b**, 3**a**, 3**b** using SART with momentum information.

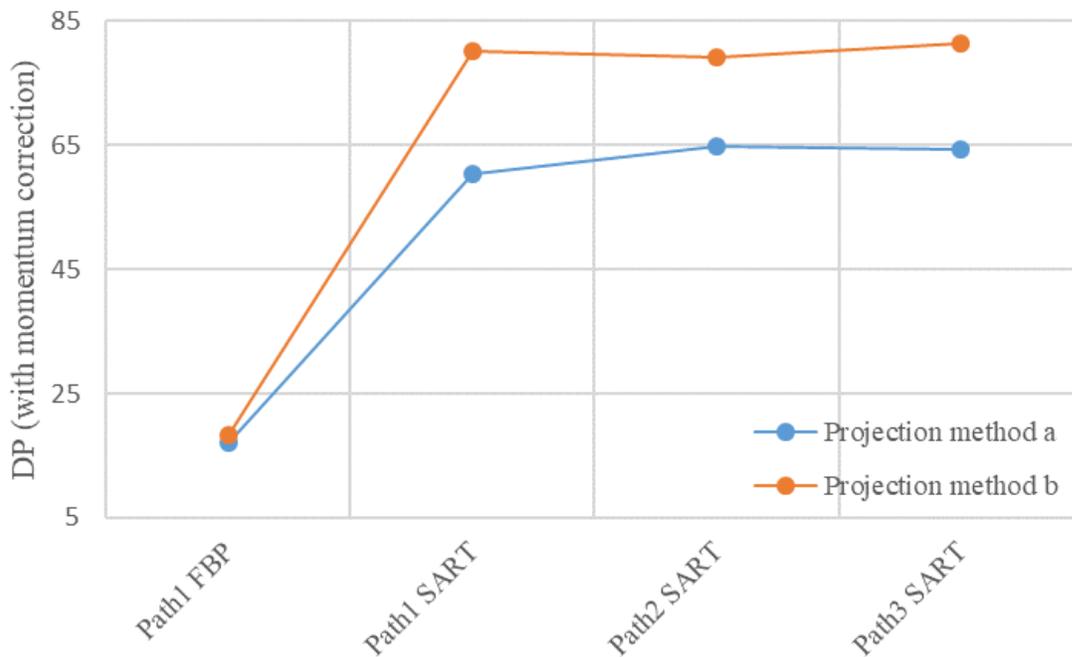

Figure 4.12    Detection power of methods 1a and 1b using FBP as a benchmark, and methods 1**a**, 1**b**, 2**a**, 2**b**, 3**a**, 3**b** using SART with momentum information.

A comparison of SNR, CNR and DP for projection methods **a** and **b** using 3 the different path models with momentum is shown in Figure 4.13 to Figure 4.15. In the case where no muon momentum is obtained, which is closer to the reality due to the high timing resolution requirement on muon detectors, remarkable differences can be observed: for projection method **a**, 3 different paths generate very similar result in terms of SNR, CNR and DP, however, when using along with projection method **b**, using PoCA path type can achieve 20.7% and 26.1% gain than path type 1 and 2; while the CNR's almost stay the same. Similar to the case where muon momentum information is available, projection method **b** outperforms projection method **a** by an average of 45.8%, 25.7% and 83.4% in terms of CNR, SNR and DP.

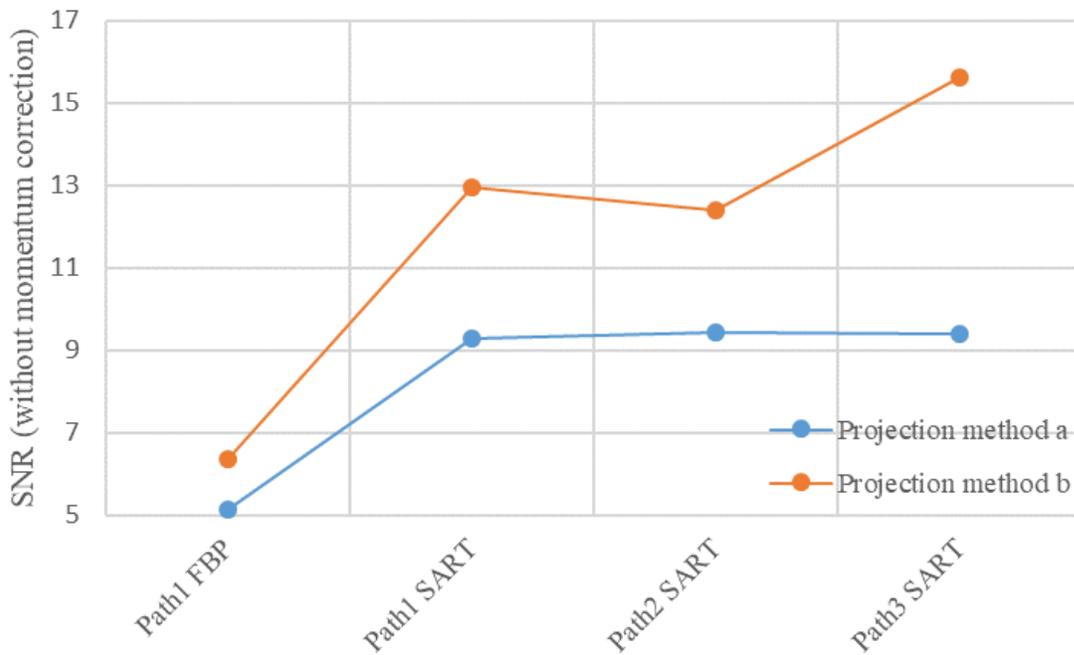

Figure 4.13    Sign to noise ratio of methods 1**a** and 1**b** using FBP as a benchmark, and methods 1**a**, 1**b**, 2**a**, 2**b**, 3**a**, 3**b** using SART without momentum information.

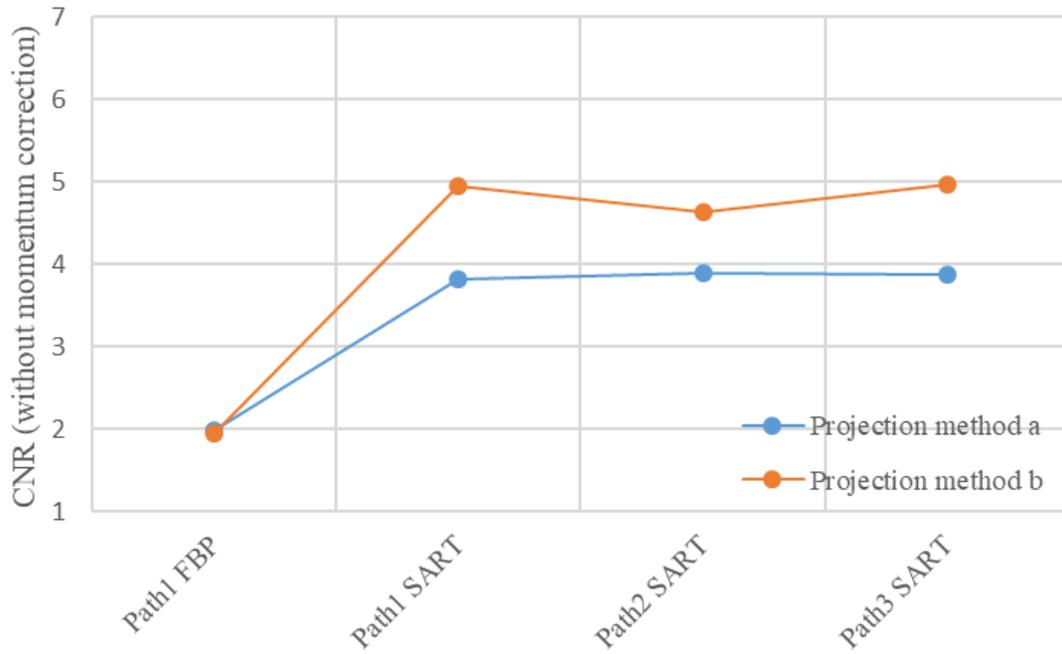

Figure 4.14    Contrast to noise ratio of methods 1**a** and 1**b** using FBP as a benchmark, and methods 1**a**, 1**b**, 2**a**, 2**b**, 3**a**, 3**b** using SART without momentum information.

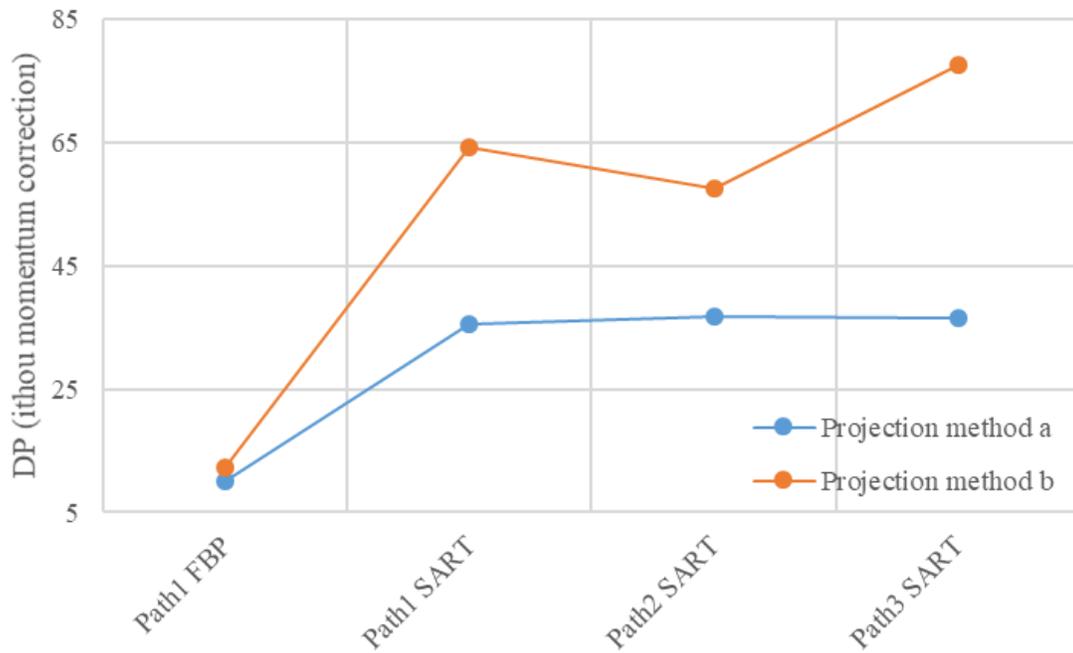

Figure 4.15    Detection power of methods 1**a** and 1**b** using FBP as a benchmark, and methods 1**a**, 1**b**, 2**a**, 2**b**, 3**a**, 3**b** using SART without momentum information.

A comparison of DP for cases with and without muon momentum information using 3 different path types is shown Figure 4.16. When muon momentum information is available and projection **a** is in use, three muon path types are expected to generate similar results, in another word, under this circumstance muon path model doesn't play a significant role. However, when muon momentum information is absent, on average when using projection method **a** would have a 42.7% detection power loss than when muon momentum information is absent; however, on average the detection power when using projection method **b** only decreases by 17.1%. Especially using projection **b** along with path type 3 only suffers a 5% detection power loss when muon momentum is absent. Thus, it is not hard to find that among the different combinations of path type and projection method using muon path type 3 along with projection method **b** is expected to yield the best result.

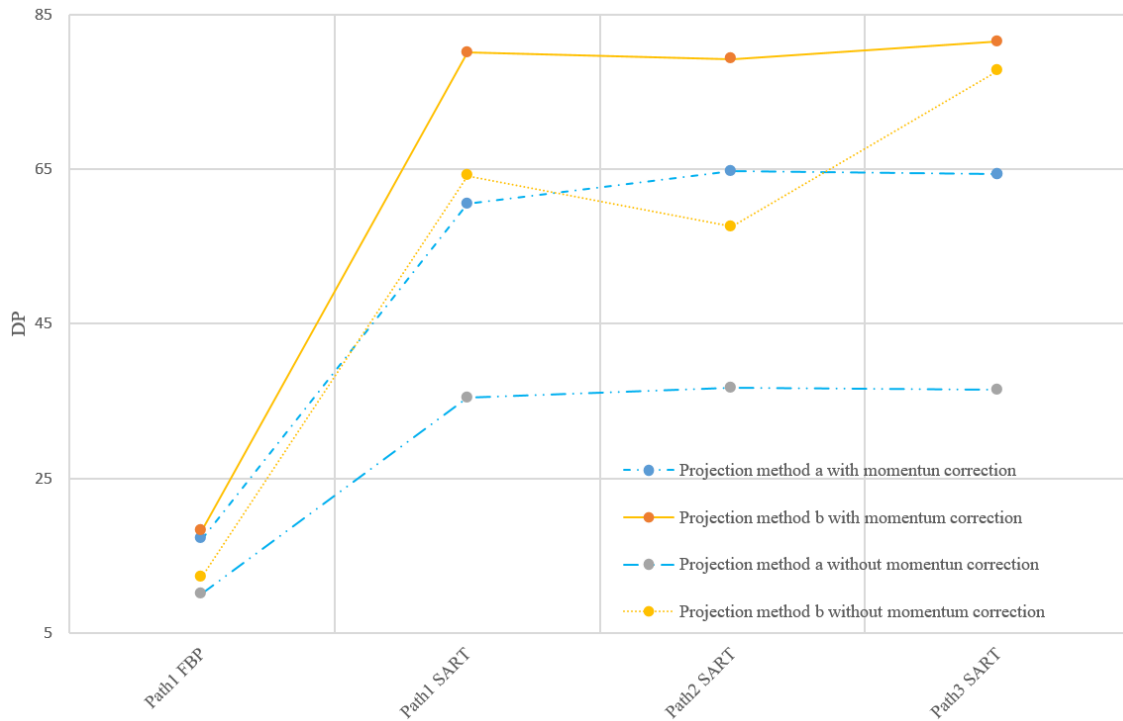

Figure 4.16     A comparison of the detection power for the different described methods with and without momentum information.

## 4.5 Detection limits

With the results presented above, it is confident that one spent nuclear assembly missing in a VSC-24 dry storage cask would be readily detected with any of the methods discussed above either with or without muon momentum information given the perfect position resolution of muon detectors and 90 views. Position measurement uncertainty of muon detectors will be incorporated in section 5.3 and its expected effects will be addressed as well. In this chapter, all the data has perfect position accuracy. If an algorithm can't resolve the missing fuel assemblies in dry storage cask with idealized data, it is much less likely to be able to recover difference between the empty slot and its surrounding spent nuclear assemblies. Thus, in order to investigate the expected detection limit of our μCT algorithms in scenarios that are relevant to international safeguards, idealized data will be continued to be used. A VSC-24 dry storage cask with partial defects in assemblies was simulated. Without loss of generality, we investigated a geometry where one half and a quarter of selected fuel assemblies were missing at different locations. In Figure 4.17 on top left, one can see the location of removed quarter and half fuel assemblies. One quarter fuel assembly was removed at the center, another two quarter fuel assemblies were removed at the edge, and one-half assembly was removed at the edge. A quantity of $1.5 \times 10^7$ muons, equivalent to ~39.5 hours of exposure, was used to reconstruct the dry cask with algorithm 3**b**. In the reconstructed images, either with or without momentum information, the location of all missing fuel assemblies, quarter and half, appears to be visible.

To analyze this observation qualitatively and quantitatively, two ways to detect missing fuel were investigated. The first way exploited the geometric symmetry of the 24 spent nuclear fuel assemblies. Six slices crossing the reconstructed image using algorithm **3b** but without momentum are shown in Figure 4.17. A comparison between slice 1 and slice 3 was used to detect a partial defect on the edge as shown in Figure 4.18 on the top. Similarly, slice 2 reveals the missing quarter fuel assembly at the center of the cask as shown in Figure 4.18 in the middle, and the comparison between slice 4 and slice 5 reveals a partial defect on the rim between two fuel assemblies as shown in Figure 4.18 at the bottom.

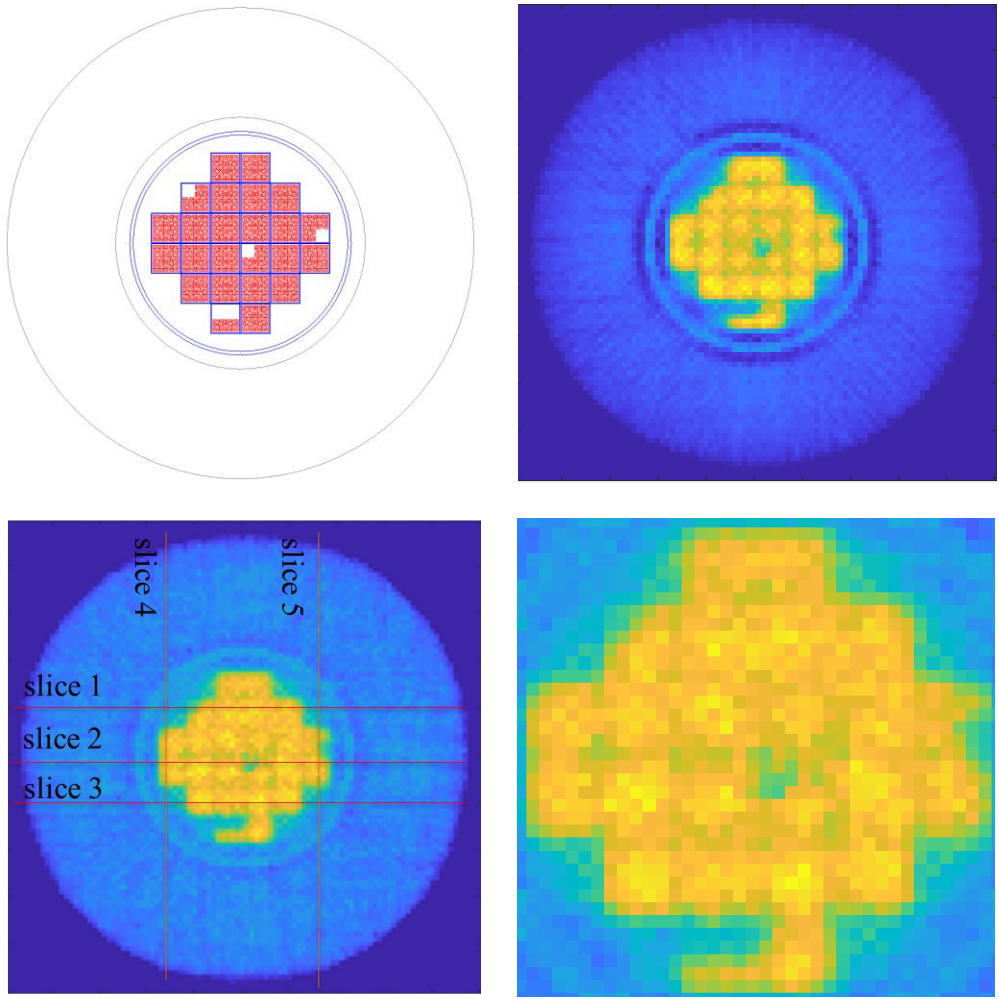

Figure 4.17 Top-down (top-left) view of the cask with a half fuel assembly and 3 quarter fuel assemblies missing. The reconstructed image using method 3**b** is shown with perfect momentum information (top-right), without momentum information (bottom-left), and in a zoomed-in view without momentum information (bottom-right).

The second way was to calculate the difference in scattering density between an empty quarter slot and the rest of the same fuel assembly. It turned out that a quarterly fuel missing at the upper left edge was the most difficult to detect scenario among these locations as shown in Figure 4.17. The estimated scattering densities in this missing quarter fuel slot and the remaining three quarters of the fuel assemblies were 7.3±0.8 and 10.2 ± 0.5 (arb. units), respectively; these values are separated by a difference of 5.8 σ. Of course, actual measurement conditions, radiation background, and detector position measurement uncertainty would reduce this difference. However, one may expect the limiting case of a quarter fuel assembly missing to be detected given sufficient measurement time to allow enough muons to be measured.

Figure 4.18    A comparison of the slices of scattering density from the reconstructed image without muon momentum showing slice 1 and 2 at the top, slice 2 in the middle and a comparison of slice 4 and 5 at the bottom. The slices are illustrated in Figure 4.17. See text for details.

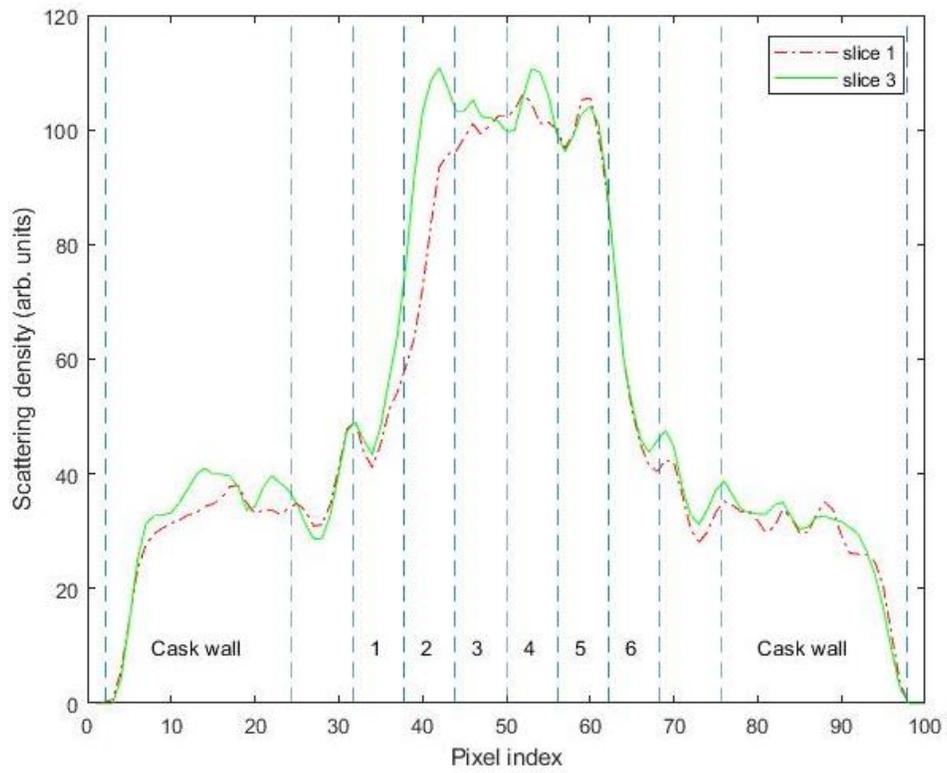

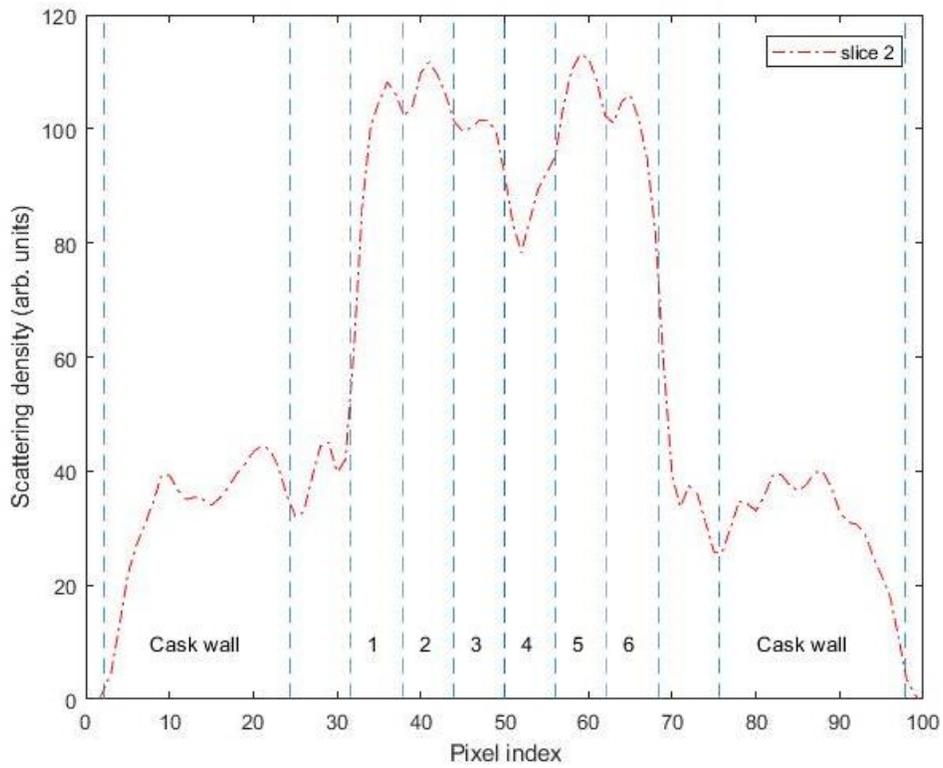

Figure 4.18 continued

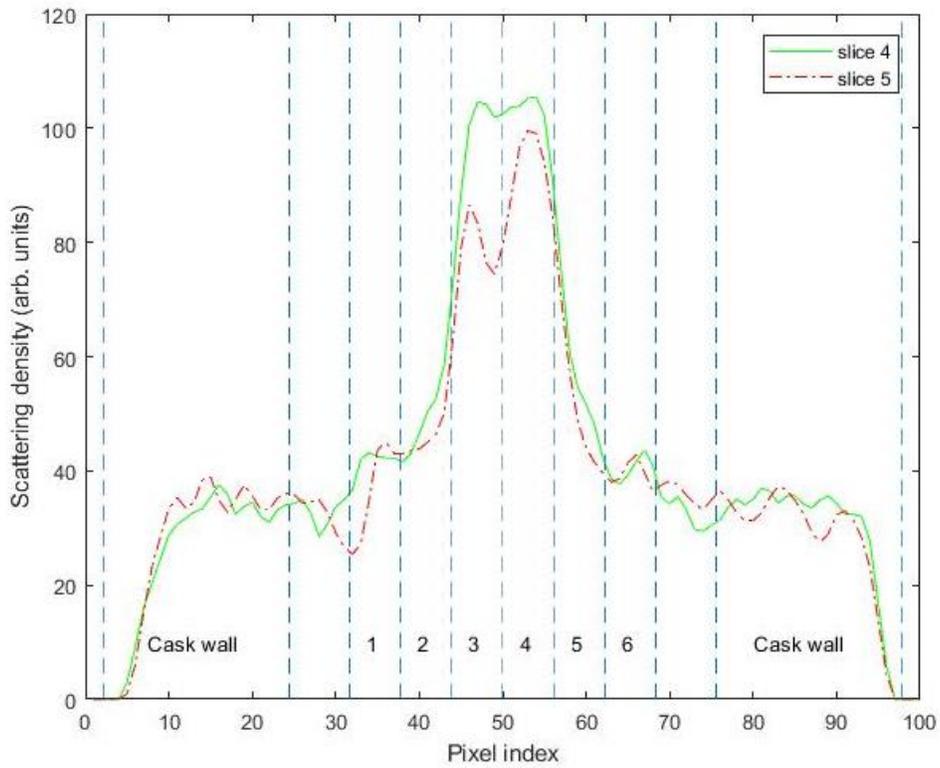

Figure 4.18 continued

## 4.6  Summary

To better represent the population of dry storage casks in use, a concrete dry storage cask with different loading profiles was chosen as the object for image reconstruction studies. Before building a GEANT4 simulation workspace for this concrete dry storage cask, a MC-10 steel walled cask with a specific fuel assembly missing pattern was simulated to validate against a physical measurement done by others. Generally, the simulated result is in reasonable enough agreement with the physical measurement to give is some predictive value. Later the MC-10 cask geometry in the validated GEANT4 workspace was replaced with a VSC-24 concrete walled dry cask with one spent nuclear fuel assembly missing in the middle. Then this Geant4 workspace was used to generate simulated data. Six different methods coupled with FBP or SART were used for image reconstruction. Method 1**a** using FBP functioned as benchmark for comparison. SNR, CNR and DP were defined used as metrics to quantify the reconstructed image quality. With 90 views, all methods are expected to be able to resolve the missing fuel assembly in middle of the VSC-24 cask. However, the PoCA trajectory model is slightly better than the straight path models. Projection method **b** is expected to significantly improve the reconstructed image quality and reduce the reliance on muon momentum compared to traditional projection method **a**. Meanwhile, SART is expected to perform about twice as good as than the FBP method. Method 3b is expected to achieve the best result among the 6 methods. Once achieving the goal of detecting one or multiple whole missing fuel assemblies, the same dry cask with a half or two quarter fuel assemblies missing were simulated and reconstructed with method 3**b**. With 90 views, either with or without muon momentum information, method 3**b** is expected to detect a quarter fuel missing with ideal data.

# 5  ENGINEERING RESTRICTIONS AND OPTIMIZATION

Almost all previous simulated muon imaging reconstruction work [6][9][42][46] [53][54] [58] used large area position sensitive detectors to collect a large enough amount of ideal muon events. Because of the low flux rate of cosmic ray muons, registering a large number of muons leads to a long exposure time. Meanwhile, the gain of image quality brought by extra muons may not be worth the corresponding extra time in a particular application. Thus, it is beneficial to investigate the relation between the reconstructed image quality and the exposure time. In this work, it is assumed that the duration of each view is constant, and the total exposure time is mainly determined by the number of views; thus, a different number of views will be used to optimize the exposure time. Due to the limited detector position resolution, the reconstructed images using perfect muon data may be not be achievable with field experimental data. Thus, in order to make our simulation results closer to reality, different detector position uncertainties will be incorporated into the reconstruction process. When it comes to imaging large objects, like a dry storage cask, a major economic concern is the cost of the large area position sensitive detectors and the associatiated readout electronics. Although a gas wire detector can reach a position resolution of 0.2 mm to 50 $\mu$m, it may be expensive to build 350 cm $\times$150 cm detectors. As a part of our work, we built large area scintillation detectors with wavelength shifting fibers described in section 2.3.1, which can reach a position resolution of 1 cm, possibly at a lower price than a gas detector of similar size. The position resolution could be lowered to 0.5 cm or less, but it may double the cost. Thus, before upgrading the detectors, it is necessary to investigate the expected effects of detector position uncertainty. This chapter is meant to serve as a guide for future field experiments in imaging dry storage casks with cosmic ray muons.

## 5.1   View sampling

Due to the large number of spent nuclear fuel dry storage casks in use and the long time (of order 1 day) to get a high confident image of a dry storage cask, it will be very inefficient and slow to monitor all dry storage casks in one storage facility let alone national wide. Also, in most cases where the seal of a spent nuclear dry storage cask is intact, one may suspect that a less stringent reconstructed image would suffice the requirement of investigation. For suspicious spent nuclear fuel dry storage casks, like the one with a damaged seal or looming fuel assembly missing pattern in the reconstructed image with a few views [54], one may imagine a concept of operations whereby more views can be recorded in order to generate a clearer image of the dry storage cask. By doing so, the scanning time for each dry storage cask could possibly be significantly shortened from a day to a few hours and muon imager system usage efficiency will be improved as well.  As discussed in Chapter 4 method 3**b** along with SART is expected to yield the best result among these methods discussed, thus method 3**b** along with SART is used here to investigate how the number of views affect the reconstructed image quality using the exact same data as in Chapter 4, but with less views.

A quantity of 5, 10, 18, 45 or 90 views were used to reconstruct the image with or without muon momentum information as shown in Figure 5.1. No matter what the number of view is, these views shall always be evenly distributed between 0 and $\pi$, for example, when there is a total of 5 views, the views are at azimuthal angles of 0, 36, 72, 108, 144 degrees. With muon momentum information, it takes about 5 views to generate a decently clear image of the spent nuclear fuel dry storage cask as shown in Figure 5.1 on top right, which is corresponding to 2 to 4 hours exposure. If this happened in the field experiment, more views shall be taken to get a clearer image of the spent nuclear fuel assemblies inside of the dry storage cask to make sure whether there are assemblies missing. Let's say another 5 views with an azimuthal angle of 18, 54, 90, 126 and 162 degrees will be added to first 5 views to reconstruct image with higher resolution as shown in the second row in Figure 5.1 on the right. At this moment, it is sure that there is one spent nuclear assembly missing at least. Thus, this dry storage cask needs to be opened for visual check, otherwise we would move on to scan another dry storage cask. For the case that no muon momentum information is obtained, it is hard to tell whether there is one spent nuclear fuel assembly missing in the dry storage cask with 5 views or even 10 views as shown in Figure 5.1 in the top row and second row on the left. It takes about 18 views without muon momentum to generate an image of the same quality as the image reconstructed with 5 views and muon momentum information. The SNR, CNR and DP of these reconstructed images shown Figure 5.1 are given in Table 5.1

Figure 5.1     Reconstructed images of dry storage cask with one spent nuclear fuel assembly missing in the middle with 5 (first row), 10 (second row), 18 (third row), 45 (fourth row) and 90 (bottom row) views. Images reconstructed without muon momentum information are in the left column and these with muon momentum information are in the right column.

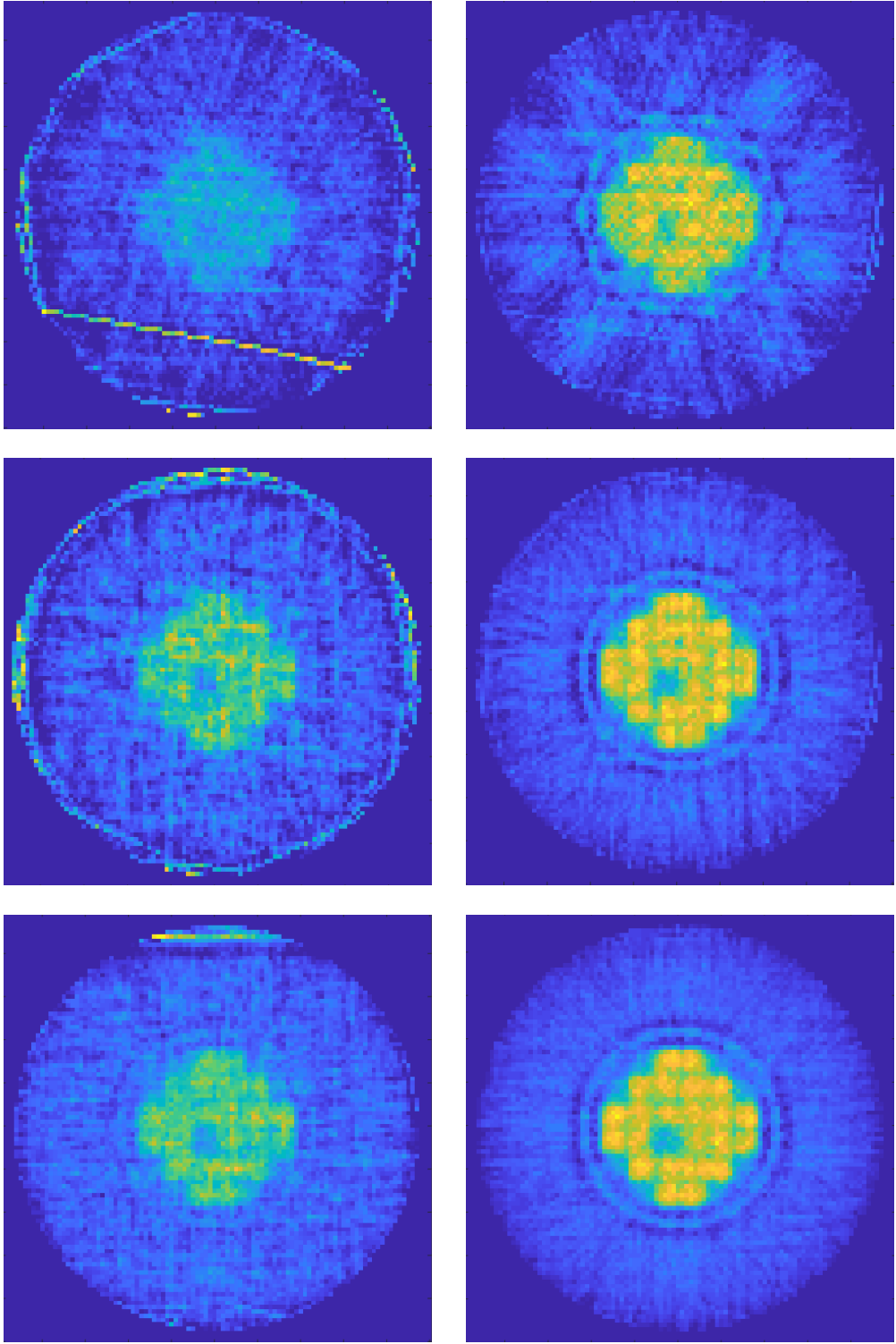

Figure 5.1 continued

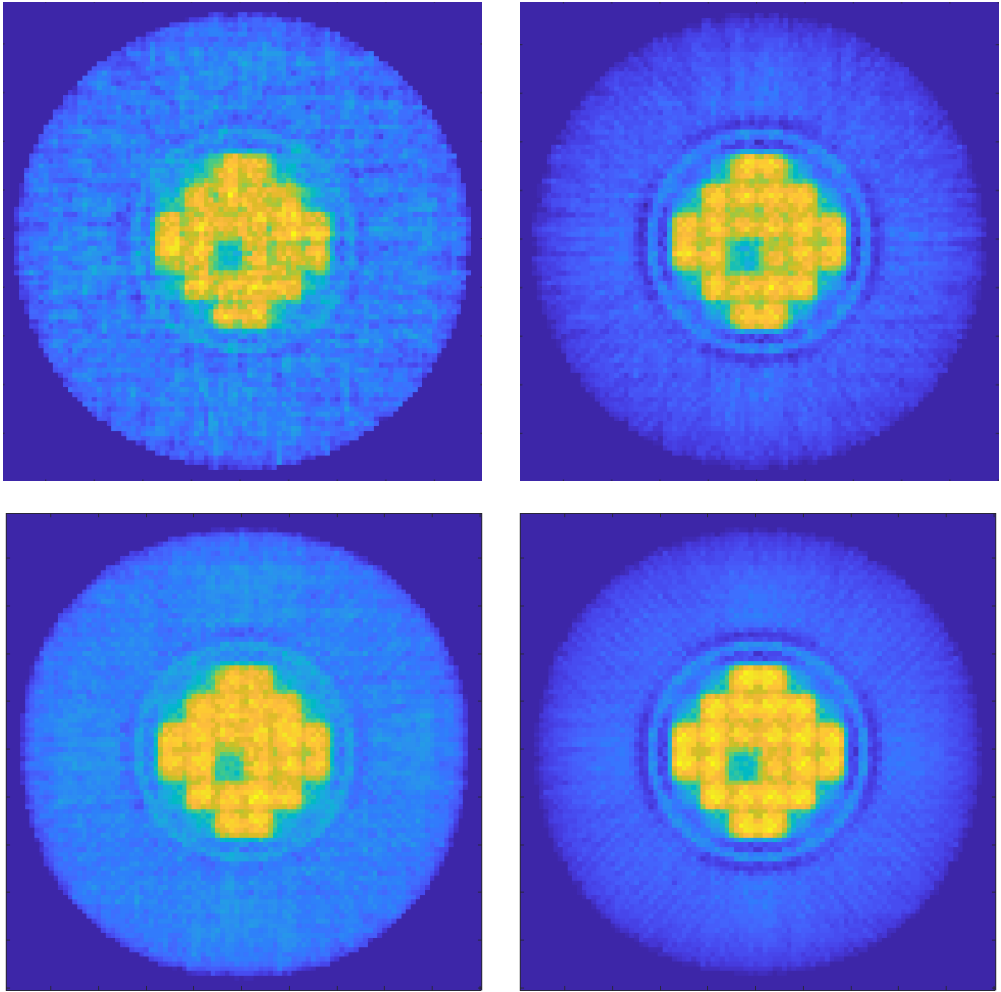

Figure 5.1 continued

Table 5.1        Expected image characteristics for different number of views

| Number of views | Without momentum | | | With momentum | | |
|---|---|---|---|---|---|---|
| | SNR | CNR | DP | SNR | CNR | DP |
| SART | | | | | | |
| 5 | 5.09 | 0.72 | 3.67 | 8.17 | 1.6 | 13.2 |
| 10 | 6.78 | 2.22 | 15.03 | 10.15 | 3.94 | 39.96 |
| 18 | 8.64 | 3.23 | 27.93 | 11.18 | 4.17 | 46.75 |
| 45 | 11.67 | 4.00 | 46.69 | 13.49 | 5.24 | 70.78 |
| 90 | 15.64 | 4.96 | 77.61 | 14.54 | 5.60 | 81.45 |

Comparisons of SNR, CNR and DP for the different number of views with or without muon momentum information using method 3**b** are shown in Figure 5.4. Again, it has been shown that using muon momentum information as a correction for scattering angle can achieve high image quality. It can be seen from Figure 5.4 that the signal to noise ratio is roughly linear proportional to the number of views used to the reconstruct the image and contrast to noise ratio increases dramatically at the beginning and the trend to slow down gradually. If the goal is to detect partial fuel missing in a spent nuclear fuel assembly, SNR and CNR could be equally important. If the goal is to detector one or more whole spent nuclear fuel assemblies missing, the contrast to noise ratio, a.k.a., separability, would become the dominant factor. As a rule of thumb, CNR needs to be bigger than 3 to detect the target from its background [55], thus it takes about 9 views with muon momentum information or 17 views without muon momentum information to detect the missing fuel assembly. Given the image size of reconstruction volume 100×100×1, according to view sampling requirements, 100×π/2=157 views per 180 degrees is required. However, for the scenario when muon momentum information is available, SNR and CNR of reconstructed image with 90 views was only improved by 6.9% and 7.8% separately relative to that of reconstructed image with 45 views, but it doubled the measurement time. Thus, when muon momentum is available, it is not wise to go above 45 views to image the dry storage cask. However, when muon momentum information is absent, SNR and CNR of reconstructed image with 90 views would be improved by 34.0% and 24.0% separately relative to that of reconstructed image with 45 views, thus it is justifiable to use more than 45 views if goal is to detect partial fuel missing in any assemblies in the spent nuclear fuel dry storage cask.

Another question that was not addressed here is how to choose the number of muons per view given the total muons to be used for image reconstruction in order to improve reconstructed image quality. When the total number of muons to be registered is given, the total measurement time is also fixed subsequently. Thus, the number of muons per view and the number of views in total shall be optimized to achieve the best possible result. Two extremities could shed some light on this point: one is that only one view is used resulting a radiography image instead of a tomography image, it is hard to tell whether there is multiple fuel assemblies missing let alone partial fuel missing in an assembly; another is that too many views are used, let's say 180 views or more, each view will have an insufficient number of muons, which will lead to a very noisy reconstructed image because it does have a good amount of samples. Instead of using an extreme case or choosing a random number of views, the most appropriate number of views shall be used. The issue may be addressed in the future.

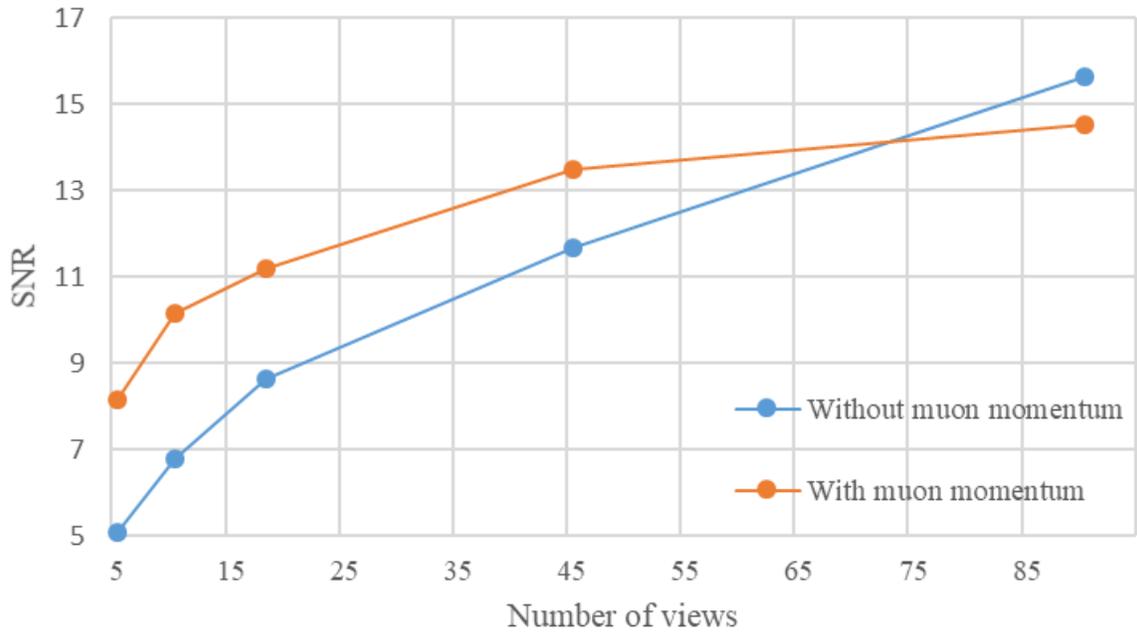

Figure 5.2    Comparison of SNR for different number of views with and without muon momentum.

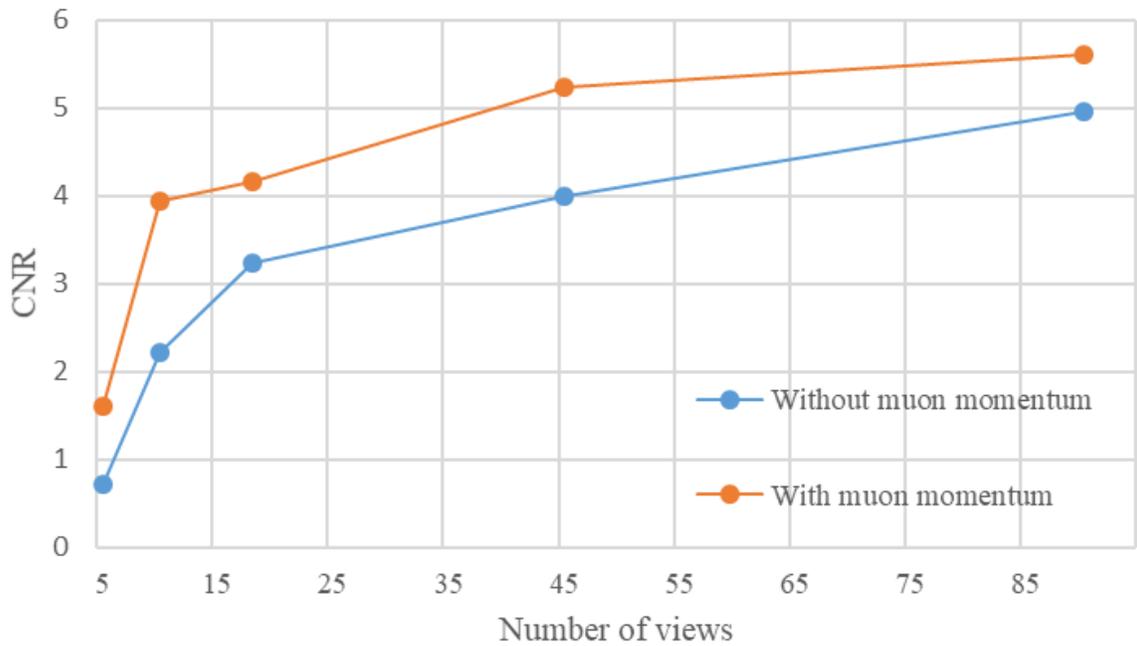

Figure 5.3    Comparison of CNR and DP for different number of views with and without muon momentum.

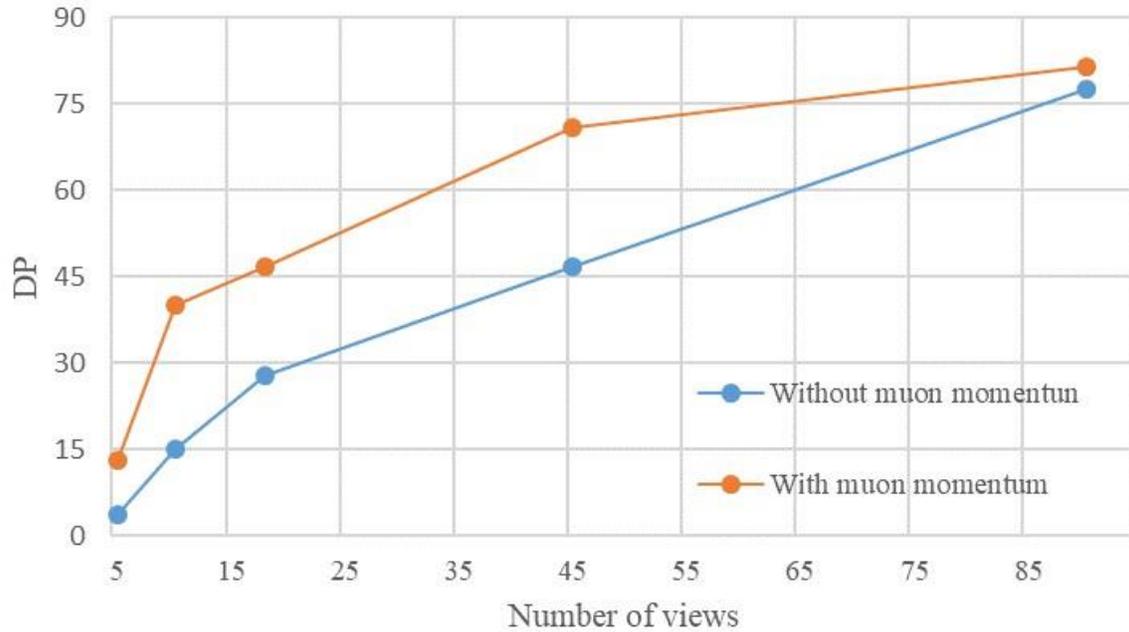

Figure 5.4          Comparison of DP for different number of views with and without muon momentum.

## 5.2   Detector geometry

Although using large area position sensitive planar detectors or ring detectors which could cover the whole cask can generate complete information of the cask wall and spent nuclear fuel assemblies inside, it is not economically practicable to build such large area detectors with readout electronics. When small detectors whose width is shorter than the diameter of cask are in use, one way to get around this restriction of detector size as suggested by [9] is placing the upper detector in one of the n positions within the range (0∘,180∘] on the rim of  cask and placing the lower detector at each of the n positions in the range (180∘, 360∘] to still yield the same sinogram information as collected with ring detectors, however, it would significantly increase the measurement time by a factor n$^2$. Due to the central symmetry of the cask wall and its smaller scattering density compared to spent nuclear fuel inside the cask, a complete sinogram information of the whole cask may not be a necessity to reconstruct the spent nuclear assemblies inside of the cask, but a complete sinogram information of the spent nuclear fuel assemblies is still needed. In order to use smaller planar muon detectors without extending the measurement time, the length of the detector needs to be no smaller than the diameter of canister. Thus, muon trackers sized of 1.6 m ×1.2 m were simulated to register muons crossing a VSC-24 dry storage cask with one spent nuclear fuel missing. The configuration is exactly similar to one described in test model configuration section except the detector size as shown in Figure 4.3 on the left. Only the data registered by muon track in a 1.6×1.2 m$^2$ area in previous section (about 3.3×10$^6$ muons) was used here to reconstruct the spent nuclear fuel with method 3**b** as shown in Figure 5.5 on the top left. The expected SNR, CNR and DP are 9.67, 4.85 and 46.89, respectively, which are worse than that of full size detectors, respectively: 14.54, 5.60 and 81.45. Even so, the reconstructed signal in the empty slot is still expected to be significantly separated from its surrounding fuel assemblies by 4.85σ, while when large area detectors were used, they were separated by 5.6σ. Thus, smaller detectors are expected to be able detect a single missing fuel assembly with good confidence.

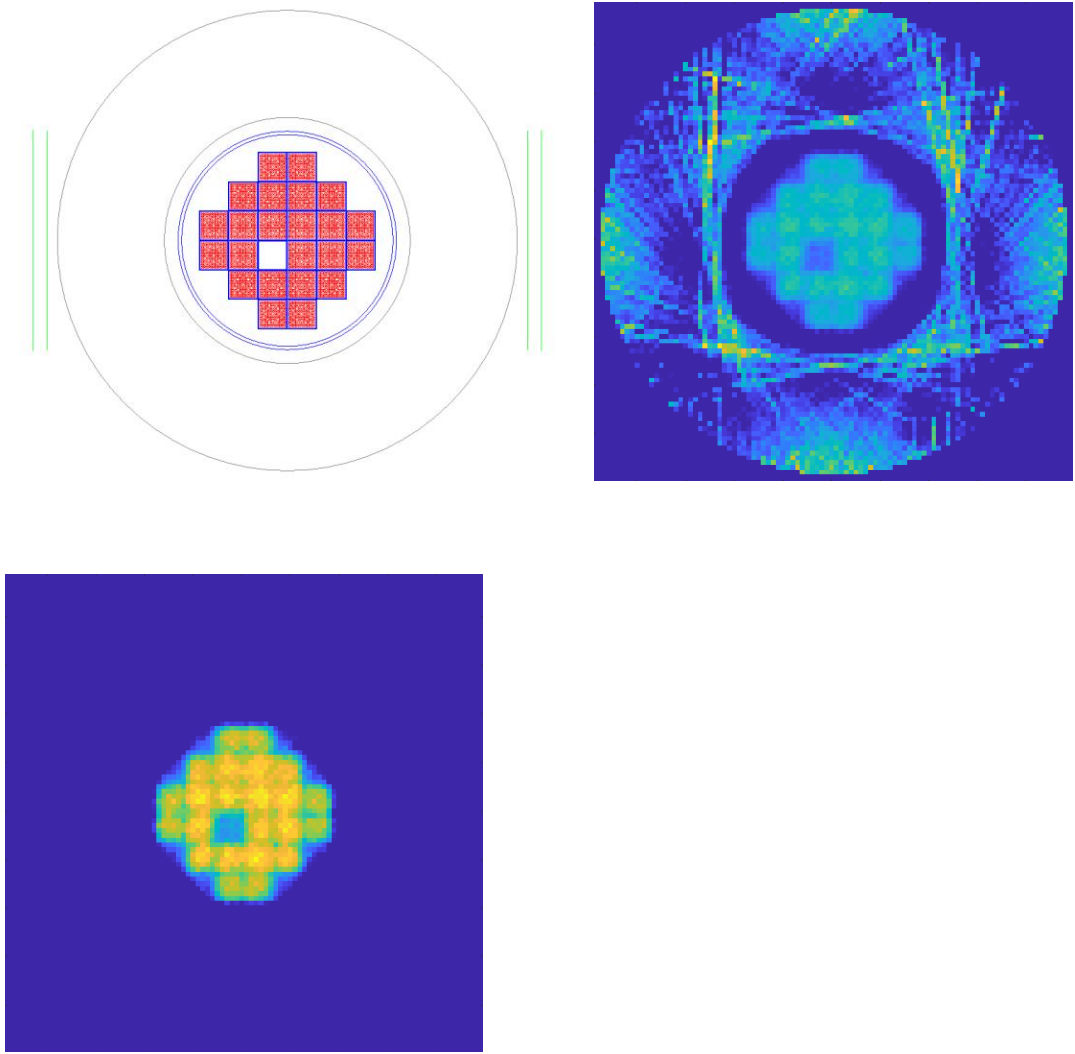

Figure 5.5    Top-down illustration of the cask and small detectors built in Geant4 (top-left), original reconstructed image (top-right) with method 3**b**, and the reconstructed image of the spent nuclear fuel after removing the image of cask (bottom-left).

## 5.3    Expected effects of measurement uncertainty

The positions and directions of incident muons and exiting muons are measured with a pair of position sensitive detectors. Each detector can register the position where the muon crosses the detector. The incident and exiting muon positions are registered by detectors nearby the object under investigation, i.e., detector 3 and 4. Assuming that there is no scattering that happened in the process of a muon traversing the detectors, thus by the connecting these two points on the adjacent muon detectors 1 and 2, the muons' incident direction can be determined. Similarly, the muon's exiting direction can be determined in the same way using the other detectors. Also, let's assume that the four muon detectors are identical and have the same position measurement uncertainty r. For a muon coming from zenith angle $\theta$, its measured direction will be within a cone with central line along the segment connecting the points on the two adjacent detectors with an angular uncertainty of $[-\Delta\theta_1, \ \Delta\theta_2]$, as shown in Figure 5.6.

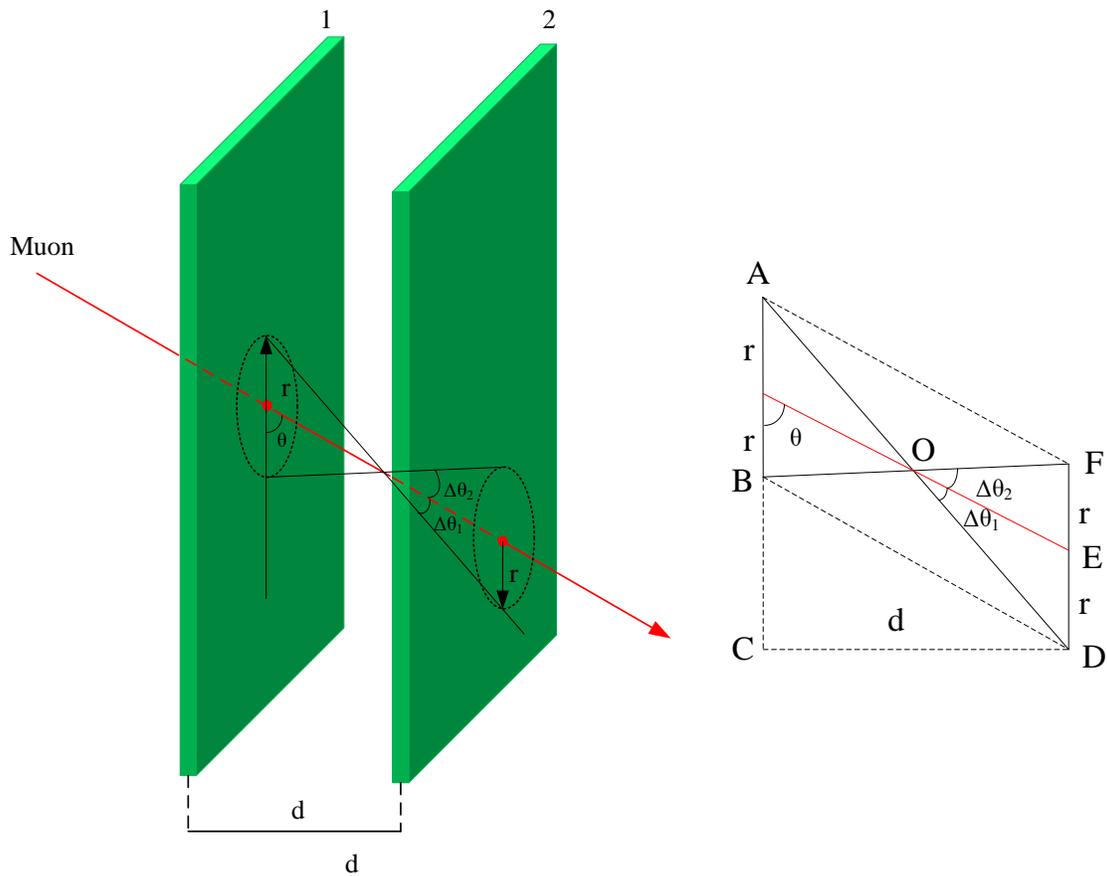

Figure 5.6    Illustration of the muon position measurement with uncertainty in 3D on the left and in 2D on right.

The concrete derivation of $\Delta\theta_1$ and $\Delta\theta_2$ is shown below from Eq. (5.1) to Eq. (5.6):

$$OA = OD = \frac{\sqrt{d^2 + (d \cdot \cot(\theta) + 2r)^2}}{2} \tag{5.1}$$

$$BD = AF = \frac{d}{\sin(\theta)} \tag{5.2}$$

$$OE = \frac{d}{2 \cdot \sin(\theta)} \tag{5.3}$$

$$OF = \sqrt{OA^2 + AF^2 - 2 \cdot OA \cdot AF \cdot \cos(\Delta\theta_1)} \tag{5.4}$$

$$\Delta\theta_1 = \arccos(\frac{OE^2 + OD^2 - DE^2}{2 \cdot OE \cdot OD}) \tag{5.5}$$

$$\Delta\theta_2 = \arccos(\frac{OE^2 + OF^2 - FE^2}{2 \cdot OE \cdot OF}) \tag{5.6}$$

Even with the same detectors and configuration, for muons coming from different zenith angles, the detectors will have different angular resolutions. Given the muon detectors separation d=60 cm and resolution r = 0.1 cm, an illustration of how angular resolution changes with zenith angle is shown in Figure 5.7. The smaller the zenith angle is, the larger the vertical distance the muon will travel, which makes the ratio of detector positional measurement uncertainty r to the distance muon traveled smaller. Thus, instead of assigning a constant angular uncertainty to the registered muons, muon detector positional measurement uncertainty r will be incorporated into crossing points on detectors 1 and 2, 3 and 4. In order to achieve high angular resolution when muon detectors are set up vertically, increasing the separation between adjacent detectors and using muons from small zenith angles can be adopted; however, these two conditions are mutually exclusive if the adjacent detectors are placed at the same latitude. Because when the separation between the adjacent two detectors is increased, more and more muons coming from a small zenith angle would not cross the next detector. One extremity is the separation is large enough that only muon with a zenith angle almost close to 90 degrees would be able to cross the muon detector pair, which has the lowest muon flux rate. Thus, a potential configuration could be that a vertical offset is added to each detector to increase muon flux rate meanwhile reducing angular uncertainty as shown in Figure 5.8. This configuration was not adopted in any GEANT4 simulation experiments in this dissertation but is expected be helpful in the field experiment in the future.

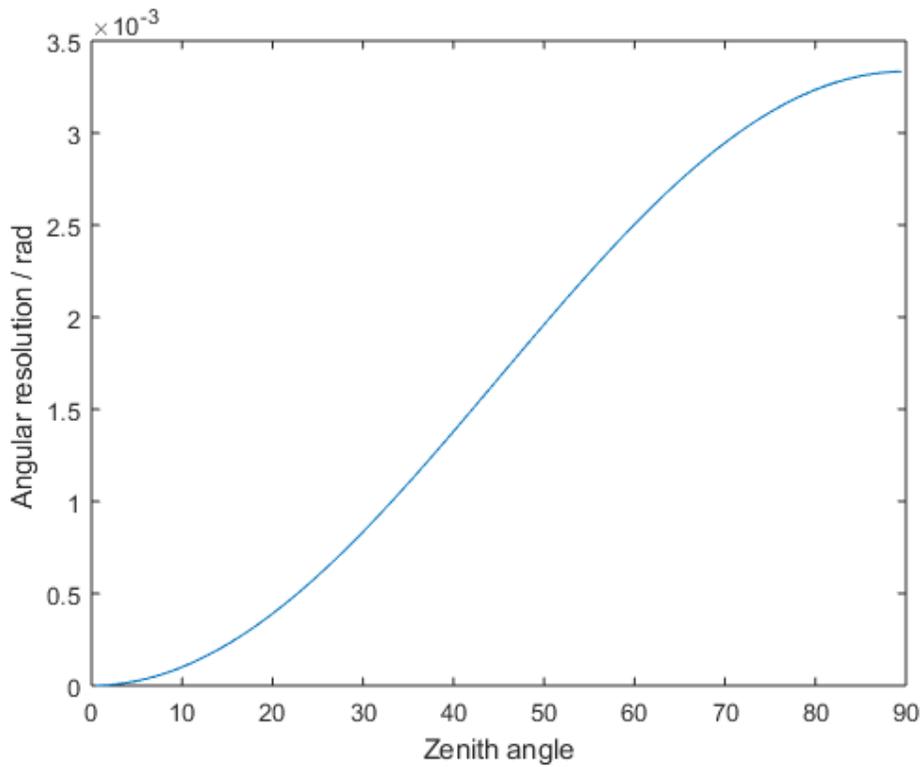

Figure 5.7        Illustration of angular resolution vs zenith angle

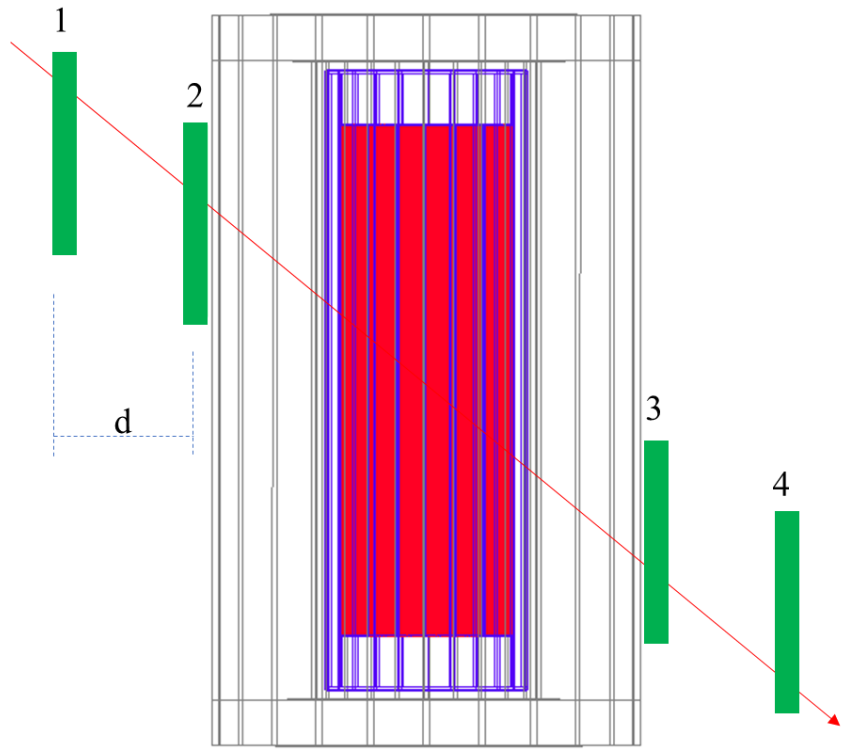

Figure 5.8      A potential configuration for increasing muon flux crossing 4 detector and reducing the angular uncertainty.

In order to understand whether it is feasible to the lower the cost of required large area position sensitive muon detector, we designed and fabricated our own scintillation detectors as described in section 2.3.2. The position resolution of our plastic scintillator detector panel is primarily determined by the gap between wave length shifting fibers embedded in the scintillator panel. Our prototype detector size is 32 cm wide and 32 cm long with 1 cm gap between each wavelength shifting fiber, theoretically 1cm position can be achieved [56]. Decreasing the gap between each wavelength shifting fiber by half would double the number of readout channel. Thus, before building an even larger area position sensitive plastic scintillator detector using the same design, it is imperative to find out the acceptable detector resolution to reconstruct a dry storage cask with the measured data. Similarly, detector separation was set to 60 cm; three different detector uncertainties $\sigma = 1$ cm, 0.5 cm, 0.1 cm, 0.01cm and perfect resolution ($\sigma = 0$ cm) were simulated. The reconstructed images using method 1**b** and 3**b** are shown in Figure 5.9 and Figure 5.10. A numerical analysis of these reconstructed images is shown in Table 5.2. If the goal is to detect a partial assembly missing in the dry storage cask, SNR could be more important than CNR given the physical meaning of SNR, the uniformity of pixel value in a fuel assembly. If the goal is to detect one or multiple whole assemblies missing, CNR better relates to the expected performance. Thus, a tradeoff shall be made between SNR and CNR when it comes to quantify the performance of different methods if certain information is known. Due to the lack of knowledge of what may be missing, SNR and CNR will equally treated. The trends of SNR, CNR and DP changing with position uncertainty σ for methods 1b and 3b with or without muon momentum are shown in Figure 5.11, Figure 5.12 and Figure 5.13. It is interesting to note that when muon momentum information is not available, the SNR of both method 1**b** and 3**b** is expected to increase with the worsening of detection resolution. On average, method 3**b** achieves a 19.4% gain in SNR relative to method 1**b.** Furthermore, the CNR of both methods 1**b** and 3**b** is expected to decrease with the worsening of detection resolution. On average, method 1**b** is expected to achieve a 7.7% gain of CNR relative to method 3**b.** When perfect muon momentum information is available, both SNR and CNR are expected to increase with the improvement of position resolution. On average, method 3**b** is expected to achieve an 8.4% gain in SNR relative to method 1**b**, and method 1**b** is expected to have a 2.2% improvement in CNR compared with method 3**b**. A rule of thumb to detect an object (compared to some background) from a reconstructed image is to achieve a CNR no less than 3 [57].

Figure 5.9    Reconstructed images of dry storage cask using method **1b** with detection position uncertainty $\sigma =$ 1 cm, 0.5 cm, 0.1 cm, 0.01cm and 0 cm (from top to bottom) without momentum (left column) and with momentum (right column).

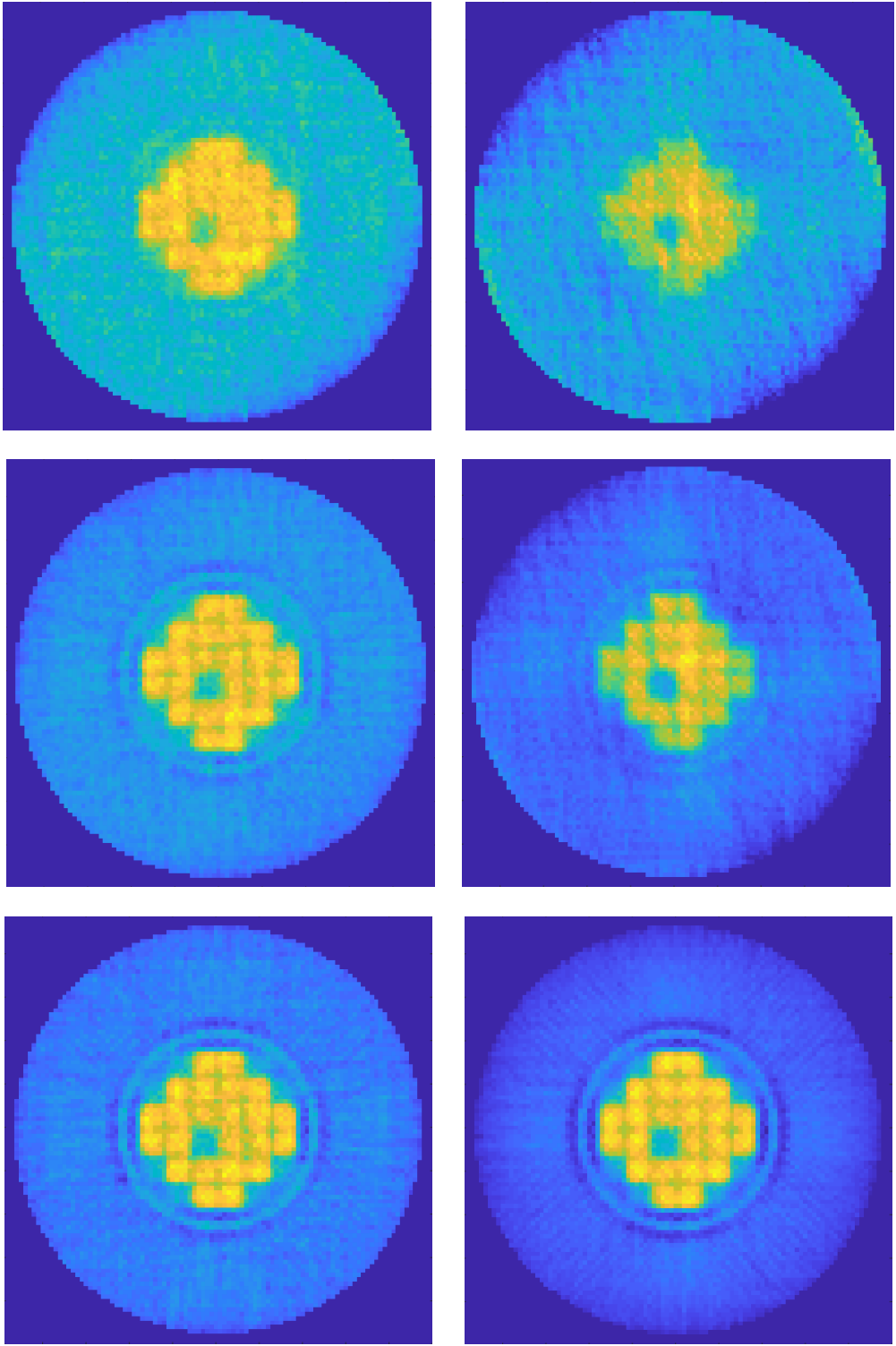

Figure 5.9 continued

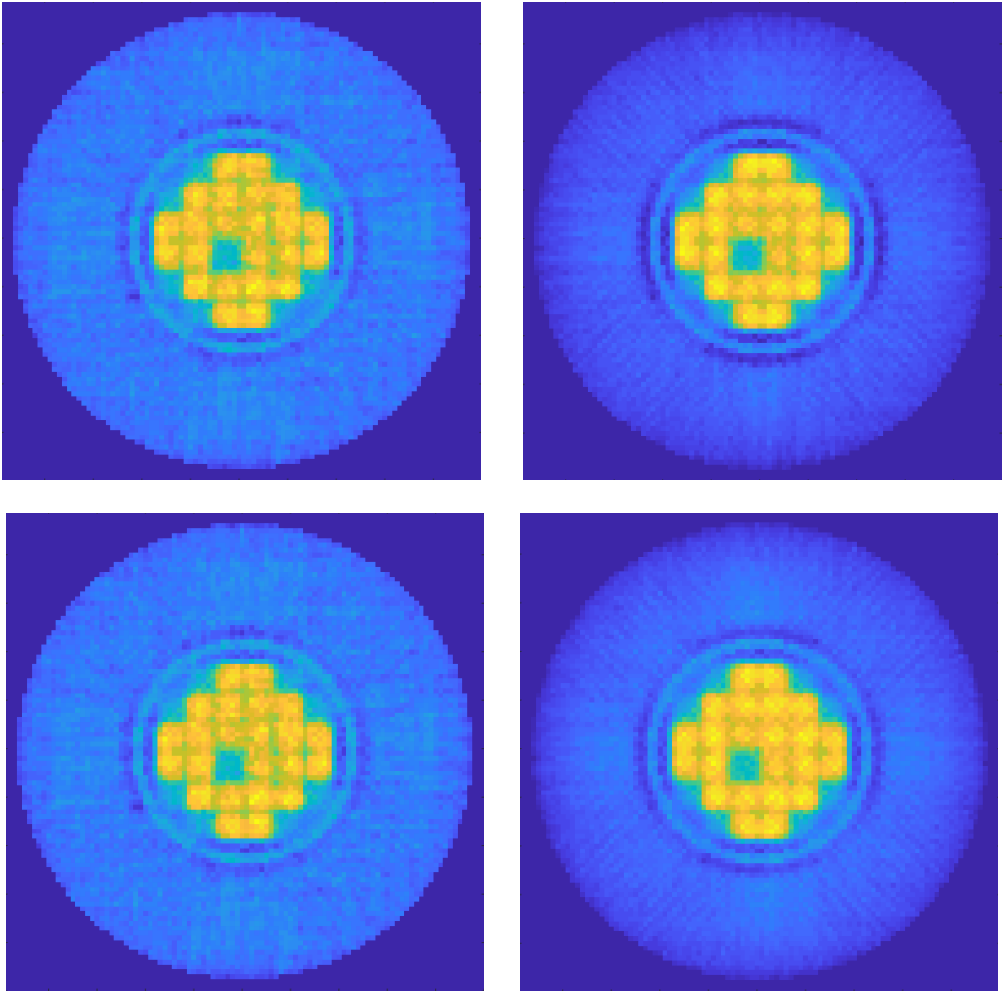

Figure 5.9 continued

Figure 5.10    Reconstructed images of dry storage cask using method **3b** with detection position uncertainty $\sigma = 1$ cm, 0.5 cm, 0.1 cm, 0.01cm and 0 cm (from top to bottom) without momentum (left column) and with momentum (right column).

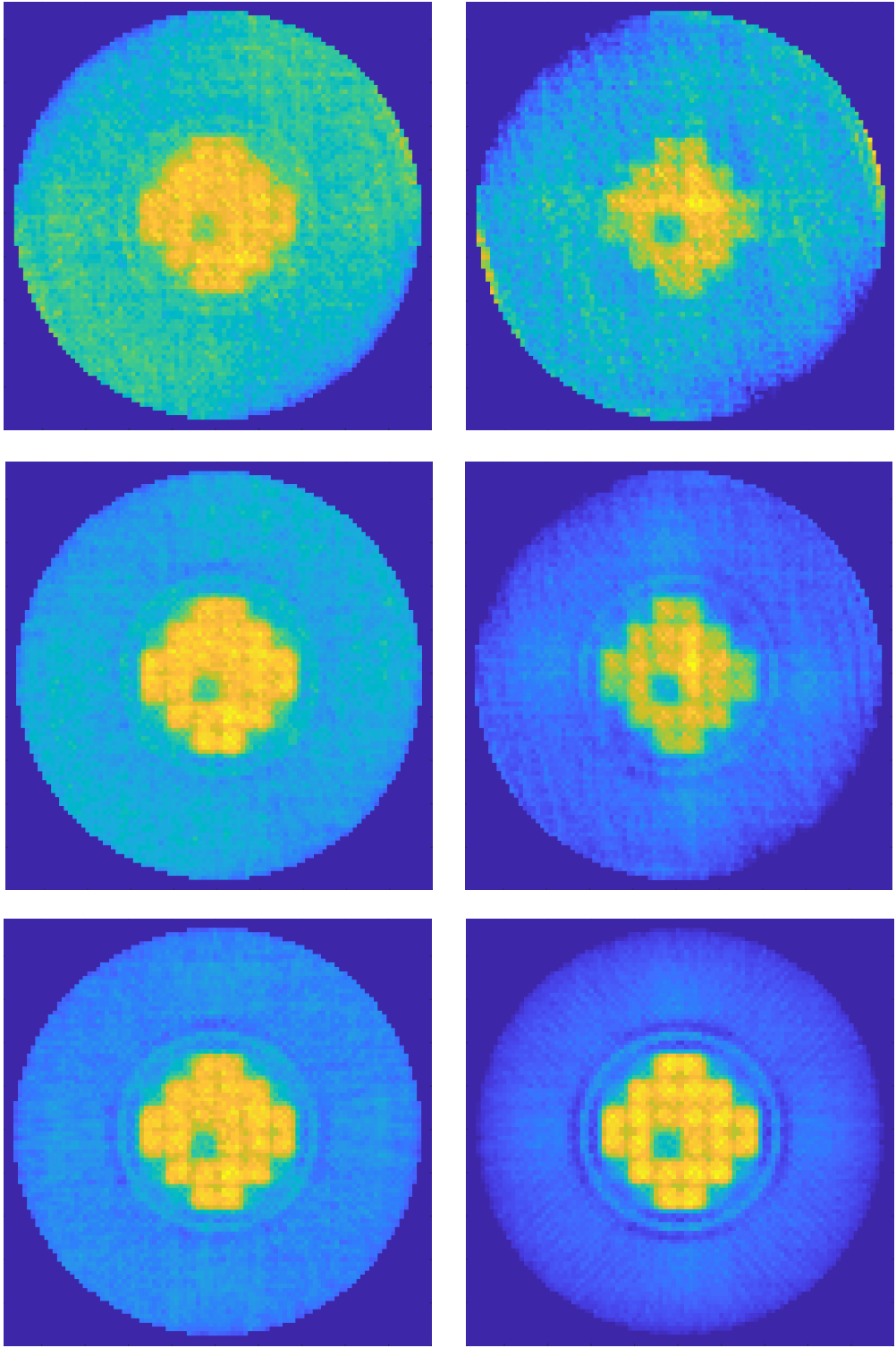

Figure 5.10 continued

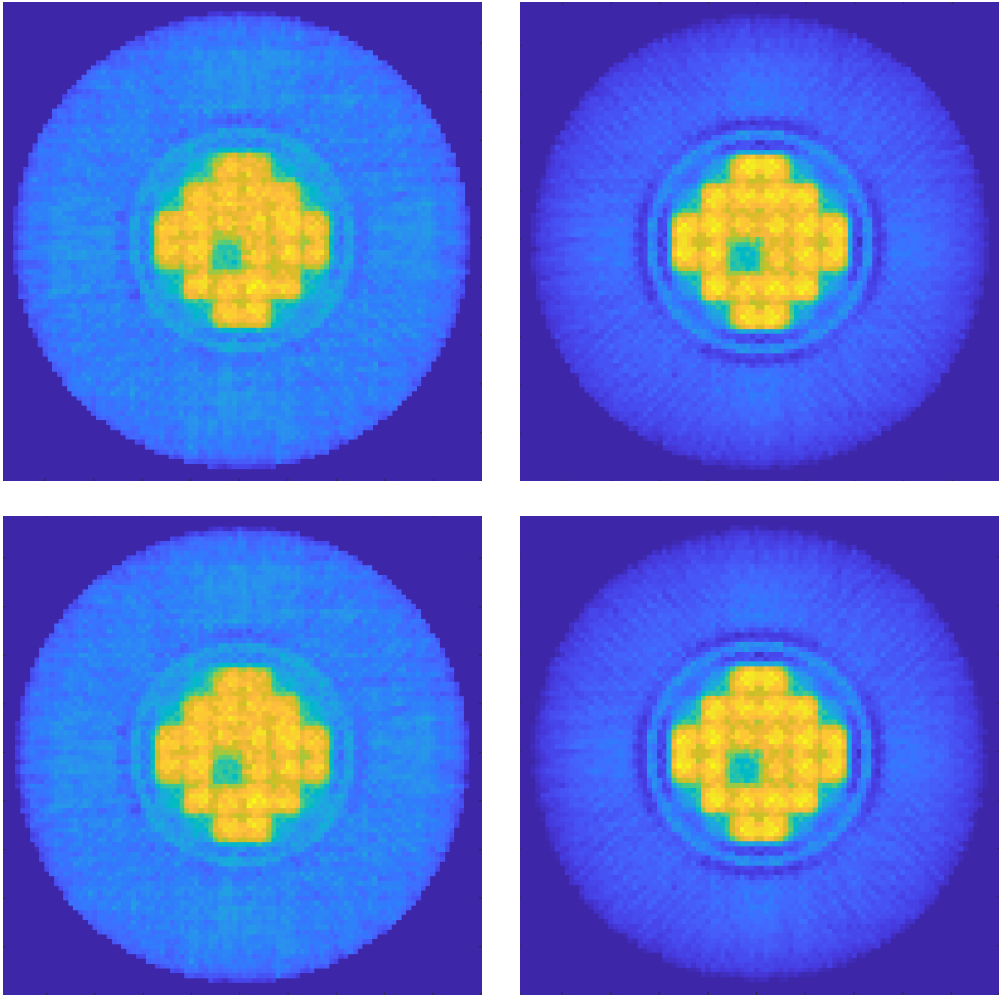

Figure 5.10 continued

Table 5.2        Expected image characteristics for different number of views

| Tracing algorithm | Measurement uncertainty (σ) cm | Without momentum | | | With momentum | | |
|---|---|---|---|---|---|---|---|
| | | SNR | CNR | DP | SNR | CNR | DP |
| | | SART | | | | | |
| **1b** | 1.0 | 13.38 | 3.56 | 47.64 | 8.60 | 2.47 | 21.28 |
| | 0.5 | 13.18 | 4.35 | 57.39 | 8.98 | 3.61 | 32.44 |
| | 0.1 | 13.21 | 5.20 | 68.76 | 13.03 | 5.50 | 71.70 |
| | 0.01 | 12.64 | 5.15 | 65.12 | 13.38 | 5.72 | 76.55 |
| | 0 | 12.57 | 5.10 | 64.13 | 13.42 | 5.74 | 77.06 |
| **3b** | 1.0 | 16.67 | 3.11 | 51.81 | 9.17 | 2.83 | 26.00 |
| | 0.5 | 16.53 | 4.04 | 66.79 | 10.01 | 3.75 | 37.54 |
| | 0.1 | 16.00 | 4.74 | 75.82 | 14.23 | 5.46 | 77.71 |
| | 0.01 | 15.73 | 4.96 | 78.04 | 14.39 | 5.56 | 79.98 |
| | 0 | 15.71 | 4.96 | 77.91 | 14.38 | 5.56 | 79.94 |

Even with a very coarse detector position uncertainty of 1 cm, method 1b and 3b are expected to yield images with a CNR larger than 3 without any muon momentum information, as shown at the top left in Figure 5.9 and Figure 5.10. This result seems to justify use of our plastic scintillator detector design, as described in section 2.3.2. However, with muon momentum they had worse CNR's. According to the previous simulation and reconstruction using perfect position resolution in section 4.4, using muon momentum information as a correction to scattering angle helps to improve reconstructed image quality, but this doesn't hold for poor position resolution, as shown here, especially when the position uncertainty is bigger than 0.5 cm. This may be attributed to inaccuracy of the muon path obtained with the detectors with a large uncertainty. Using muon momentum to correct the scattering angles along the incorrect muon path can be expected to further worsen the reconstructed image. However, with the improvement of position uncertainty, muon momentum information is gradually expected to favorably impact the reconstructed image quality.

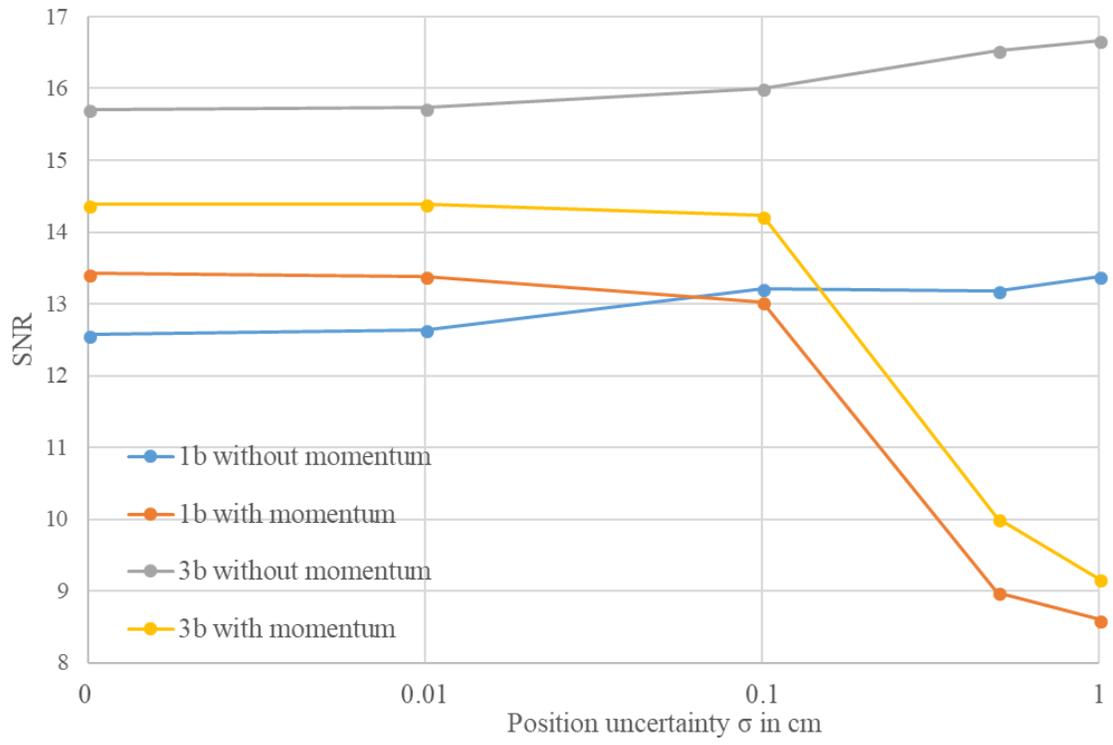

Figure 5.11        SNR vs position uncertainty

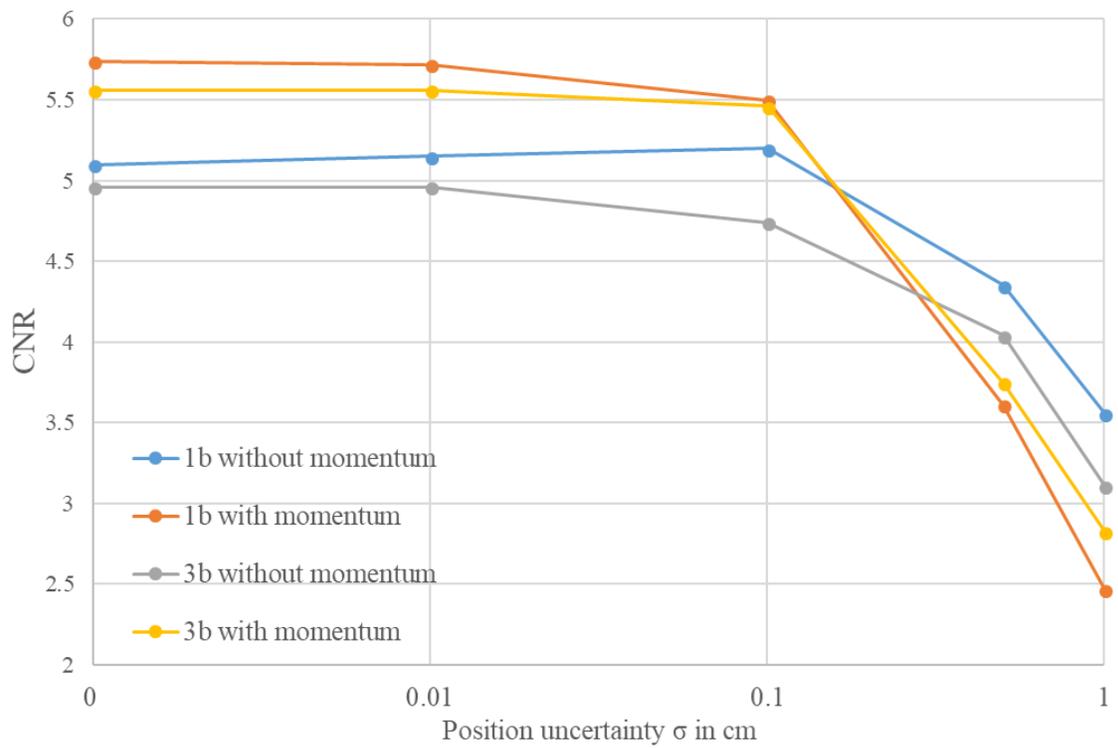

Figure 5.12        CNR vs position uncertainty

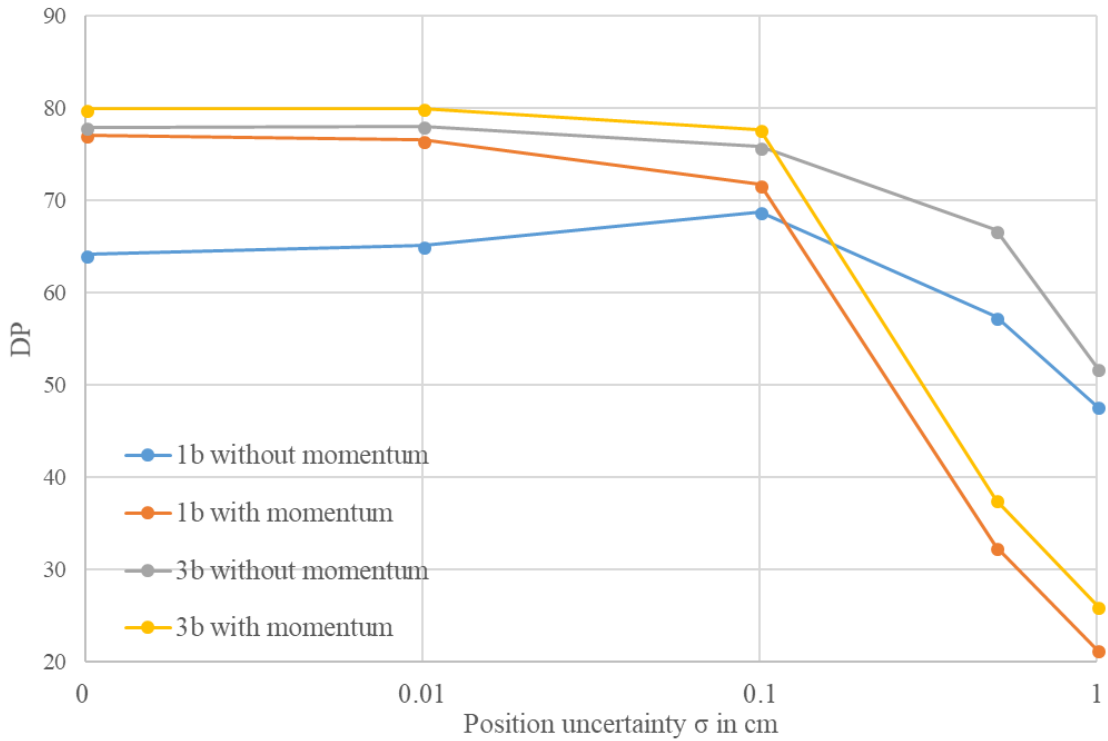

Figure 5.13        DP vs position uncertainty

The critical uncertainty is about 0.2 cm. It can be seen from Figure 5.12 that with the improvement of position uncertainty, CNR is expected to increase less than linearly; thus, the gain in reconstructed image quality may not be worth the extra cost of high resolution detectors, especially when the position uncertainty is smaller than 0.1 cm.

In conclusion, for poor detector position resolution, a straight path approach is expected to work better than the PoCA path and using muon momentum information to correct the scattering angle is expected to negatively impact the reconstructed image quality. However, with the improvement in detector position uncertainty, the PoCA path model gradually outperforms the straight path model; meanwhile, having muon momentum information positively impacts reconstructed image quality. In other words, method 1**b** without muon momentum can be expected to work best for poor resolution situations, while method 3**b** with momentum can be expected to achieve the best results if position resolution is high and muon momentum information is available.

## 5.4   Replacement with dummy material

A potential trickier scenario is where one or more spent nuclear fuel assemblies are removed and replaced with dummy material in order to appear identical, which may make the reconstructed image look similar to that of an intact dry storage cask with no spent nuclear fuel missing or replaced. This is especially true of transmission tomography using x-ray, neutron or even muons, because the transmission ratio is less sensitive to atomic number than the variance of the scattering angle caused by MCS, which has a stronger dependence on atomic number of the object it traversed. Thus, the situation where one spent nuclear assembly was stolen and replaced with a dummy assembly with different material as shown in at the top left Figure 5.14, was simulated to investigate the capability of our algorithms using muon scattering angle. As discussed in Chapter 4 method 3**b** along with SART is expected to work best among these methods discussed for idealized data, thus method 3**b** along with SART is used here to detect dummy assemblies. Similarly, the metrics SNR, CNR and DP are used to quantify the result. The exact same setup was adopted here to generate simulated data as in section 4.3. The simulated VSC-24 dry storage cask with one of middle assemblies replaced with iron or lead dummy assembly is shown in Figure 5.14 on the top. The reconstructed images of the dry storage cask with an iron or lead dummy assembly are shown in the middle and bottom rows in Figure 5.14. Reconstructed images without muon momentum as a correction are on the left and with muon momentum information are on the right Figure 5.14.

The SNR, CNR and DP of the reconstructed images without and with momentum for the scenario where one of the middle four assemblies was replaced with iron are expected to be 14.16, 3.73, 53.17 and 14.42, 3.80, 54.53. In both cases, either with or without momentum, the dummy Fe assembly is expected to be separated from its surrounding assemblies by about 3.7 σ. However, when the dummy material is lead, the SNR, CNR and DP of the reconstructed images without and with momentum for this scenario are 15.61, 0.338, 5.275 and 15.76, 0.019, 0.30. Due to the extremely low CNR 0.338, it is impossible to tell the lead assembly from its surrounding spent nuclear fuel assemblies. It is also expected that for the same setup, other dummy assemblies made up of tungsten cannot be differentiated either.

The nature of detecting a dummy assembly inside the dry storage cask is to differentiate materials represented by different radiation lengths with cosmic ray muon tomography technique. When the value of the radiation lengths is expressed in $g \cdot cm^{-2}$, radiation length (or scattering density) is monotonically decreasing with the increase of atomic number, which is the rationale used to differentiate substances. The radiation length of tungsten (Z=74), lead (Z=82) and uranium (Z=92) are 6.76, 6.37 and 6.00 $g \cdot cm^{-2}$, respectively. These three simple substances could be differentiated with the cosmic ray muon computed tomography technique described in this dissertation. However, when it comes to composite materials or compounds, it will be much harder (or perhaps unlikely) to tell two different material apart. For the calculation of radiation length of composite material and compound, see Appendix A. $UO_2$ has a radiation length of 6.65 $g \cdot cm^{-2}$, which is quite close to that of tungsten or lead metals. Besides, with the concrete cask and steel canister, the collapse of registered muon data down two 2D plane would make it even intractable to differentiate $UO_2$ from tungsten or lead metals.

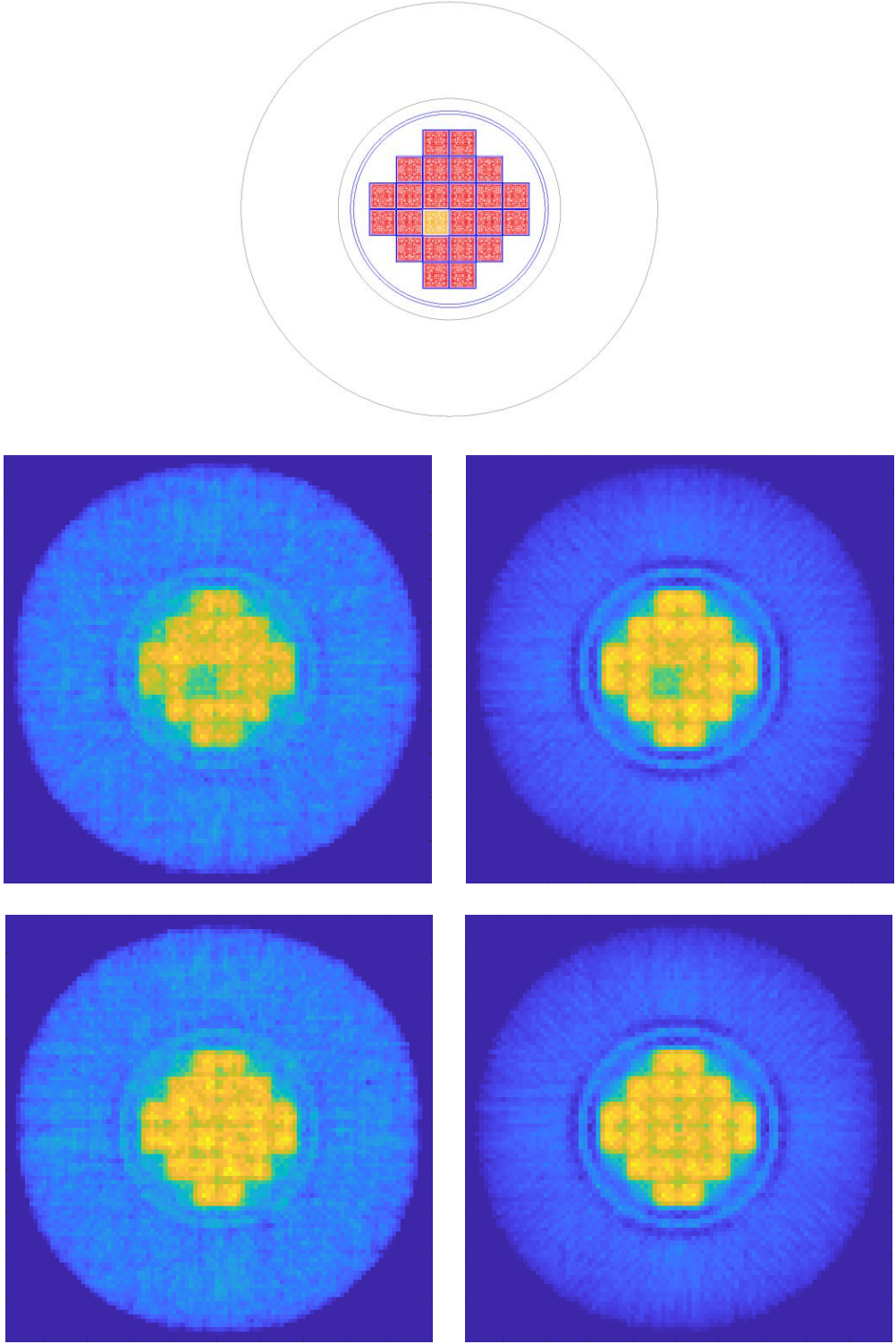

Figure 5.14    Simulated dry storage cask with one spent nuclear fuel assembly replaced with Fe or W dummy assembly (top row), reconstructed images with a Fe dummy assembly (middle row) and W dummy assembly (bottom row).  Reconstructed images in the left and right columns are without and with muon momentum, separately.

## 5.5   Measurement time calculation

The time needed to register a certain number of muons crossing four muon position-sensitive detectors and a spent nuclear fuel dry storage cask is dependent on detector size (height h and width w ), orientation $\vec{A}$, the solid angle of the muon detectors $\Omega$, and the distance d between the center of the dry storage cask and the center of muon detector. This is because the muon spectrum follows an angular distribution roughly proportional to $\cos^2(\theta)$ where $\theta$ is zenith angle. It reaches the maximum flux rate at vertical direction, i.e., when $\theta$ is at zero degree; unfortunately, these muons coming down from the vertical direction barely cross the two pairs of muon detectors vertically placed on the lateral side of the dry storage cask. One configuration of dry storage cask and muon detectors as shown in Figure 5.15 is used to help calculate the useful muon flux rate crossing four muon detectors.

The angular distribution of cosmic ray muons can be approximated by

$$\frac{dn}{d\Omega} = \frac{3}{\pi}\cos^2\theta \quad (\frac{muons}{min \cdot sr \cdot cm^2}) \tag{5.7}$$

And the detector angle can be calculated with

$$\Omega = \frac{\vec{A} \cdot \vec{n}}{d^2} = \frac{h \cdot w \cdot \sin\varphi}{d^2} \quad (sr) \tag{5.8}$$

The muon flux rate crossing these two pair of detectors is

$$N = \frac{dn}{d\Omega} \cdot \Omega \cdot (hw) \cdot (\vec{A} \cdot \vec{n}) = \frac{3}{\pi}\cos^2\theta \cdot \frac{(h \cdot w \cdot \sin\varphi)^2}{d^2} \quad (\frac{muons}{min}) \tag{5.9}$$

From this derived Eq. (5.9) when $\theta$ is equal to 0° and $\varphi$ is equal to 90°, the useful muon flux rate reaches its maximum flux rate, which matches the real situation.

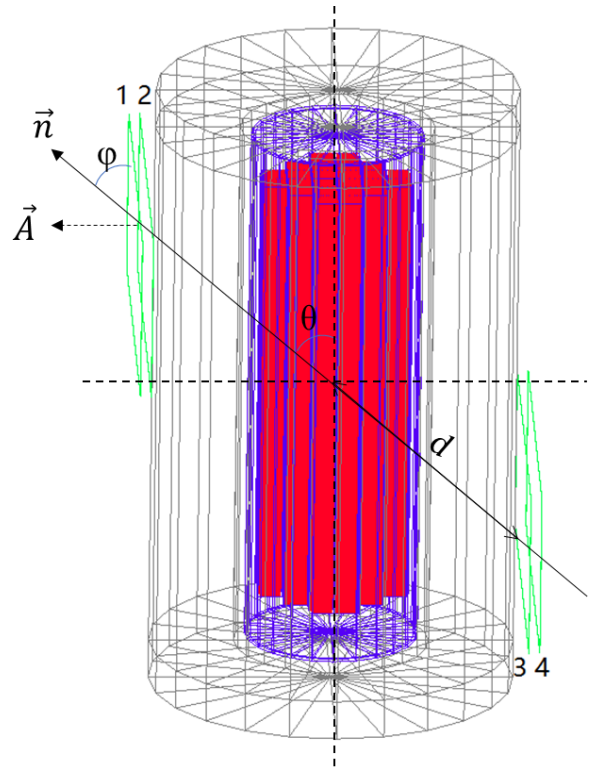

Figure 5.15    Detector geometry, zenith angle and detector plane vectors used for the calculation of measurement time.

The muon detectors used in Chapter 3 to detect one fuel assembly missing have a height of 150 cm and width of 350 cm. The distance d is 190.4 cm and both θ and φ are equal to 66.8 degrees. The expected time needed to register $10^6$ muons with various setup is shown in Table 5.3. To collect $1.7 \times 10^7$ muons with full size detectors and a set of vertical offsets of 150, 200 or 250 cm, it is expected to take 1.3, 1.04 or 0.98 days. Due to the spent nuclear fuel length of 366 cm, the total length of the vertical offset and detector height shall not exceed the spent nuclear fuel assembly length of 366 cm to maintain uniformity along the vertical direction. Otherwise, it is very likely to misclassify the partial spent nuclear fuel missing cask as a full cask. Because of this restriction, the maximum vertical offset that can be applied to the full-size muon detectors with height of 150 cm is about 200 cm.

Table 5.3　　　　Useful muon flux rate for various detector geometry and orientation

|  | Height cm | Width cm | Vertical offset cm | Zenith angle deg | Detector angle deg | Muons/min |
|---|---|---|---|---|---|---|
| Full size | 350 | 150 | 150 | 66.8 | 66.8 | 9518 |
|  |  |  | 200 | 60.3 | 60.3 | 12022 |
|  |  |  | 250 | 54.6 | 54.6 | 12731 |
| Small size | 160 | 120 | 150 | 66.8 | 66.8 | 1273 |
|  |  |  | 200 | 60.3 | 60.3 | 1608 |
|  |  |  | 250 | 54.6 | 54.6 | 1702 |

Thus, the shortest time needed to achieve to the results shown in Chapter 4 using $1.7 \times 10^7$ muons is about 1.04 days. This does not consider the number of muons rejected due to electronics considerations such as the coincidence time window or any dead time.

## 5.6　Summary

To make the simulated results closer to the reality, different engineering restrictions and optimization were considered in this chapter including view sampling, detector geometry, expected effects of measurement uncertainty, and replacement with dummy material.

In order to shorten the exposure time, the relation between reconstructed image quality and the number of views were investigated. A quantity of 5, 10, 18, 45 or 90 views were used to reconstruct the image with or without muon momentum information. It turned out that one needs 9 views with muon momentum information or 17 views without muon momentum information in order to be able to expect to detect a missing fuel assembly. Due to the central symmetry of the cask wall and its smaller scattering density compared to spent nuclear fuel inside the cask, a complete sinogram information of the whole cask is not a necessity to reconstruct the spent nuclear assemblies inside of the cask.  Instead of using full size detectors (3.5 m ×1.5 m), smaller detectors of 1.6 m×1.2 m were used in simulation to register about $3.3 \times 10^6$ muons for reconstruction. These smaller detectors are still expected to be able to enable one to separate the empty fuel slot from its surrounding fuel assemblies by 4.85σ, but at a cost of a 33.5% loss in SNR and a 13.4% loss in CNR compared with the larger detectors. Five different position uncertainties, 0 cm, 0.01 cm, 0.1 cm, 0.5 cm and 1 cm were incorporated into the reconstruction process. Overall, method 1**b** is expected to achieve better CNR and method 3**b** is expected to achieve better CNR. When position resolution is poorer (i.e., closer to 1 cm), a straight path model is expected to work better than use of the PoCA path and using muon momentum to correct the scattering angle only harms the reconstructed image quality. However, with the improvement of position uncertainty, the PoCA path model gradually outperforms the straight path model; meanwhile, muon momentum favorably impacts reconstructed image quality. Finally, trickier scenarios where one spent nuclear fuel assembly was stolen and replaced with dummy assemblies were considered. Use of method 3b is expected to enable one to tell an iron dummy assembly from its surrounding spent nuclear assemblies, either with or without perfect muon information available, by about 3.7σ. Even so, the developed methods cannot be expected to differentiate dummy assemblies made of lead or tungsten from real spent nuclear fuel assemblies due to their similar radiation lengths.

# 6   CONCLUSIONS AND FUTURE WORK

## 6.1   Summary of results

In traditional X-ray computed tomography (CT), the negative natural logarithm of the transmission rate of photons is equal to the linear integral of the attenuation coefficient of the objects crossed by the X-rays. Similarly, the variance of muon scattering angles is also equal to a linear integral of the scattering density of the objects along the track of muons. Due to the same form, the X-ray computed tomography technique can be transferred to muon imaging. Thus, a new imaging modality, muon computed tomography, which uses the variance of muon scattering angle, has been born. However, muon CT faces some unique challenges including: 1) long measurement time due to low cosmic ray muon flux rate, 2) more accurate muon path models are needed, and 3) the difficulty in precisely measuring muon momentum or lowing the reliance of reconstructed image quality on muon momentum.

In this work, three different muon path models, 1) a straight path along the muon incident trajectory, 2) a straight path along the muon incident direction crossing the PoCA point and the so-called 3) PoCA path; along with two different projection methods (called projection methods **a** and **b**); and two different reconstruction methods (FBP and SART) were investigated for use in muon CT imaging spent nuclear fuel dry storage casks simulated in a validated Geant4 workspace, both in the ideal case and with relevant engineering restrictions considered. The work started with imaging a VSC-24 spent nuclear fuel dry storage cask with one fuel assembly missing in the middle using idealized GEANT4 simulated data. With or without muon momentum information, all methods generated clear reconstructed images of the dry cask, and method **3b** yielded the best expected result. When perfect muon momentum information is available, using projection method b, CNR, SNR and detection capability can be expected to be improved by 17.1%, 8.4% and 27.1% compared to the case where... When muon momentum information is not available, using projection method b, CNR, SNR and detection capability can be expected to be significantly improved by 45.8%, 25.6% and 83.6% compared to the case.... It is also expected that the SART image reconstruction method is able to achieve about twice the performance compared with the FBP method. To further investigate the detection capability of method **3b**, a scenario where a dry cask with quarter assemblies and half assemblies missing was simulated in GEANT4 to generate data for reconstruction. This method is still expected to detect the quarter fuel assemblies missing, either with or without muon momentum information. Even in the case where no muon momentum is obtained, the separation is 5.8 σ.

Later, engineering restrictions including muon detector position measurement uncertainty, reduced measurement time, and using smaller detectors were simulated as well. Five different detector position uncertainties ($\sigma = 0$ cm, 0.01 cm, 0.1 cm, 0.5 cm and 1 cm) were incorporated into the reconstruction process, and it turned out that when detector position uncertainty is large, using the muon momentum to correct the estimated muon scattering angle is expected to disserve the reconstructed image quality. Meanwhile, when detector position uncertainty is large, use of a straight path along the muon incident trajectory is expected to work better than use of the so-called PoCA trajectory. With the improvement of detector position uncertainty, muon momentum information is expected to favorably impact the reconstructed image quality and the PoCA trajectory model is expected to outperform the straight path model. Even when using muon detectors with a positional measurement uncertainty of 1 cm and without any muon momentum information, one can expect to be able to detect a single missing fuel assembly in a VCS-24 dry storage cask within one day exposure time. If the goal is to detect one or more whole assemblies missing, 9 views with muon momentum information or 18 views without muon momentum information are expected to suffice, which corresponds to 2 to 4 hours of exposure, respectively.

Lastly, more intricate cases in which some assemblies were stolen and replaced with dummy material were simulated. Scenarios where one of the middle four spent nuclear assemblies was replaced with a dummy assembly made of steel or lead or tungsten were simulated in GEANT4 and reconstructed with method 3**b**. A steel dummy assembly is expected to be easily distinguished from the surrounding real spent nuclear fuel assemblies in a dry storage cask by about 3.7 σ, but not lead or tungsten dummy assemblies because their radiation lengths are very close to that of the spent nuclear fuel.

## 6.2   Future research

### 6.2.1   The influence of muon momentum on its scattering angle

For muons crossing an object, muons with larger momentum trend to scatter less than muons with smaller momentum and consequently they will have a relatively smaller variance of scattering angle. It is because that muons with larger momentum travel faster and spend less time in the vicinity of target nucleus, thus they experience less multiple Coulomb scattering events, which leads to a smaller scattering angle. Thus, it would be necessary to correct the scattering angles with corresponding muon momentum. Previous methods used to deal with the heterogeneity of muon momentum are described in APPENDIX A. Although there is no precise mathematical relation available between the value of muon momentum and its scattering angle and distribution, it is clear that muon momentum would affect the magnitude and distribution of muon scattering angle after traversing an object. A phenomenologically based formula Eq. (6.5) may be used to correct the influence of heterogeneity of muon momentum on scattering angle and variance when muon momentum is available:

$$\theta' = \frac{p}{p_0} \theta \qquad\qquad (6.1)$$

where $p$ and $\theta$ are the momentum and scattering angle of the muon, and $p_0$ is a nominal momentum usually chosen to be 3 GeV/c.

In this dissertation, Eq. (6.1) has been used as a correction for scattering angle and it has achieved significant improvement in reconstructed image quality which can be seen from the comparison of Figure 4.8 with Figure 4.9. Eq. (6.1) may be further improved by delving the multiple Coulomb scattering theory or by Monte Carlo simulation, which is left for future work.

### 6.2.2   Use of muon most probably trajectory

With idealized simulated data, algorithms using PoCA trajectory achieved a better result than the same algorithms using a straight path, either the straight path along muon incident trajectory or a straight path along muon incident direction crossing PoCA point. The difference in performance between these trajectory models is expected to be even larger when it comes to fully 3D muon CT or fully 3D muon tomography as described in the following section. The muon most probable trajectory model is presented in section 2.4. Due to the collapse of 3D data down to a 2D plane, the gain of improved trajectory model reflected in reconstructed image is limited. Besides, the incorporation of the most probable trajectory in the reconstruction process can significant increase computation burden. However, it is expected to be worth when a fully 3D reconstruction is carried out.

### 6.2.3 Fully 3D muon CT

Cosmic ray muons are unlike manmade X-rays beam or neutron beam found in traditional transmission based topographical applications, which are usually found in a certainty more predictable shapes, like a parallel beam, a fan beam or a cone beam, or at least they can be tuned to one of these shapes by applying a certain collimator. However, muons are the decay product of pions which are created by cosmic rays interacting with gaseous atoms in the upper atmosphere and they shower on the earth from 0 to 90 degree zenith angle, almost isotopic in horizontal directions. Besides, with cosmic ray muons it is almost impossible to apply a collimator, like those found in X-ray or slow neutron tomography, to generate a certain shape of beam due to its extremely high energy. In this dissertation, even though muon detectors were placed on the lateral side of a dry storage cask with a vertical offset and rotated around the cask to register muons crossing the image volume enveloped by muon detectors. the reconstructed images are 2 dimensional. Because the scattering angles of one or two muons are insufficient to represent the distribution of muon scattering angles after crossing a same object along the same path, thus all registered muons were collapsed down to a horizontal plane with some corrections to form quasi-parallel beam after re-sortting. This method worked well in this dissertation due to the uniformity along vertical direction of the object under interrogation. Otherwise, artifacts would be expected to be seen in the reconstructed images. For these scenarios where the object under interrogation is not uniform along vertical direction, a simple PoCA method and statistical reconstruction [58] with or without muon momentum may be used to create fully 3-dimensional tomographic image of the objects under interrogation. But they are very likely to yield a poor resolution compared to the resolution obtained in Chapter 4. Simple PoCA based imaging reconstruction is illustrated in Figure 6.1.

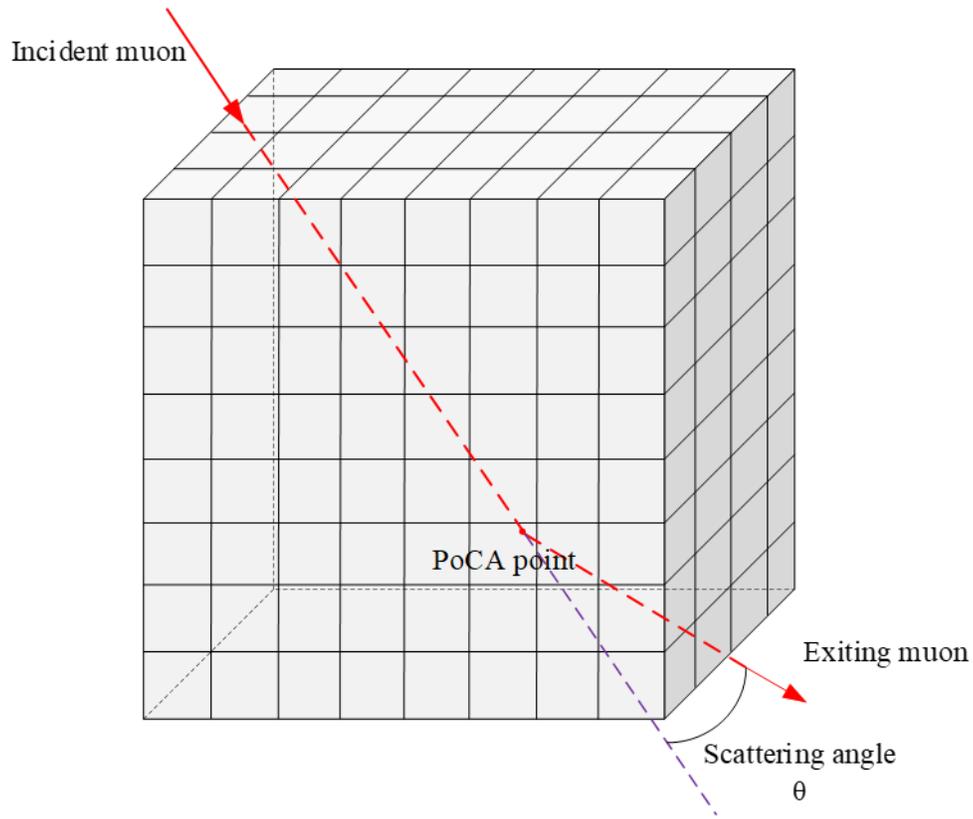

Figure 6.1　　　Illustration of PoCA based 3D muon imaging technique.

Forward extrapolate muon incident trajectory and backward extrapolate muon exiting trajectory to find the PoCA point, if these two extending lines intersects in 3-dimensional space, then the intersection is PoCA point, otherwise the middle point of the common perpendicular line to these two extending line will be defined as PoCA point. The traditional treatment is to discretize the image reconstruction volume into voxels and store the scattering angle of each muon from one projection in the voxel where its PoCA point is located, at last calculate the variance of scattering angles in each voxel to yield a 3-dimensional image. Since it was already shown that by back projecting a muon's scattering angle into each pixel or voxel crossed by its trajectory then calculating the variance of the scattering angle in each pixel or voxel, the reconstructed image quality is expected to be significantly improved and have a weaker reliance on muon momentum information. Thus, one way to improve the traditional simple PoCA imaging method is adopting this projection method, i.e., putting each scattering angle into the voxels crossed by it PoCA trajectory, which is very likely to achieve the same improvements as demonstrated in dissertation on muon computed tomography.

### 6.2.4   Regularization methods

Algebraic based reconstruction techniques could be more powerful than transform based technique (like FBP) when the object is under sampled or the projections are not evenly taken over 180 degrees, especially when the energy propagation paths between incident point and exiting points are subject to bending due to reflection or scatter, which is the case in muon imaging. The nature of algebraic based reconstruction techniques is to treat each narrow beam as a numerical equation then solve these equations. Typical regularization methods for algebraic based reconstruction techniques include Tikhonov regularization, total variation regularization, L1 and L2 regularization. They are expected to yield less noisy reconstructed images with clearer boundaries. Given the core of this dissertation to develop muon computed tomography framework, incorporation of regularization is left for future research.

# LIST OF REFERENCES

# APPENDICES

# APPENDIX A    CORRECTIONS FOR POLYENERGETIC MUON

For a monoenergetic muon beam crossing an object, the scattering angles follow a gaussian distribution with zero mean value and a standard deviation as shown in Eq. (6.1) and Eq. (6.2).

$$f_\theta(\theta) \cong \frac{1}{\sqrt{2\pi}\sigma_\theta} e^{-\frac{\theta^2}{2\sigma_\theta^2}} \qquad (A.1)$$

$$\sigma_\theta \cong \frac{15}{\beta c p} \sqrt{\frac{L}{L_{rad}}} \qquad (A.2)$$

Three different ways may be used to deal with heterogeneity of muon momentum:

The first way, also the coarsest way, is to indiscriminately treat the heterogeneous muon as monoenergetic, however, this would blur the reconstructed images as have been shown in Figure 4.7 and make it impossible to accurately identify materials although it may still be able to differentiate low Z and high Z materials.

Secondly, for polyenergetic muon beam, the muon scattering angles in each group within a certainty range of energy or momentum follow a gaussian distributions with zero mean value and different variances. For the $i^{th}$ group of muons, the variance can be expressed as [59]:

$$\sigma_{\theta i} \cong \frac{15}{\beta c p_i} \sqrt{\frac{L}{L_{rad}}} \qquad (A.3)$$

Unfortunately, the resulting summation of these Gaussian distribution is not separable because they have the same mean value. Otherwise, the measurement muon scattering angles could be divided into different energy group, then use mean energy in each group to correct corresponding scattering angles.

Thirdly, to alleviate the problem caused by the absence of muon momentum, one can place an a known object in the image reconstruction volume as points of reference as shown in Eq. (6.4) [60]

$$\lambda_{cal}{}^{x} = \lambda_{meas}{}^{x} \frac{\lambda_{pred}{}^{ref}}{\lambda_{meas}{}^{ref}} = \lambda_{meas}{}^{x} C_{F} \qquad (A.4)$$

where $\lambda_{meas}$ and $\lambda_{meas}{}^{ref}$ are the uncalibrated scattering densities of the test sample and a reference object, $\lambda_{pred}{}^{ref}$ is the predicted theoretical scattering density of the reference object, and $C_{F} = \frac{\lambda_{pred}{}^{ref}}{\lambda_{meas}{}^{ref}}$ is the calibration factor. This still yields a coarse result unless the scattering density of the reference object is similar to the sample object, which is not realistic because in reality in most cases the objects under interrogation are unknown. It may be useful to detect known objects, for example uranium metal in cargo.

# APPENDIX B    SCATTERING DENSITY OF COMPOUND

For simple substance, the radiation length can be calculated with the following formula

$$X_0 = \frac{716.4A}{Z(Z+1)\ln\frac{287}{\sqrt{Z}}} \quad g \cdot cm^{-2} \tag{B.1}$$

It is often expressed in $g \cdot cm^{-2}$ to eradicate the dependence on physical status, i.e., mass density. A comparison of experimentally measured and estimated radiation length for a series simple substance used in this dissertation is shown in Table B.1.

For composite material, the radiation length can be approximated with [61]

$$\frac{W_0}{X_0} = \sum_{i=1}^{n} \frac{W_i}{X_i} \tag{B.2}$$

where $W_0$ is the total mass of the composite material in gram, $X_0$ is the estimated radiation length of the object in $g \cdot cm^{-2}$, $W_i$ and $X_i$ are the mass and radiation length of $i^{th}$ component in gram and $g \cdot cm^{-2}$, and n is the number of components.

Table B.1    Experimentally measured vs analytically calculated radiation length for different simple substances

| Simple substance | Z | A (g/mol) | Measured Radiation length (g/cm²) | Estimated Radiation length (g/cm²) | Error (%) |
|---|---|---|---|---|---|
| O | 8 | 15.99 | 34.24 | 34.46 | 0.64 |
| Al | 13 | 26.98 | 24.01 | 24.26 | 1.06 |
| Fe | 26 | 55.85 | 13.84 | 14.14 | 2.17 |
| Cu | 29 | 63.55 | 12.86 | 13.16 | 2.34 |
| W | 74 | 183.84 | 6.76 | 6.77 | 0.08 |
| Pb | 82 | 207.2 | 6.37 | 6.31 | 0.93 |
| U | 92 | 238.03 | 6.00 | 5.86 | 2.26 |

For compound, the radiation length can be approximated with

$$\frac{A_0 N_0}{X_0} = \sum_{i=1}^{n} \frac{A_i N_i}{X_i} \tag{B.3}$$

where $A_0$ and $N_0$ are the average atomic mass and total number of atoms of the compound, $A_i$ and $N_i$ are the atomic mass and number of atoms of $i^{th}$ element in the compound

For example, the radiation length of $UO_2$ is calculated as:

$$N_0 = 1 + 2 = 3 \text{ moles} \tag{B.4}$$

$$A_0 = (238.03 \times 1 + 15.99 \times 2)/3 = 90.003 \ g/mol \tag{B.5}$$

With the equation for compound:

$$\frac{90.003 \times 3}{X_0} = \frac{238.03 \times 1}{6.00} + \frac{15.99 \times 2}{34.24} \tag{B.6}$$

$$X_0 = 6.6496 \ g/cm^{-2} \tag{B.7}$$

The experimentally measured radiation length of UO2 is 6.65 $g/cm^{-2}$ and the error is 0.007%

# VITA

Zhenzhi Liu acquired his bachelor degree in nuclear engineering and technology field from Harbin Engineering University, China. He graduated as one of the top 3 percent best undergraduate students. Instead of being content with the status quo, Zhengzhi chose to take the challenges of going to a foreign country to pursue higher education. Soon, he matriculated at the University of Tennessee. After joining Prof. Hayward's research group, he focused on a novel and meaningful research on monitoring of used fuel storage containers with cosmic ray muon in partnership with Oregon State University. On his way to earn a Ph.D. degree, he received a concurrent master degree in science. During graduate study, he did intern at Oregon State University and Oak Ridge National Laboratory. Finally, he received his Ph.D. in August 2018.